\documentclass[12pt,a4paper]{article}%
\usepackage{amssymb,enumitem,booktabs,subfig,xcolor,todonotes,microtype,setspace}
\usepackage[top=1.4in, bottom=1.4in, left=1.35in, right=1.35in]{geometry}
\usepackage{natbib}
\usepackage{url}
\usepackage[pdftex,colorlinks=true]{hyperref}
\usepackage{amsmath}
\usepackage{amsfonts}
\usepackage{epsfig}
\usepackage{graphics}
\usepackage{multirow}
\usepackage{graphicx}
\usepackage{amssymb}
\usepackage{palatino,eulervm}
\usepackage{rotating}
\usepackage{dcolumn}
\usepackage{amsthm,thmtools}%
\newcolumntype{L}{D{.}{.}{2,3}}
\definecolor{darkblue}{rgb}{0,0,.6}
\hypersetup{citecolor=darkblue,linkcolor=darkblue,urlcolor=darkblue}
\DeclareMathOperator*{\argmin}{arg\,min}

\providecommand{\U}[1]{\protect\rule{.1in}{.1in}}

\graphicspath{{plots/}}

\allowdisplaybreaks[4]
\declaretheorem{theorem}
\declaretheorem{lemma}
\makeatletter
\def\th@newremark{\th@remark\thm@headfont{\bfseries}}
\makeatletter
\theoremstyle{newremark}

\newtheorem{prop}{Proposition}
\newtheorem{corollary}{Corollary}
\newtheorem{assumption}{Assumption}
\declaretheoremstyle[
  spaceabove=6pt, spacebelow=6pt,
  headfont=\bfseries,
  notefont=\mdseries, notebraces={(}{)},
bodyfont=\normalfont,
  postheadspace=0.5em,
  ]{mystyle}

\begin{document}

\title{Estimating Time-Varying Networks for High-Dimensional Time Series}
\author{{\normalsize Jia Chen\thanks{Department of Economics and Related Studies,
University of York, UK. Jia Chen's research was partially supported by the
ESRC (Grant Reference: ES/T01573X/1).},\ \ \ Degui Li\thanks{Department of
Mathematics, University of York, UK. },\ \ \ Yuning Li\thanks{Department of
Economics and Related Studies, University of York, UK.},\ \ \ Oliver
Linton\thanks{Faculty of Economics, University of Cambridge, Cambridge, UK. }}\\
{\normalsize\em University of York and University of Cambridge}}
\date{{\normalsize Version: \today}}
\maketitle

\centerline{\bf Abstract}

We explore time-varying networks for high-dimensional locally stationary time
series, using the large VAR model framework with both the transition and
(error) precision matrices evolving smoothly over time. Two types of
time-varying graphs are investigated: one containing directed edges of Granger
causality linkages, and the other containing undirected edges of partial
correlation linkages. Under the sparse structural assumption, we propose a
penalised local linear method with time-varying weighted group LASSO to
jointly estimate the transition matrices and identify their significant
entries, and a time-varying CLIME method to estimate the precision matrices.
The estimated transition and precision matrices are then used to determine the
time-varying network structures. Under some mild conditions, we derive the
theoretical properties of the proposed estimates including the consistency and
oracle properties. In addition, we extend the methodology and theory to cover
highly-correlated large-scale time series, for which the sparsity assumption
becomes invalid and we allow for common factors before estimating the
factor-adjusted time-varying networks. We provide extensive simulation studies
and an empirical application to a large U.S. macroeconomic dataset to
illustrate the finite-sample performance of our methods.

\bigskip



\newpage


\section{Introduction}

\label{sec1} \renewcommand{\theequation}{1.\arabic{equation}} \setcounter{equation}{0}

In recent years, the network analysis has become an effective tool to explore
inter-connections among a large number of variables, with applications to
various disciplines such as: epidemiology, economics, finance, and social
networks \citep[e.g.,][]{N02, BKT13, DY14, DY15, HSS14, S17, BB19, ZCLW19}.
The so-called graphical model is commonly used in the network analysis to
visualise the connectedness of a large panel with vertices representing
variables in the panel and the presence of an edge indicating appropriate
(conditional) dependence between the variables. In the past decades, most of
the existing literature on statistical estimation and inference of network
data limits attention to the \emph{static} network, which is assumed to be
invariant over time \citep[e.g.,][]{YL07, FFW09, LW13, BSM15, ZLWL22}.
However, such an assumption may be too restrictive and often fails in
practical applications where the underlying data generating mechanism is
dynamic. There have been some attempts in the recent literature to relax the
static network assumption, allowing the connectivity structure to exhibit
time-varying features. For example, \cite{KSAX10} and \cite{ZLW10} study
dynamic network models with smooth time-varying structural changes; whereas
\cite{WYR21} consider change-point detection and estimation in dynamic
networks. However, most of the aforementioned literature typically assumes
that the network data are independent, which often becomes invalid in
practice. We aim to relax this restrictive assumption and model large-scale
network data under a general temporal dependence structure.

Vector autoregression (VAR) is a fundamental modelling tool for multivariate
time series data \citep[e.g.,][]{Lu06}. In recent years, there has been
increasing interest in extending the finite-dimensional VAR to the
high-dimensional setting. Under appropriate sparsity restrictions on the
transition (or autoregressive coefficient) matrices, various regularised
methods have been proposed to estimate high-dimensional VAR models and
identify non-zero entries in the transition matrices
\citep[e.g.,][]{BM15, HLL15, KC15, DZZ16}. \cite{ZPLLW17} introduce a network
VAR model by incorporating the adjacency matrix to capture the network effect
and estimate the model via ordinary least squares. More recently, \cite{CFZ20}
and \cite{MPS22} further study high-dimensional VAR and network VAR with
latent common factors, allowing strong cross-sectional dependence in large
panel time series. The methodology and theory developed in these papers
heavily rely on the stationarity assumption with both transition and
volatility matrices being time-invariant.

The stable VAR model cannot capture smooth structural changes and breaks in
the underlying data generating process, two typical dynamic features in time
series data collected over a long time span. To address this problem,
\cite{DQC17} consider a time-varying VAR model for high-dimensional time
series (allowing the number of variables to diverge at a sub-exponential rate
of the sample size), and estimate the time-varying transition matrices by
combining the kernel smoothing with $\ell_{1}$-regularisation, whereas
\cite{SS22} simultaneously detect breaks and estimate transition matrices in
high-dimensional VAR via a three-stage procedure using the total variation
penalty. \cite{XCW20} detect structural breaks and estimate smooth changes
(between breaks) in the covariance and precision matrices of high-dimensional
time series (covering VAR as a special case). In the present paper, we aim to
jointly estimate the time-varying transition and precision matrices in the
high-dimensional sparse VAR under the local stationarity framework. Motivated
by the stable network time series analysis in \cite{BB19}, we use the
estimated transition and precision matrices to further construct two
time-varying networks: one containing directed edges of Granger causality
linkages, and the other containing undirected edges of partial correlation
linkages.

The proposed time-varying network via VAR is naturally connected to the
locally stationary models, which have been systematically studied in the
literature for low-dimensional time series. \cite{D97} is among the first to
introduce a locally stationary time series model via a time-varying spectral
representation. \cite{DS06} study a time-varying ARCH model and propose a
kernel-weighted quasi-maximum likelihood estimation method. \cite{HL10}
further consider a time-varying version of GARCH model and introduce a
semiparametric method to estimate both the parametric and nonparametric
components involved. \cite{Vo12} and \cite{ZW12} study nonparametric
kernel-based estimation and inference in a general class of locally stationary
time series. \cite{KL12} extend the locally stationary model framework to the
diffusion process. \cite{YGP20} develop a kernel estimation method and theory
for time-varying vector moving average models. The present paper complements
the locally stationary time series literature by further exploring the
high-dimensional dynamic network structure.

We study the time-varying VAR and network models for large-scale time series,
allowing the number of variables to be much larger than the time series
length. Under the sparsity assumption on the transition and precision matrices
with smooth structural changes, we introduce a three-stage estimation
procedure: (i) preliminary local linear estimation of the transition matrices
and their derivatives with time-varying LASSO; (ii) joint local linear
estimation and feature selection of the time-varying transition matrices with
weighted group LASSO; (iii) estimation of the precision matrix via
time-varying CLIME. To guarantee the oracle property, the weights of LASSO in
the second estimation stage are constructed via a local linear approximation
to the SCAD penalty \citep[e.g.,][]{ZL08} using the consistent preliminary
estimates obtained in the first stage. Our penalised estimation methodology
for the time-varying transition matrices is connected to various nonparametric
screening and shrinkage methods developed for high-dimensional
functional-coefficient models \citep[e.g.,][]{WX09, L12, FMD14, LLW14, LKZ15},
whereas the time-varying CLIME is a natural extension of the conventional
CLIME for static precision matrix estimation \citep[e.g.,][]{CLL11}. The
theoretical properties of the techniques developed in the aforementioned
literature (such as the oracle property and minimax optimal convergence rates)
rely on the independent data assumption. Extension of the methodology and
theory to the high-dimensional locally stationary time series is non-trivial,
requiring new technical tools such as the concentration inequality for
time-varying VAR. Under some regularity conditions, we show that the proposed
local linear estimates with weighted group LASSO equal to the infeasible
oracle estimates with prior information on the significant entries of
time-varying transition matrices, and the precision matrix estimate with
time-varying CLIME is uniformly consistent with sensible convergence rates
under various matrix norms. The estimated transition matrices are used to
consistently estimate the uniform network structure with directed Granger
causality linkages, whereas the estimated precision matrix is used to
construct the network structure with undirected partial correlation linkages.

We further consider highly-correlated large-scale time series, for which the
sparsity model assumption is no longer valid in which case the methodology and
theory need to be substantially modified. The approximate factor model
\citep[e.g.,][]{CR83} or its time-varying version \citep[e.g.,][]{SW17} is
employed to accommodate the strong cross-sectional dependence among a large
number of time series. In particular, we assume that the high-dimensional
idiosyncratic error process in the approximate factor model satisfies the
time-varying VAR structure with the sparsity restriction imposed on its
transition and precision matrices. The latent common and idiosyncratic
components need to be estimated consistently. With the approximated
idiosyncratic error vectors, the penalised local linear estimation method with
weighted group LASSO and time-varying CLIME are applied to estimate the
time-varying transition and precision matrices. Subsequently, the
factor-adjusted time-varying network estimates with directed Granger causality
and undirected partial correlation linkages are obtained. Our paper thus
substantially extends the recent work on the factor-adjusted stable VAR model
estimation \citep[e.g.,][]{FMM21, BCO22, KM22}.

Our simulation studies demonstrate that the proposed methodology can
accurately estimate the time-varying Granger and partial correlation networks
when the number of time series variables is comparable to the sample size. In
particular, for the time-varying transition matrix estimation, the penalised
local linear method with weighted group LASSO outperforms the conventional
local linear method (which often fails in the high-dimensional time series
setting) and produces numerical results similar to those of the oracle
estimation. For the time-varying error precision matrix estimation, the
numerical performance of the proposed time-varying CLIME is comparable to that
of the time-varying graphical LASSO. We further apply the developed
methodology to the FRED-MD macroeconomic dataset and estimate both the Granger
causality and partial correlation networks via the proposed time-varying VAR model.

The rest of the paper is organised as follows. Section \ref{sec2} introduces
the time-varying VAR and network model structures. Section \ref{sec3} presents
the estimation procedures for the time-varying transition and precision
matrices and Section \ref{sec4} gives the asymptotic properties of the
developed estimates. Section \ref{sec5} considers the factor-adjusted
time-varying VAR model and network estimation. Sections \ref{sec6} and
\ref{sec7} report simulation studies and an empirical application,
respectively. Section \ref{sec8} concludes the paper. A supplemental document
contains proofs of the main theorems, some technical lemmas with proofs,
verification of a key assumption and discussions on tuning parameter
selection. Throughout the paper, we let $\vert\cdot\vert_{0}$, $\vert
\cdot\vert_{1}$, $\Vert\cdot\Vert$ and $\vert\cdot\vert_{\max}$ denote the
$L_{0}$, $L_{1}$, $L_{2}$ (Euclidean) and maximum norms of a vector,
respectively. Let ${\mathbf{I}}_{d}$ and ${\mathbf{O}}_{d\times d}$ be a
$d\times d$ identity matrix and null matrix, respectively. For a $d\times d$
matrix ${\mathbf{W}}=(w_{ij})_{d\times d}$, we let $\Vert{\mathbf{W}}%
\Vert=\lambda_{\max}^{1/2}\left( {\mathbf{W}}^{^{\intercal}}{\mathbf{W}%
}\right) $ be the operator norm, $\Vert{\mathbf{W}}\Vert_{F}=\left[
\mathsf{Tr}\left( {\mathbf{W}}^{^{\intercal}}{\mathbf{W}}\right) \right]
^{1/2}$ the Frobenius norm, $\Vert{\mathbf{W}}\Vert_{1}=\max_{1\leq j\leq
d}\sum_{i=1}^{d} |w_{ij}|$, $\Vert{\mathbf{W}}\Vert_{\max}=\max_{1\leq i\leq
d}\max_{1\leq j\leq d} |w_{ij}|$, and $\vert{\mathbf{W}}\vert_{1}=\sum
_{i=1}^{d}\sum_{j=1}^{d} |w_{ij}|$, where $\lambda_{\max}(\cdot)$ is the
maximum eigenvalue of a matrix and $\mathsf{Tr}(\cdot)$ is the trace. Denote
the determinant of a square matrix as $\mathsf{det}(\cdot)$. Let $a_{n}\sim
b_{n}$, $a_{n}\propto b_{n}$ and $a_{n}\gg b_{n}$ denote that $a_{n}%
/b_{n}\rightarrow1$, $0<\underline{c}\leq a_{n}/b_{n}\leq\overline{c}<\infty$
and $b_{n}/a_{n}\rightarrow0$, respectively.


\section{Time-varying VAR and network models}

\label{sec2} \renewcommand{\theequation}{2.\arabic{equation}} \setcounter{equation}{0}

In this section, we first introduce a locally stationary VAR model with
time-varying transition and precision matrices, and then define two types of
time-varying network structures with Granger causality and partial correlation
linkages, respectively. Section \ref{sec5} will further generalise them to the
factor-adjusted time-varying VAR and network setting.

\subsection{Time-varying VAR models}

\label{sec2.1}

Suppose that $(X_{t}:t=1,\mathcal{\ldots},n)$ with $X_{t}=(x_{t,1}%
,\mathcal{\ldots},x_{t,d})^{^{\intercal}}$ is a sequence of $d$-dimensional
random vectors generated by a time-varying VAR model of order $p$:
\begin{equation}
X_{t}=\sum_{k=1}^{p}{\mathbf{A}}_{t,k}X_{t-k}+e_{t}\ \ \mathrm{with}%
\ \ e_{t}={\boldsymbol{\Sigma}}_{t}^{1/2}\varepsilon_{t}%
,\ \ t=1,\mathcal{\ldots},n,\label{eq2.1}%
\end{equation}
where ${\mathbf{A}}_{t,k}={\mathbf{A}}_{k}(t/n)$, $k=1,\mathcal{\ldots},p$,
are $d\times d$ time-varying transition matrices with each entry being a
smooth deterministic function of scaled times, ${\boldsymbol{\Sigma}}%
_{t}={\boldsymbol{\Sigma}}(t/n)$ is a $d\times d$ time-varying volatility
matrix, and $(\varepsilon_{t})$ is a sequence of independent and identically
distributed (i.i.d.) $d$-dimensional random vectors with zero mean and
identity covariance matrix. Define ${\boldsymbol{\Omega}}_{t}%
={\boldsymbol{\Omega}}(t/n)$ as the inverse of ${\boldsymbol{\Sigma}}_{t}$,
the time-varying precision matrix. We consider the ultra large time series
setting, i.e., the dimension $d$ is allowed to diverge at an exponential rate
of the sample size $n$. The time-varying VAR model (\ref{eq2.1}) is a natural
extension of the finite-dimensional time-varying VAR to high-dimensional time
series. If ${\boldsymbol{\Sigma}}_{t}$ is replaced by a time-invariant
covariance matrix, (\ref{eq2.1}) becomes the same model as that considered by
\cite{DQC17}. Furthermore, when both ${\mathbf{A}}_{t,k}$,
$k=1,\mathcal{\ldots},p$, and ${\boldsymbol{\Sigma}}_{t}$ are time-invariant
constant matrices, (\ref{eq2.1}) becomes the high-dimensional stable VAR:
\begin{equation}
X_{t}=\sum_{k=1}^{p}{\mathbf{A}}_{k}X_{t-k}+{\boldsymbol{\Sigma}}%
^{1/2}\varepsilon_{t},\label{eq2.2}%
\end{equation}
which has been extensively studied in the recent literature
\citep[e.g.,][]{BM15, HLL15, KC15, BB19, LZ21}. Throughout the paper, we
assume that the following conditions are satisfied.

\begin{assumption}
\label{ass:1}

\emph{(i)\ Uniformly over $\tau\in[0, 1]$, it holds that $\mathsf{det}\left(
{\mathbf{I}}_{d}-\sum_{k=1}^{p}{\mathbf{A}}_{k}(\tau)z^{k}\right) \neq0$ for
any $z\in{\mathbb{C}}$ with modulus no larger than one, where ${\mathbb{C}}$
denotes the set of complex numbers. Each entry in ${\mathbf{A}}_{k}(\cdot)$ is
second-order continuously differentiable over $[0,1]$. }

\emph{(ii)\ The precision matrix ${\boldsymbol{\Omega}}(\tau)$ is positive
definite uniformly over $\tau\in[0, 1]$, and the operator norm of
${\boldsymbol{\Sigma}}(\tau)$ is uniformly bounded over $\tau\in[0, 1]$.
Furthermore, each entry in ${\boldsymbol{\Sigma}}(\tau)$ and
${\boldsymbol{\Omega}}(\tau)$ is second-order continuously differentiable over
$[0,1]$.}

\emph{(iii)\ For any $d$-dimensional vector $u$ satisfying $\Vert u\Vert=1$,
$\mathsf{E}\left[ \exp\left\{ \iota_{1}(u^{^{\intercal}}\varepsilon_{t}%
)^{2}\right\} \right] \leq C_{0}<\infty$, where $\iota_{1}$ and $C_{0}$ are
positive constants.}
\end{assumption}

The first condition in Assumption \ref{ass:1}(i) is a natural extension of the
stability assumption imposed on the constant transition matrices
\citep[e.g.,][]{Lu06}, indicating that the time-varying VAR process is locally
stationary/stable and leading to the following Wold representation
\begin{equation}
\label{eq2.3}X_{t}=\sum_{k=0}^{\infty}{\boldsymbol{\Phi}}_{t,k} e_{t-k},
\end{equation}
with the coefficient matrices ${\boldsymbol{\Phi}}_{t,k}$ being absolutely
summable (in appropriate matrix norm). For example, when $p=1$, we have
${\boldsymbol{\Phi}}_{t,0}={\mathbf{I}}_{d}$ and ${\boldsymbol{\Phi}}%
_{t,k}=\Pi_{j=1}^{k}{\mathbf{A}}_{t-j+1,1}$ for $k\geq1$. Assume that, for $k$
sufficiently large,
\begin{equation}
\label{eq2.4}\max_{0\leq t\leq n}\Vert{\boldsymbol{\Phi}}_{t,k}\Vert\leq
C_{1}\rho^{k},
\end{equation}
where $C_{1}$ is a positive constant and $0<\rho<1$. A similar assumption can
be found in \cite{DQC17}. In some special model settings, (\ref{eq2.4}) may be
violated, and we refer the interested readers to the discussions in
\cite{BM15} and \cite{LZ21}. In fact, the condition (\ref{eq2.4}) may be
removed by imposing some high-level conditions (e.g., the sub-Gaussian
condition on $x_{t,i}$ proved in Lemma B.1). The smoothness conditions in
Assumption \ref{ass:1}(i)(ii) are common in kernel-based local estimation
method and theory. The sub-Gaussian moment condition in Assumption
\ref{ass:1}(iii) is not uncommon in the literature of high-dimensional feature
selection and covariance/precision matrix estimation \citep[e.g.,][]{W19}, and
is weaker than the Gaussian assumption frequently used in the high-dimensional
VAR literature \citep[e.g.,][]{BM15, KC15}.


\subsection{Time-varying network structures}

Write ${\mathbf{A}}_{t,k}=\left(  a_{k,ij|t}\right)  _{d\times d}$,
${\boldsymbol{\Omega}}_{t}=\left(  \omega_{ij|t}\right)  _{d\times d}$,
${\mathbf{A}}_{k}(\tau)=\left(  a_{k,ij}(\tau)\right)  _{d\times d}$ and
${\boldsymbol{\Omega}}(\tau)=\left(  \omega_{ij}(\tau)\right)  _{d\times d}$,
where $1\leq t\leq n$ and $0\leq\tau\leq1$. We define the network structure
via a time-varying graph ${\mathbb{G}}_{t}=({\mathbb{V}},{\mathbb{E}}_{t})$,
where ${\mathbb{V}}=\{1,2,\mathcal{\ldots},d\}$ denotes a set of vertices, and
${\mathbb{E}}_{t}=\left\{  (i,j)\in{\mathbb{V}}\times{\mathbb{V}}%
:\ c_{ij|t}\neq0,\ i\neq j\right\}  $ denotes a time-varying set of edges. The
choice of $c_{ij|t}$ is determined by the definition of linkage. The
construction of ${\mathbb{G}}_{t}$ is similar to that in \cite{KSAX10} and
\cite{ZLW10} for independent network data. Following the stable network
analysis in \cite{BB19} and \cite{BCO22}, we next consider two types of
time-varying linkages: the directed Granger causality linkage and undirected
partial correlation linkage.

The definition of Granger causality is first introduced by \cite{G69} to
investigate the causal relations in small economic time series systems. In the
context of stable VAR (with order $p$), we say that $x_{t,j}$ Granger causes
$x_{t,i}$ if there exists $k\in\{1,2,\mathcal{\ldots},p\}$ such that
$x_{t-k,j}$ improves predictability of $x_{t,i}$ by reducing the forecasting
error. It is a natural idea to use the stable transition matrices
${\mathbf{A}}_{k}=\left(  a_{k,ij}\right)  _{d\times d}$ in (\ref{eq2.2}) to
determine the Granger causality structure, i.e., if there exists at least one
$k$ such that $a_{k,ij}\neq0$, then $x_{t,j}$ Granger causes $x_{t,i}$. We may
extend the stable Granger causality structure to a more general time-varying
version using (\ref{eq2.1}). At a given time point $t$, we say that lags of
$x_{t,j}$ Granger cause $x_{t,i}$ if there exists at least one $k$ such that
$a_{k,ij|t}\neq0$. Hence, for given $\tau\in(0,1)$, we define the time-varying
local graph ${\mathbb{G}}_{\tau}^{G}=\left(  {\mathbb{V}},{\mathbb{E}}_{\tau
}^{G}\right)  $ with
\begin{equation}
{\mathbb{E}}_{\tau}^{G}=\left\{  (i,j)\in{\mathbb{V}}\times{\mathbb{V}%
}:\ \exists\ k\in\{1,2,\mathcal{\ldots},p\},\ a_{k,ij}(\tau)\neq0\right\}
.\label{eq2.5}%
\end{equation}

The partial correlation is a commonly-used conditional dependence measure for
network time series. We next extend it to the time-varying setting using
${\boldsymbol{\Omega}}_{t}={\boldsymbol{\Omega}}(t/n)$ in (\ref{eq2.1}). Let
$\rho_{ij|t}=\mathsf{cor}(e_{t,i}, e_{t,j} | e_{t,k}, k\neq i,j)$ be the
time-varying (contemporaneous) partial correlation between the innovations
$e_{t,i}$ and $e_{t,j}$, where $e_{t,i}$ is the $i$-th element of $e_{t}$.
Following \cite{D72}, we may show that $\rho_{ij|t}\neq0$ is equivalent to
$\omega_{ij|t}\neq0$ for $i\neq j$. Hence, we can construct the set of edges
by collecting the index pairs of the non-zero entries in the time-varying
precision matrix. For $\tau\in(0, 1)$, define the local graph ${\mathbb{G}%
}_{\tau}^{P}=\left( {\mathbb{V}}, {\mathbb{E}}_{\tau}^{P}\right) $ with
\begin{equation}
\label{eq2.6}{\mathbb{E}}_{\tau}^{P}=\left\{ (i,j)\in{\mathbb{V}}%
\times{\mathbb{V}}:\ \omega_{ij}(\tau)\neq0,\ i\neq j\right\} .
\end{equation}

In practice, the primary interest often lies in the full network structures
over the entire time interval. This requires the construction of a uniform
version of ${\mathbb{G}}_{\tau}^{G}$ and ${\mathbb{G}}_{\tau}^{P}$. Denote the
uniform graphs by ${\mathbb{G}}^{G}=\left(  {\mathbb{V}},{\mathbb{E}}%
^{G}\right)  $ and ${\mathbb{G}}^{P}=\left(  {\mathbb{V}},{\mathbb{E}}%
^{P}\right)  $, with
\begin{equation}
{\mathbb{E}}^{G}=\left\{  (i,j)\in{\mathbb{V}}\times{\mathbb{V}}%
:\ \exists\ k\in\{1,2,\mathcal{\ldots},p\}\ \mathrm{and}\ \tau\in
(0,1),\ a_{k,ij}(\tau)\neq0\right\}  \label{eq2.7}%
\end{equation}
and
\begin{equation}
{\mathbb{E}}^{P}=\left\{  (i,j)\in{\mathbb{V}}\times{\mathbb{V}}%
:\ \exists\ \tau\in(0,1),\ \omega_{ij}(\tau)\neq0,\ i\neq j\right\}
.\label{eq2.8}%
\end{equation}
It is easy to verify that ${\mathbb{E}}_{\tau}^{G}\subset{\mathbb{E}}^{G}$ and
${\mathbb{E}}_{\tau}^{P}\subset{\mathbb{E}}^{P}$ for any $\tau\in(0,1)$.
Section \ref{sec3.4} below defines the discrete versions of the above uniform
networks and provide their estimates.


\section{Methodology}

\label{sec3} \renewcommand{\theequation}{3.\arabic{equation}} \setcounter{equation}{0}

Let $A_{k,i}^{^{\intercal}}(\cdot)$ and $C_{i}^{^{\intercal}}(\cdot)$ be the
$i$-th row of ${\mathbf{A}}_{k}(\cdot)$ and ${\boldsymbol{\Omega}}%
^{-1/2}(\cdot)$, respectively,
\begin{equation}
{\boldsymbol{\alpha}}_{i\bullet}(\cdot)=\left[  A_{1,i}^{^{\intercal}}%
(\cdot),\mathcal{\ldots},A_{p,i}^{^{\intercal}}(\cdot)\right]  ^{^{\intercal}%
},\ \ {\mathbf{X}}_{t}=\left(  X_{t}^{^{\intercal}},\mathcal{\ldots}%
,X_{t-p+1}^{^{\intercal}}\right)  ^{^{\intercal}},\label{eq3.1}%
\end{equation}
and $\tau_{t}=t/n$. The time-varying VAR model (\ref{eq2.1}) can be
equivalently written as
\begin{equation}
x_{t,i}={\boldsymbol{\alpha}}_{i\bullet}^{^{\intercal}}(\tau_{t}){\mathbf{X}%
}_{t-1}+e_{t,i}\ \ \mathrm{with}\ \ e_{t,i}=C_{i}^{^{\intercal}}(\tau
_{t})\varepsilon_{t},\ \ i=1,\mathcal{\ldots},d,\label{eq3.2}%
\end{equation}
which is a high-dimensional time-varying coefficient autoregressive model with
a scalar response and $pd$ candidate predictors for each $i$. As the dimension
of the predictors is allowed to be ultra large, we need to impose an
appropriate sparsity restriction on the vector of time-varying parameters
${\boldsymbol{\alpha}}_{i\bullet}(\cdot)$ to limit the number of its
significant elements. High-dimensional varying-coefficient models have been
systematically studied in the literature and various nonparametric screening
and shrinkage methods have been proposed to select the significant covariates,
estimate the coefficient functions and identify the model structure under the
independent data assumption
\citep[e.g.,][]{WLH08, WX09, L12, CHLP14, FMD14, LLW14, LKZ15}. In this
section, under the high-dimensional locally stationary time series framework,
we propose a three-stage procedure to estimate the Granger causality and
partial correlation network structures: (i) first obtain preliminary local
linear estimates of ${\boldsymbol{\alpha}}_{i\bullet}(\cdot)$ (and its
derivatives) using time-varying LASSO, which serves as a first-stage screening
of the predictors in ${\mathbf{X}}_{t-1}$; (ii) conduct local linear
estimation and feature selection using weighted group LASSO, where the weights
are constructed via a local linear approximation to the SCAD penalty using the
preliminary estimates of ${\boldsymbol{\alpha}}_{i\bullet}(\cdot)$ from Stage
(i); (iii) estimate the error precision matrix ${\boldsymbol{\Omega}}(\cdot)$
via the time-varying CLIME method. The estimated transition and precision
matrices are finally used to construct the uniform network structures.

\subsection{Preliminary time-varying LASSO estimation}

\label{sec3.1}

For $\tau\in(0,1)$, under the smoothness condition on the transition matrices
in Assumption \ref{ass:1}(i), we have the following local linear approximation
to ${\boldsymbol{\alpha}}_{i\bullet}(\tau_{t})$:
\[
{\boldsymbol{\alpha}}_{i\bullet}(\tau_{t})\approx{\boldsymbol{\alpha}%
}_{i\bullet}(\tau)+{\boldsymbol{\alpha}}_{i\bullet}^{\prime}(\tau)(\tau
_{t}-\tau),\ \ i=1,\mathcal{\ldots},d,
\]
when $\tau_{t}$ falls within a small neighbourhood of $\tau$, where
${\boldsymbol{\alpha}}_{i\bullet}^{\prime}(\cdot)$ is a $(pd)$-dimensional
vector of the first-order derivatives of the elements in ${\boldsymbol{\alpha
}}_{i\bullet}(\cdot)$. Hence, for each $i\in\{1,2,\mathcal{\ldots},d\}$ and a
given $\tau\in(0,1)$, we define the following local linear objective function
\citep[e.g.,][]{FG96}:
\begin{equation}
\mathcal{L}_{i}({\boldsymbol{\alpha}},{\boldsymbol{\beta}}\ |\ \tau)=\frac
{1}{n}\sum\limits_{t=1}^{n}\left\{  x_{t,i}-\left[  {\boldsymbol{\alpha}%
}+{\boldsymbol{\beta}}(\tau_{t}-\tau)\right]  ^{^{\intercal}}{\mathbf{X}%
}_{t-1}\right\}  ^{2}K_{h}(\tau_{t}-\tau),\label{eq3.3}%
\end{equation}
where $K_{h}(\cdot)=\frac{1}{h}K(\cdot/h)$ with $K(\cdot)$ being a kernel
function and $h$ being a bandwidth or smoothing parameter. The estimates of
${\boldsymbol{\alpha}}_{i\bullet}(\tau)$ and ${\boldsymbol{\alpha}}_{i\bullet
}^{\prime}(\tau)$ are obtained by minimising $\mathcal{L}_{i}%
({\boldsymbol{\alpha}},{\boldsymbol{\beta}}\ |\ \tau)$ with respect to
${\boldsymbol{\alpha}}$ and ${\boldsymbol{\beta}}$. However, this local linear
estimation is only feasible when the dimension of the predictors is fixed or
significantly smaller than the sample size $n$ \citep[e.g.,][]{C07, LCG11}. In
our high-dimensional setting, as the number of predictors may exceed $n$, it
is challenging to obtain satisfactory estimation by directly minimising
$\mathcal{L}_{i}({\boldsymbol{\alpha}},{\boldsymbol{\beta}}\ |\ \tau)$. To
address this issue, we assume that the number of significant components in
${\boldsymbol{\alpha}}_{i\bullet}(\tau)$ is much smaller than $n$ and then
incorporate a LASSO penalty term in the local linear objective function
(\ref{eq3.3}).

The LASSO estimation was first introduced by \cite{T96} in the context of
linear regression and has become one of the most commonly-used tools in
high-dimensional variable and feature selection. We next adopt a time-varying
version of the LASSO estimation. Define
\begin{equation}
\label{eq3.4}\mathcal{L}_{i}^{\ast}({\boldsymbol{\alpha}}, {\boldsymbol{\beta
}}\ |\ \tau)=\mathcal{L}_{i}({\boldsymbol{\alpha}}, {\boldsymbol{\beta}%
}\ |\ \tau)+\lambda_{1} \left( \vert{\boldsymbol{\alpha}}\vert_{1}%
+h\vert{\boldsymbol{\beta}}\vert_{1}\right) ,
\end{equation}
where $\lambda_{1}$ is a tuning parameter. Let $\widetilde{\boldsymbol{\alpha
}}_{i\bullet}(\tau)$ and $\widetilde{\boldsymbol{\alpha}}_{i\bullet}^{\prime
}(\tau)$ be the solution to the minimisation of $\mathcal{L}_{i}^{\ast
}({\boldsymbol{\alpha}}, {\boldsymbol{\beta}}\ |\ \tau)$ with respect to
${\boldsymbol{\alpha}}$ and ${\boldsymbol{\beta}}$. We call them the
preliminary time-varying LASSO estimates. This LASSO estimation may not
accurately identify the true significant predictors, but can remove a large
number of irrelevant predictors and hence, serves as a preliminary screening
step. Furthermore, the first-stage estimates will be used to construct weights
in the weighted group LASSO in the second stage to more precisely estimate the
time-varying parameters and accurately select the significant predictors.

\subsection{Penalised local linear estimation with weighted group LASSO}

\label{sec3.2}

In order to estimate the uniform Granger causality network, we next introduce
a \emph{global} penalised method to simultaneously estimate the time-varying
parameters at $\tau_{t}$, $t=1,\mathcal{\ldots},n$, and identify the non-zero
index sets $\mathcal{J}_{i}=\bigcup_{t=1}^{n}\mathcal{J}_{i}(\tau_{t})$ and
$\mathcal{J}_{i}^{\prime}=\bigcup_{t=1}^{n}\mathcal{J}_{i}^{\prime}(\tau_{t}%
)$, where
\[
\mathcal{J}_{i}(\tau)=\left\{  1\leq j\leq pd:\ \alpha_{i,j}(\tau
)\neq0\right\}  \ \ \mathrm{and}\ \ \mathcal{J}_{i}^{\prime}(\tau)=\left\{
1\leq j\leq pd:\ \alpha_{i,j}^{\prime}(\tau)\neq0\right\}
\]
with $\alpha_{i,j}(\cdot)$ and $\alpha_{i,j}^{\prime}(\cdot)$ being the $j$-th
element of ${\boldsymbol{\alpha}}_{i\bullet}(\cdot)$ and ${\boldsymbol{\alpha
}}_{i\bullet}^{\prime}(\cdot)$, respectively. For each $i$, note that
identifying the zero elements in ${\boldsymbol{\alpha}}_{i\bullet}^{\prime
}(\tau_{t})$ (uniformly over $t$) is equivalent to identifying the indices
$j$, $1\leq j\leq pd$, such that $D_{i,j}=0$, where
\[
D_{i,j}^{2}=\sum\limits_{t=1}^{n}\left[  \alpha_{i,j}(\tau_{t})-\frac{1}%
{n}\sum\limits_{s=1}^{n}\alpha_{i,j}(\tau_{s})\right]  ^{2}.
\]
In practice, $D_{i,j}^{2}$ can be estimated by
\[
\widetilde{D}_{i,j}^{2}=\sum\limits_{t=1}^{n}\left[  \widetilde{\alpha}%
_{i,j}(\tau_{t})-\frac{1}{n}\sum\limits_{s=1}^{n}\widetilde{\alpha}_{i,j}%
(\tau_{s})\right]  ^{2},
\]
using the preliminary time-varying LASSO estimates $\widetilde{\alpha}%
_{i,j}(\tau_{t})$, $t=1,\ldots,n$. Let ${\mathbf{A}}=({\boldsymbol{\alpha}%
}_{\bullet1},\mathcal{\ldots},{\boldsymbol{\alpha}}_{\bullet n})^{^{\intercal
}}$ with ${\boldsymbol{\alpha}}_{\bullet t}=(\alpha_{1|t},\mathcal{\ldots
},\alpha_{pd|t})^{^{\intercal}}$, and ${\mathbf{B}}=({\boldsymbol{\beta}%
}_{\bullet1},\mathcal{\ldots},{\boldsymbol{\beta}}_{\bullet n})^{^{\intercal}%
}$ with ${\boldsymbol{\beta}}_{\bullet t}=(\beta_{1|t},\mathcal{\ldots}%
,\beta_{pd|t})^{^{\intercal}}$. We define a global version of the penalised
objective function with weighted group LASSO:
\begin{equation}
\mathcal{Q}_{i}({\mathbf{A}},{\mathbf{B}})=\sum_{t=1}^{n}\mathcal{L}%
_{i}({\boldsymbol{\alpha}}_{\bullet t},{\boldsymbol{\beta}}_{\bullet
t}\ |\ \tau_{t})+\sum_{j=1}^{pd}p_{\lambda_{2}}^{\prime}\left(  \left\Vert
\widetilde{\boldsymbol{\alpha}}_{i,j}\right\Vert \right)  \Vert
{\boldsymbol{\alpha}}_{j}\Vert+\sum_{j=1}^{pd}p_{\lambda_{2}}^{\prime}\left(
\widetilde{D}_{i,j}\right)  \Vert h{\boldsymbol{\beta}}_{j}\Vert,\label{eq3.5}%
\end{equation}
where
\[
\widetilde{\boldsymbol{\alpha}}_{i,j}=\left[  \widetilde{\alpha}_{i,j}%
(\tau_{1}),\mathcal{\ldots},\widetilde{\alpha}_{i,j}(\tau_{n})\right]
^{^{\intercal}},\ \ {\boldsymbol{\alpha}}_{j}=\left(  \alpha_{j|1}%
,\mathcal{\ldots},\alpha_{j|n}\right)  ^{^{\intercal}},\ \ {\boldsymbol{\beta
}}_{j}=\left(  \beta_{j|1},\mathcal{\ldots},\beta_{j|n}\right)  ^{^{\intercal
}},
\]
while $\lambda_{2}$ is a tuning parameter and $p_{\lambda}^{\prime}(\cdot)$ is
the derivative of the SCAD penalty function:
\[
p_{\lambda}^{\prime}(z)=\lambda\left[  I(z\leq\lambda)+\frac{(a_{0}%
\lambda-z)_{+}}{(a_{0}-1)\lambda}I(z>\lambda)\right]  ,
\]
with $a_{0}=3.7$ as suggested in \cite{FL01} and $I(\cdot)$ being the
indicator function. The penalty terms in (\ref{eq3.5}) are motivated by the
local linear approximation to the SCAD penalty function \citep{ZL08}. The
terms $p_{\lambda_{2}}^{\prime}\left(  \left\Vert \widetilde
{\boldsymbol{\alpha}}_{i,j}\right\Vert \right)  $ and $p_{\lambda_{2}}%
^{\prime}\left(  \widetilde{D}_{i,j}\right)  $ in (\ref{eq3.5}) serve as the
weights for the group LASSO, and their values are determined by the
preliminary estimates in Section \ref{sec3.1}, i.e., the corresponding weight
is heavy when $\left\Vert \widetilde{\boldsymbol{\alpha}}_{i,j}\right\Vert $
or $\widetilde{D}_{i,j}$ is close to zero, whereas it is light or equal to
zero when $\left\Vert \widetilde{\boldsymbol{\alpha}}_{i,j}\right\Vert $ or
$\widetilde{D}_{i,j}$ is large. An advantage of using $\widetilde{D}_{i,j}$ in
the second penalty term over the $L_{2}$-norm of $\widetilde
{\boldsymbol{\alpha}}_{j}^{\prime}=\left[  \widetilde{\alpha}_{i,j}^{\prime
}(\tau_{1}),\mathcal{\ldots},\widetilde{\alpha}_{i,j}^{\prime}(\tau
_{n})\right]  ^{^{\intercal}}$ is that the estimates of the time-varying
parameters involved in $\widetilde{D}_{i,j}$ often perform more stably than
their derivative counterparts.

Let $\widehat{\mathbf{A}}_{i}$ and $\widehat{\mathbf{B}}_{i}$ be the minimiser
of $\mathcal{Q}_{i}({\mathbf{A}},{\mathbf{B}})$ with respect to ${\mathbf{A}}$
and ${\mathbf{B}}$, where
\begin{align}
& \widehat{\mathbf{A}}_{i}=\left(  \widehat{\boldsymbol{\alpha}}%
_{i,1},\mathcal{\ldots},\widehat{\boldsymbol{\alpha}}_{i,pd}\right)
\ \ \mathrm{with}\ \ \widehat{\boldsymbol{\alpha}}_{i,j}=\left[
\widehat{\alpha}_{i,j}(\tau_{1}),\mathcal{\ldots},\widehat{\alpha}_{i,j}%
(\tau_{n})\right]  ^{^{\intercal}},\nonumber\\
& \widehat{\mathbf{B}}_{i}=\left(  \widehat{\boldsymbol{\alpha}}_{i,1}%
^{\prime},\mathcal{\ldots},\widehat{\boldsymbol{\alpha}}_{i,pd}^{\prime
}\right)  \ \ \mathrm{with}\ \ \widehat{\boldsymbol{\alpha}}_{i,j}^{\prime
}=\left[  \widehat{\alpha}_{i,j}^{\prime}(\tau_{1}),\mathcal{\ldots}%
,\widehat{\alpha}_{i,j}^{\prime}(\tau_{n})\right]  ^{^{\intercal}}.\nonumber
\end{align}
The index set $\mathcal{J}_{i}$ is estimated by $\widehat{\mathcal{J}}%
_{i}=\left\{  j:\ \widehat{\boldsymbol{\alpha}}_{i,j}\neq{\mathbf{0}}%
_{n}\right\}  $, and $\mathcal{J}_{i}^{\prime}$ is estimated by $\widehat
{\mathcal{J}}_{i}^{\prime}=\left\{  j:\ \widehat{\boldsymbol{\alpha}}%
_{i,j}^{\prime}\neq{\mathbf{0}}_{n}\right\}  $, where ${\mathbf{0}}_{k}$ is a
$k$-dimensional vector of zeros. A similar shrinkage estimation method is used
by \cite{LKZ15} and \cite{CLWZ21} to identify a high-dimensional semi-varying
coefficient model structure for independent data. So far as we know, there is
no work on such a penalised technique and its relevant theory for
high-dimensional locally stationary time series data.

\subsection{Estimation of the time-varying precision matrix}

\label{sec3.3}

In this section, we study the estimation of ${\boldsymbol{\Omega}}(\cdot)$ in
model (\ref{eq2.1}), which is crucial to uncover the time-varying and uniform
network structures of partial correlations. Estimation of large static
precision matrices has been extensively studied under the sparsity assumption,
and various estimation techniques, such as the penalised likelihood, graphical
Danzig selector and CLIME, have been proposed in the literature
\citep[e.g.,][]{LF09, Y10, CLL11}. \cite{XCW20} further introduce a
time-varying CLIME method for high-dimensional locally stationary time series
which are observable. Note that in this paper, ${\boldsymbol{\Omega}}(\cdot)$
is the time-varying precision matrix for the high-dimensional unobservable
error vector $e_{t}$ and hence, its estimation requires substantial
modification of the time-varying CLIME methodology and theory.

With $\widehat{\boldsymbol{\alpha}}_{i\bullet}(\cdot)$, $i=1,\mathcal{\ldots
},d$, from Section \ref{sec3.2}, we can then extract estimates of the
time-varying transition matrices, denoted by $\widehat{\mathbf{A}}_{k}%
(\tau_{t})$, $t=1,\mathcal{\ldots},n$, $k=1,\mathcal{\ldots},p$, and
approximate $e_{t}$ by
\begin{equation}
\widehat{e}_{t}=\left(  \widehat{e}_{t,1},\mathcal{\ldots},\widehat{e}%
_{t,d}\right)  ^{^{\intercal}}=X_{t}-\sum_{k=1}^{p}\widehat{\mathbf{A}}%
_{k}(\tau_{t})X_{t-k},\ \ t=1,\mathcal{\ldots},n.\label{eq3.6}%
\end{equation}
The approximation accuracy depends on the uniform prediction rates of the
time-varying weighted group LASSO estimates. In order to apply the
time-varying CLIME, we assume that ${\boldsymbol{\Omega}}(\cdot)$ satisfies a
uniform sparsity assumption, a natural extension of the classic sparsity
assumption to the locally stationary time series setting. Specifically, we
assume $\left\{  {\boldsymbol{\Omega}}(\tau):0\leq\tau\leq1\right\}
\in\mathcal{S}(q,\xi_{d})$, where {\small
\begin{equation}
\mathcal{S}(q,\xi_{d})=\left\{  {\mathbf{W}}(\tau)=\left[  w_{ij}%
(\tau)\right]  _{d\times d},0\leq\tau\leq1:\ {\mathbf{W}}(\tau)\succ
0,\ \sup_{0\leq\tau\leq1}\Vert{\mathbf{W}}(\tau)\Vert_{1}\leq C_{2}%
,\ \sup_{0\leq\tau\leq1}\max_{1\leq i\leq d}\sum_{j=1}^{d}|w_{ij}(\tau
)|^{q}\leq\xi_{d}\right\}  ,\label{eq3.7}%
\end{equation}
} where $0\leq q<1$, \textquotedblleft${\mathbf{W}}\succ0$" denotes that
${\mathbf{W}}$ is positive definite, and $C_{2}$ is a bounded positive
constant. Define
\begin{equation}
\widehat{\boldsymbol{\Sigma}}(\tau)=\left[  \widehat{\sigma}_{ij}%
(\tau)\right]  _{d\times d}\ \ \mathrm{with}\ \ \widehat{\sigma}_{ij}%
(\tau)=\sum_{t=1}^{n}\varpi_{n,t}(\tau)\widehat{e}_{t,i}\widehat{e}_{t,j}%
/\sum_{t=1}^{n}\varpi_{n,t}(\tau),\label{eq3.8}%
\end{equation}
where the weight function $\varpi_{n,t}(\cdot)$ is constructed via the local
linear smoothing:
\[
\varpi_{n,t}(\tau)=K\left(  \frac{\tau_{t}-\tau}{b}\right)  s_{n,2}%
(\tau)-K_{1}\left(  \frac{\tau_{t}-\tau}{b}\right)  s_{n,1}(\tau),
\]
in which $s_{n,j}(\tau)=\sum_{t=1}^{n}K_{j}\left(  \frac{\tau_{t}-\tau}%
{b}\right)  $, $K_{j}(x)=x^{j}K(x)$, and $b$ is a bandwidth. With the uniform
sparsity assumption (\ref{eq3.7}), we estimate ${\boldsymbol{\Omega}}(\tau)$
via the time-varying CLIME method:
\begin{equation}
\widetilde{\boldsymbol{\Omega}}(\tau)=\left[  \widetilde{\omega}_{ij}%
(\tau)\right]  _{d\times d}=\argmin_{\boldsymbol{\Omega}}|{\boldsymbol{\Omega
}}|_{1}\ \ \ \ \ \mathrm{subject\ to}\ \ \left\Vert \widehat
{\boldsymbol{\Sigma}}(\tau){\boldsymbol{\Omega}}-{\mathbf{I}}_{d}\right\Vert
_{\mathrm{max}}\leq\lambda_{3},\label{eq3.9}%
\end{equation}
where $\lambda_{3}$ is a tuning parameter. As the underlying time-varying
precision matrix is symmetric, the matrix estimate obtained from (\ref{eq3.9})
needs to be symmetrised to obtain the final estimate, denoted as
$\widehat{\boldsymbol{\Omega}}(\tau)=\left[  \widehat{\omega}_{ij}%
(\tau)\right]  _{d\times d}$, where
\begin{equation}
\widehat{\omega}_{ij}(\tau)=\widehat{\omega}_{ji}(\tau)=\widetilde{\omega
}_{ij}(\tau)I\left(  |\widetilde{\omega}_{ij}(\tau)|\leq|\widetilde{\omega
}_{ji}(\tau)|\right)  +\widetilde{\omega}_{ji}(\tau)I\left(  |\widetilde
{\omega}_{ij}(\tau)|>|\widetilde{\omega}_{ji}(\tau)|\right)  .\label{eq3.10}%
\end{equation}

\subsection{Estimation of uniform time-varying networks}

\label{sec3.4}

In practice, when the sample size $n$ is sufficiently large, it is often
sensible to approximate the uniform edge sets, ${\mathbb{E}}^{G}$ and
${\mathbb{E}}^{P}$, by the following discrete versions:
\begin{equation}
{\mathbb{E}}_{n}^{G}=\left\{  (i,j)\in{\mathbb{V}}\times{\mathbb{V}}%
:\ \exists\ k\in\{1,2,\mathcal{\ldots},p\}\ \ \mathrm{and}\ \ t\in
\{1,\mathcal{\ldots},n\},\ a_{k,ij}(\tau_{t})\neq0\right\}  \label{eq3.11}%
\end{equation}
and
\begin{equation}
{\mathbb{E}}_{n}^{P}=\left\{  (i,j)\in{\mathbb{V}}\times{\mathbb{V}}%
:\ \exists\ t\in\{1,\mathcal{\ldots},n\},\ \omega_{ij}(\tau_{t})\neq0,\ i\neq
j\right\}  .\label{eq3.12}%
\end{equation}
Hence, we next estimate ${\mathbb{E}}_{n}^{G}$ and ${\mathbb{E}}_{n}^{P}$
instead of ${\mathbb{E}}^{G}$ and ${\mathbb{E}}^{P}$. With the time-varying
transition and precision matrix estimates in Sections \ref{sec3.2} and
\ref{sec3.3}, we can estimate ${\mathbb{E}}_{n}^{G}$ by
\begin{equation}
\widehat{\mathbb{E}}_{n}^{G}=\left\{  (i,j)\in{\mathbb{V}}\times{\mathbb{V}%
}:\ \exists\ k\in\{1,2,\mathcal{\ldots},p\},\ \sum_{t=1}^{n}\widehat{a}%
_{k,ij}^{2}(\tau_{t})>0\right\}  ,\label{eq3.13}%
\end{equation}
where $\widehat{a}_{k,ij}(\tau_{t})$ is the $(i,j)$-entry of $\widehat
{\mathbf{A}}_{k}(\tau_{t})$, and estimate ${\mathbb{E}}_{n}^{P}$ by
\begin{equation}
\widehat{\mathbb{E}}_{n}^{P}=\left\{  (i,j)\in{\mathbb{V}}\times{\mathbb{V}%
}:\ \exists\ t\in\{1,\mathcal{\ldots},n\},\ \left\vert \widehat{\omega}%
_{ij}(\tau_{t})\right\vert \geq\lambda_{3},\ i\neq j\right\}  ,\label{eq3.14}%
\end{equation}
where $\lambda_{3}$ is the tuning parameter used in the time-varying CLIME.


\section{Main theoretical results}

\label{sec4} \renewcommand{\theequation}{4.\arabic{equation}} \setcounter{equation}{0}

To ease the notational burden, throughout this section, we focus on the
time-varying VAR(1) model:
\begin{equation}
X_{t}={\mathbf{A}}(\tau_{t})X_{t-1}+{\boldsymbol{\Sigma}}_{t}^{1/2}%
\varepsilon_{t},\label{eq4.1}%
\end{equation}
where ${\mathbf{A}}(\tau)=\left[  \alpha_{ij}(\tau)\right]  _{d\times d}$. For
a general time-varying VAR($p$) model (\ref{eq2.1}), it can be equivalently
re-written as a $(pd)$-dimensional VAR(1) model as follows:
\[
{\mathbf{X}}_{t}={\mathbf{A}}_{t}^{\ast}{\mathbf{X}}_{t-1}+{\mathbf{e}}_{t},
\]
where ${\mathbf{X}}_{t}$ is defined in (\ref{eq3.1}), ${\mathbf{e}}%
_{t}=\left(  e_{t}^{^{\intercal}},0_{d}^{^{\intercal}},\mathcal{\ldots}%
,0_{d}^{^{\intercal}}\right)  ^{^{\intercal}}$, and ${\mathbf{A}}_{t}^{\ast}$
is a $(pd)\times(pd)$ time-varying transition matrix:
\[
{\mathbf{A}}_{t}^{\ast}=\left(
\begin{array}
[c]{ccccc}%
{\mathbf{A}}_{t,1} & {\mathbf{A}}_{t,2} & \mathcal{\ldots} & {\mathbf{A}%
}_{t,p-1} & {\mathbf{A}}_{t,p}\\
{\mathbf{I}}_{d} & {\mathbf{O}}_{d\times d} & \mathcal{\ldots} & {\mathbf{O}%
}_{d\times d} & {\mathbf{O}}_{d\times d}\\
\vdots & \vdots & \vdots & \vdots & \vdots\\
{\mathbf{O}}_{d\times d} & {\mathbf{O}}_{d\times d} & \mathcal{\ldots} &
{\mathbf{I}}_{d} & {\mathbf{O}}_{d\times d}%
\end{array}
\right)  .
\]

\subsection{Uniform consistency of the time-varying LASSO estimates}

\label{sec4.1}

Define {\small
\begin{equation}
\label{eq4.2}{\boldsymbol{\Psi}}(\tau)=\left[
\begin{array}
[c]{cc}%
{\boldsymbol{\Psi}}_{0}(\tau) & {\boldsymbol{\Psi}}_{1}(\tau)\\
{\boldsymbol{\Psi}}_{1}(\tau) & {\boldsymbol{\Psi}}_{2}(\tau)
\end{array}
\right] \ \ \mathrm{with} \ \ {\boldsymbol{\Psi}}_{k}(\tau)=\frac{1}{n}%
\sum\limits_{t=1}^{n}\left( \frac{\tau_{t}-\tau}{h}\right) ^{k} X_{t-1}%
X_{t-1}^{^{\intercal}}K_{h}(\tau_{t} - \tau),\ \ k=0,1,2,
\end{equation}
} and
\[
\mathcal{B}_{i}(\tau)=\left\{ \left( u_{1}^{^{\intercal}}, u_{2}^{^{\intercal
}}\right) ^{^{\intercal}}: \|u_{1}\|^{2}+\|u_{2}\|^{2}=1,\ \sum_{j=1}%
^{d}\left( |u_{1,j}|+|u_{2,j}|\right) \leq3 \left( \sum_{j\in\mathcal{J}%
_{i}(\tau)}|u_{1,j}|+\sum_{j\in\mathcal{J}_{i}^{\prime}(\tau)}|u_{2,j}|\right)
\right\} ,
\]
where $\mathcal{J}_{i}(\tau)$ and $\mathcal{J}_{i}^{\prime}(\tau)$ are defined
as in Section \ref{sec3.2} but with $p=1$. To derive the uniform consistency
property of the preliminary time-varying LASSO estimates defined in Section
\ref{sec3.1}, we need the following assumptions, some of which may be weakened
at the cost of lengthier proofs.

\begin{assumption}
\label{ass:2}

\emph{(i)\ The kernel $K(\cdot)$ is a bounded, continuous and symmetric
probability density function with a compact support $[-1,1]$.}

\emph{(ii)\ The bandwidth $h$ satisfies
\[
nh/\log^{2} (n\vee d)\rightarrow\infty\ \ \mbox{and}\ \ sh^{2}\log(n\vee
d)\rightarrow0,
\]
where $s=\max_{1\leq i\leq d}s_{i}$ with $s_{i}$ being the cardinality of the
index set $\mathcal{J}_{i}$. }
\end{assumption}

\begin{assumption}
\label{ass:3}

\emph{(i)\ The tuning parameter $\lambda_{1}$ satisfies }
\[
\zeta_{n,d}:=\log(n\vee d)\left[ (nh)^{-1/2}+sh^{2}\right] =o(\lambda
_{1})\ \ \mbox{and}\ \ \sqrt{s}\lambda_{1}/h\rightarrow0.
\]


\emph{(ii)\ There exists a positive constant $\kappa_{0}$ such that, with
probability approaching one (w.p.a.1),}
\begin{equation}
\label{eq4.3}\min_{1\leq i\leq d}\min_{1\leq t\leq n}\inf_{u\in\mathcal{B}%
_{i}(\tau_{t})}u^{^{\intercal}} {\boldsymbol{\Psi}}(\tau_{t})u\geq\kappa_{0}.
\end{equation}

\end{assumption}

Assumption \ref{ass:2}(i) is a mild restriction which can be satisfied by some
commonly-used kernels such as the uniform kernel and the Epanechnikov kernel.
The compact support assumption on the kernel function is not essential and can
be replaced by appropriate tail conditions. The bandwidth conditions in
Assumption \ref{ass:2}(ii) are crucial for deriving the uniform convergence
properties of the kernel-based quantities. When $s$ is bounded and $d$
diverges at a polynomial rate of $n$, the conditions can be simplified to
$nh/\log^{2} n\rightarrow\infty$ and $h^{2}\log n\rightarrow0$. Assumption
\ref{ass:3}(ii) can be seen as a uniform version of the so-called restricted
eigenvalue condition widely used in high-dimensional linear regression models
\citep[e.g.,][]{BRT09,BM15}. Appendix D in the supplement provides sufficient
conditions for the high-dimensional locally stationary Gaussian time series to
satisfy Assumption \ref{ass:3}(ii). Furthermore, with the Hanson-Wright
inequality for time-varying (non-Gaussian) VAR processes
\citep[e.g., Proposition 6.2 in][]{ZW21}, we may show that $\max_{1\leq t\leq
n}\left\Vert {\boldsymbol{\Psi}}(\tau_{t})-\mathsf{E}[{\boldsymbol{\Psi}}%
(\tau_{t})]\right\Vert _{\max}=O_{P}\left( \sqrt{\log(n\vee d)/(nh)}\right) $.
Then, using Lemma D.1 in Appendix D and assuming $s\sqrt{\log(n\vee
d)/(nh)}=o(1)$, a sufficient condition for (\ref{eq4.3}) is
\[
\min_{1\leq i\leq d}\min_{1\leq t\leq n}\inf_{u\in\mathcal{B}_{i}(\tau_{t}%
)}u^{^{\intercal}} \mathsf{E}\left[ {\boldsymbol{\Psi}}(\tau_{t})\right]
u\geq\kappa_{0}.
\]

\renewcommand{\thetheorem}{4.\arabic{theorem}} \setcounter{theorem}{0}

\begin{theorem}\label{thm:4.1}
Suppose that Assumptions \ref{ass:1}--\ref{ass:3} are satisfied. Then we have
\begin{equation}\label{eq4.4}
\max_{1\leq i\leq d}\max_{1\leq t\leq n}\left\Vert \widetilde{\boldsymbol\alpha}_{i\bullet}(\tau_t)-{\boldsymbol\alpha}_{i\bullet}(\tau_t)\right\Vert =O_P\left(\sqrt{s}\lambda_1\right).
\end{equation}
\end{theorem}

Theorem \ref{thm:4.1} shows that the preliminary time-varying LASSO estimates
of the transition matrices are uniformly consistent with the convergence rates
relying on $s$ and $\lambda_{1}$. Although the dimension of variates $d$ is
allowed to diverge at an exponential rate of $n$, the number of significant
elements in ${\boldsymbol{\alpha}}_{i\bullet}(\cdot)$ cannot diverge too fast
in order to guarantee the consistency property. Furthermore, the uniform
convergence result (\ref{eq4.4}) can be strengthened to
\begin{equation}
\label{eq4.5}\max_{1\leq i\leq d}\sup_{0\leq\tau\leq1}\left\Vert
\widetilde{\boldsymbol{\alpha}}_{i\bullet}(\tau)-{\boldsymbol{\alpha}%
}_{i\bullet}(\tau)\right\Vert =O_{P}\left( \sqrt{s}\lambda_{1}\right) .
\end{equation}
A similar uniform convergence property holds for the first-order derivative
function estimates, see (A.1) in the proof of Theorem \ref{thm:4.1}.

\subsection{The oracle property of the weighted group LASSO estimates}

\label{sec4.2}


Denote the complement of $\mathcal{J}_{i}$ and $\mathcal{J}_{i}^{\prime}$ as
$\overline{\mathcal{J}}_{i}$ and $\overline{\mathcal{J}}_{i}^{\prime}$,
respectively, i.e., $\overline{\mathcal{J}}_{i}=\bigcap_{t=1}^{n}\left\{
j:\ \alpha_{i,j}(\tau_{t})=0\right\}  $ and $\overline{\mathcal{J}}%
_{i}^{\prime}=\bigcap_{t=1}^{n}\left\{  j:\ \alpha_{i,j}^{\prime}(\tau
_{t})=0\right\}  $. Let ${\mathbf{A}}^{o}=\left(  {\boldsymbol{\alpha}%
}_{\bullet1}^{o},\mathcal{\ldots},{\boldsymbol{\alpha}}_{\bullet n}%
^{o}\right)  ^{^{\intercal}}$ and ${\mathbf{B}}^{o}=\left(  {\boldsymbol{\beta
}}_{\bullet1}^{o},\mathcal{\ldots},{\boldsymbol{\beta}}_{\bullet n}%
^{o}\right)  ^{^{\intercal}}$, where ${\boldsymbol{\alpha}}_{\bullet t}%
^{o}=(\alpha_{1|t}^{o},\mathcal{\ldots},\alpha_{d|t}^{o})^{^{\intercal}}$ with
$\alpha_{j|t}^{o}=0$ for $j\in\overline{\mathcal{J}}_{i}$ and
${\boldsymbol{\beta}}_{\bullet t}^{o}=(\beta_{1|t}^{o},\mathcal{\ldots}%
,\beta_{d|t}^{o})^{^{\intercal}}$ with $\beta_{j|t}^{o}=0$ for $j\in
\overline{\mathcal{J}}_{i}^{\prime}$. Define the (infeasible) oracle
estimates:
\begin{align}
& \widehat{\mathbf{A}}_{i}^{o}=\left(  \widehat{\boldsymbol{\alpha}}_{i,1}%
^{o},\mathcal{\ldots},\widehat{\boldsymbol{\alpha}}_{i,d}^{o}\right)
\ \ \mathrm{with}\ \ \widehat{\boldsymbol{\alpha}}_{i,j}^{o}=\left[
\widehat{\alpha}_{i,j}^{o}(\tau_{1}),\mathcal{\ldots},\widehat{\alpha}%
_{i,j}^{o}(\tau_{n})\right]  ^{^{\intercal}},\label{eq4.6}\\
& \widehat{\mathbf{B}}_{i}^{o}=\left(  \widehat{\boldsymbol{\alpha}}%
_{i,1}^{\prime o},\mathcal{\ldots},\widehat{\boldsymbol{\alpha}}_{i,d}^{\prime
o}\right)  \ \ \mathrm{with}\ \ \widehat{\boldsymbol{\alpha}}_{i,j}^{\prime
o}=\left[  \widehat{\alpha}_{i,j}^{\prime o}(\tau_{1}),\mathcal{\ldots
},\widehat{\alpha}_{i,j}^{\prime o}(\tau_{n})\right]  ^{^{\intercal}%
},\label{eq4.7}%
\end{align}
as the values of ${\mathbf{A}}^{o}$ and ${\mathbf{B}}^{o}$ that minimise
$\mathcal{Q}_{i}({\mathbf{A}}^{o},{\mathbf{B}}^{o})$. We need to impose the
following condition on the tuning parameter $\lambda_{2}$ and the lower bounds
for the significant time-varying coefficients in the transition matrix.

\begin{assumption}
\label{ass:4}

\emph{(i)\ The tuning parameter $\lambda_{2}$ satisfies
\[
\sqrt{n}s\log(n\vee d)\zeta_{n,d}+\sqrt{ns}\lambda_{1}=o(\lambda_{2}),
\]
where $\zeta_{n,d}$ is defined in Assumption \ref{ass:3}(i).}

\emph{(ii)\ It holds that
\[
\min_{1\leq i\leq d}\min_{j\in\mathcal{J}_{i}}\left( \sum_{t=1}^{n}%
\alpha_{i,j}^{2}(\tau_{t})\right) ^{\frac{1}{2}}\geq(a_{0}+1)\lambda
_{2}\ \ \mbox{and}\ \ \min_{1\leq i\leq d}\min_{j\in\mathcal{J}_{i}^{\prime}%
}D_{i,j}\geq(a_{0}+1)\lambda_{2},
\]
where $a_{0}=3.7$ is defined in the SCAD penalty.}
\end{assumption}

When $s$ is a fixed positive integer, $h\propto n^{-1/5}$, $\lambda_{1}\propto
n^{-2/5+\eta_{0}}$ with $0<\eta_{0}<1/5$, and $d\sim\exp\left\{ n^{\eta_{1}%
}\right\} $ with $0<\eta_{1}<\eta_{0}$, it is easy to verify Assumption
\ref{ass:4}(i) by setting $\lambda_{2}\propto n^{1/2-\eta_{2}}$ with
$0<\eta_{2}<2/5-[\eta_{0}\vee(2\eta_{1})]$. Assumption \ref{ass:4}(ii) imposes
restrictions on the lower bounds for the time-varying coefficient functions
and their deviations from the means. These restrictions are weaker than
Assumption 6(ii) in \cite{LKZ15} and Assumption 8 in \cite{CLWZ21}, and they
ensure that the significant coefficient functions and derivatives can be
detected \emph{w.p.a.1}.

\begin{theorem}\label{thm:4.2}
Suppose that Assumptions \ref{ass:1}--\ref{ass:4} are satisfied. The minimiser to the objective function of the weighted group LASSO, ${\cal Q}_{i}({\mathbf A}, {\mathbf B})$, exists and equals the oracle estimates defined in (\ref{eq4.6}) and (\ref{eq4.7}) w.p.a.1. In addition, we have the following mean squared convergence result:
\begin{equation}\label{eq4.8}
\max_{1\leq i\leq d}\frac{1}{n}\sum_{t=1}^n\sum_{j=1}^d\left[ \widehat{\alpha}_{ij}(\tau_t)-\alpha_{ij}(\tau_t)\right]^2=O_P\left(s\zeta_{n,d}^2\right),
\end{equation}
where $s$ is defined in Assumption \ref{ass:2}(ii) and $\zeta_{n,d}$ is defined in Assumption \ref{ass:3}(i).
\end{theorem}

\medskip

Since the penalised local linear estimates are identical to the infeasible
oracle estimates defined in (\ref{eq4.6}) and (\ref{eq4.7}) \emph{w.p.a.1},
the sparsity property holds for the global model selection procedures proposed
in Section \ref{sec3.2}, i.e., the zero elements in the time-varying
transition matrix can be estimated exactly as zeros. Following the proof of
Theorem \ref{thm:4.2}, we may verify properties (i)--(iv) for the folded
concave penalty function discussed in \cite{FXZ14} \emph{w.p.a.1}. Hence,
Theorem \ref{thm:4.2} may be regarded as a generalisation of Theorem 1 in
\cite{FXZ14} and Theorem 3.1 in \cite{LKZ15} to high-dimensional locally
stationary time series.

\medskip

With the oracle property in Theorem \ref{thm:4.2}, it is straightforward to
derive the following consistency property of the network estimates for the
directed edges of Granger causality linkages.

\renewcommand{\thecorollary}{4.\arabic{corollary}}\setcounter{corollary}{0}

\begin{corollary}
\label{cor:4.1}

\emph{Under the assumptions of Theorem \ref{thm:4.2}, we have}
\begin{equation}
\label{eq4.9}\mathsf{P}\left( \widehat{\mathbb{E}}_{n}^{G}={\mathbb{E}}%
_{n}^{G}\right) \rightarrow1.
\end{equation}

\end{corollary}

\subsection{Uniform consistency of the time-varying CLIME estimates}

\label{sec4.3}

To derive the uniform consistency property of the time-varying CLIME
estimates, we need the following conditions on the tuning parameters $b$ and
$\lambda_{3}$.

\begin{assumption}
\label{ass:5}

\emph{(i)\ The bandwidth $b$ satisfies
\[
b\rightarrow0\ \ \ \mbox{and}\ \ \ nb/[\log(n\vee d)]^{3}\rightarrow\infty.
\]
In addition, $s\zeta_{n,d}\sqrt{\log(n\vee d)}\rightarrow0$, where
$\zeta_{n,d}$ is defined in Assumption \ref{ass:3}(i).}

\emph{(ii)\ There exists a sufficiently large constant $C_{3}$ such that
$\lambda_{3}=C_{3}\left( \nu_{n,d}^{\diamond}+\nu_{n,d}^{\ast}\right) $, where
}
\[
\nu_{n,d}^{\diamond}=\left[ \frac{\log(n\vee d)}{nb}\right] ^{1/2}%
+b^{2}\ \ \ \mbox{and}\ \ \ \nu_{n,d}^{\ast}=s\zeta_{n,d}\sqrt{\log(n\vee
d)}.
\]

\end{assumption}

The following theorem gives the uniform convergence rates of the time-varying
precision matrix estimate $\widehat{\boldsymbol{\Omega}}(\tau)$ under various
matrix norms.

\begin{theorem}\label{thm:4.3}
Suppose Assumptions \ref{ass:1}--\ref{ass:5} are satisfied and $\left\{{\boldsymbol\Omega}(\tau): 0\leq \tau\leq 1\right\}\in{\cal S}(q, \xi_d)$. Then we have
\begin{eqnarray}
&&\sup_{0\leq \tau\leq 1}\left\Vert\widehat{\boldsymbol\Omega}(\tau)-{\boldsymbol\Omega}(\tau)\right\Vert_{\max}=O_P\left(\nu_{n,d}^\diamond+\nu_{n,d}^\ast\right),\label{eq4.10}\\
&&\sup_{0\leq \tau\leq 1}\left\Vert \widehat{\boldsymbol\Omega}(\tau)-{\boldsymbol\Omega}(\tau)\right\Vert =O_P\left( \xi_d(\nu_{n,d}^\diamond+\nu_{n,d}^\ast)^{1-q}\right),\label{eq4.11}\\
&&\sup_{0\leq\tau\leq1}\frac{1}{d}\left\Vert \widehat{\boldsymbol\Omega}(\tau)-{\boldsymbol\Omega}(\tau)\right\Vert _{F}^2=O_P\left( \xi_d(\nu_{n,d}^\diamond+\nu_{n,d}^\ast)^{2-q}\right),\label{eq4.12}
\end{eqnarray}
where $\xi_d$ is defined in (\ref{eq3.7}), $\nu_{n,d}^\diamond$ and $\nu_{n,d}^\ast$ are defined in Assumption \ref{ass:5}(ii).
\end{theorem}

\medskip

The uniform convergence rates in Theorem \ref{thm:4.3} rely on $\nu
_{n,d}^{\diamond}$ and $\nu_{n,d}^{\ast}$. The first rate $\nu_{n,d}%
^{\diamond}$ is the conventional uniform convergence rate for nonparametric
kernel-based quantities, whereas the second rate $\nu_{n,d}^{\ast}$ is from
the approximation errors of $\widehat{e}_{t}$ to the latent VAR errors $e_{t}%
$. Note that the dimension $d$ affects the uniform convergence rates via
$\xi_{d}$ and $\log(n\vee d)$, and the uniform consistency property holds in
the ultra-high dimensional setting when $d$ diverges at an exponential rate of
$n$. Theorem \ref{thm:4.3} can be seen as an extension of Theorem 1 in
\cite{CLL11} to the high-dimensional locally stationary time series setting.

\medskip

From Theorem \ref{thm:4.3}, we readily have the following consistency property
for the network estimates of the undirected edges of partial correlation linkages.

\begin{corollary}
\label{cor:4.2}

\emph{Under the assumptions of Theorem \ref{thm:4.3}, if $\min_{(i,j)\in
{\mathbb{E}}^{P}} \min_{1\leq t\leq n}\vert\omega_{ij}(\tau_{t})\vert
\gg\lambda_{3}$, we have}
\begin{equation}
\label{eq4.13}\mathsf{P}\left( \widehat{\mathbb{E}}_{n}^{P}={\mathbb{E}}%
_{n}^{P}\right) \rightarrow1.
\end{equation}

\end{corollary}


\section{Factor-adjusted time-varying VAR and networks}

\label{sec5} \renewcommand{\theequation}{5.\arabic{equation}} \setcounter{equation}{0}

In this section, we let $(Z_{t}:t=1,\mathcal{\ldots},n)$ with $Z_{t}%
=(z_{t,1},\mathcal{\ldots},z_{t,d})^{^{\intercal}}$ be an observed sequence of
$d$-dimensional random vectors. To accommodate strong cross-sectional
dependence which is not uncommon for large-scale time series collected in
practice, we assume that $Z_{t}$ is generated by an approximate factor model:
\begin{equation}
Z_{t}={\boldsymbol{\Lambda}}F_{t}+X_{t},\ \ t=1,\mathcal{\ldots}%
,n,\label{eq5.1}%
\end{equation}
where ${\boldsymbol{\Lambda}}=(\Lambda_{1},\mathcal{\ldots},\Lambda
_{d})^{^{\intercal}}$ is a $d\times k$ matrix of factor loadings, $F_{t}$ is a
$k$-dimensional vector of latent factors and $(X_{t})$ is assumed to satisfy
the time-varying VAR model (\ref{eq2.1}). More generally, we may assume the
following time-varying factor model structure:
\begin{equation}
Z_{t}={\boldsymbol{\Lambda}}_{t}F_{t}+X_{t},\ \ t=1,\mathcal{\ldots
},n,\label{eq5.2}%
\end{equation}
where ${\boldsymbol{\Lambda}}_{t}={\boldsymbol{\Lambda}}(t/n)$ is a
time-varying factor loading matrix with each entry being a smooth function of
scaled time. The approximate factor model and its time-varying generalisation
have been extensively studied in the literature
\citep[e.g.,][]{CR83, BN02, SW02, MHvS11, SW17}. The primary interest of this
section is to estimate the time-varying networks for the idiosyncratic error
vector $X_{t}$. Even though the components of $Z_{t}$ may be highly
correlated, those of $X_{t}$ are often only weakly correlated. Hence, it is
sensible to impose the sparsity assumption on the time-varying transition and
precision matrices of the idiosyncratic error process, making it possible to
apply the estimation methodology proposed in Section \ref{sec3}. However, this
is non-trivial as neither the common components (${\boldsymbol{\Lambda}}F_{t}$
or ${\boldsymbol{\Lambda}}_{t}F_{t}$) nor the idiosyncratic error components
are observable. Motivated by recent work on bridging factor and sparse models
for high-dimensional data \citep[e.g.,][]{FMM21, KM22}, we next use the
principal component analysis (PCA) or its localised version to remove the
common components driven by latent factors in the observed time series data.

Let ${\mathbf{Z}}=\left(  Z_{1},\mathcal{\ldots},Z_{n}\right)  ^{^{\intercal}%
}$, ${\mathbf{F}}=\left(  F_{1},\mathcal{\ldots},F_{n}\right)  ^{^{\intercal}%
}$ and ${\mathbf{X}}=\left(  X_{1},\mathcal{\ldots},X_{n}\right)
^{^{\intercal}}$. For the conventional factor model (\ref{eq5.1}), we conduct
an eigenanalysis on the $n\times n$ matrix ${\mathbf{Z}}{\mathbf{Z}%
}^{^{\intercal}}$. The estimate of ${\mathbf{F}}$, denoted as $\widehat
{\mathbf{F}}=\left(  \widehat{F}_{1},\mathcal{\ldots},\widehat{F}_{n}\right)
^{^{\intercal}}$, is obtained as the $n\times k$ matrix consisting of the
eigenvectors (multiplied by $\sqrt{n}$) corresponding to the $k$ largest
eigenvalues of ${\mathbf{Z}}{\mathbf{Z}}^{^{\intercal}}$. The factor loading
matrix is estimated by $\widehat{\boldsymbol{\Lambda}}=\left(  \widehat
{\Lambda}_{1},\mathcal{\ldots},\widehat{\Lambda}_{d}\right)  ^{^{\intercal}%
}={\mathbf{Z}}^{^{\intercal}}\widehat{\mathbf{F}}/n$. Consequently, the common
component ${\boldsymbol{\Lambda}}F_{t}$is estimated by $\widehat
{\boldsymbol{\Lambda}}\widehat{F}_{t}$ and the idiosyncratic error component
$X_{t}$ is estimated by
\begin{equation}
\widehat{X}_{t}=Z_{t}-\widehat{\boldsymbol{\Lambda}}\widehat{F}_{t}%
,\ \ t=1,\mathcal{\ldots},n.\label{eq5.3}%
\end{equation}
For the time-varying factor model (\ref{eq5.2}), the above PCA estimation
procedure needs some amendments. Specifically, let
\[
K_{t,h_{\ast}}(\tau)=\frac{K_{h_{\ast}}(\tau_{t}-\tau)}{\sum_{s=1}%
^{n}K_{h_{\ast}}(\tau_{s}-\tau)},\ \ 0<\tau<1,
\]
where $h_{\ast}$ is a bandwidth and $K_{h_{\ast}}(\cdot)$ is defined as in
Section \ref{sec3.1}, and define the localised data matrix:
\[
{\mathbf{Z}}(\tau)=\left[  Z_{1}(\tau),\mathcal{\ldots},Z_{n}(\tau)\right]
^{^{\intercal}}\ \ \mathrm{with}\ \ Z_{t}(\tau)=Z_{t}K_{t,h_{\ast}}^{1/2}%
(\tau).
\]
Through an eigenanalysis on the matrix ${\mathbf{Z}}(\tau){\mathbf{Z}%
}^{^{\intercal}}(\tau)$, we can obtain the local PCA estimates of the factors
and factor-loading matrix, denoted by $\widehat{\mathbf{F}}(\tau)=\left[
\widehat{F}_{1}(\tau),\mathcal{\ldots},\widehat{F}_{n}(\tau)\right]
^{^{\intercal}}$ and $\widehat{\boldsymbol{\Lambda}}(\tau)$, respectively.
Then, the idiosyncratic error vector $X_{t}$ is approximated by
\begin{equation}
\widehat{X}_{t}=Z_{t}-\widehat{\boldsymbol{\Lambda}}(\tau_{t})\widehat{F}%
(\tau_{t}),\ \ t=1,\mathcal{\ldots},n,\label{eq5.4}%
\end{equation}
where we've kept the same notation $\widehat{X}_{t}$ as in (\ref{eq5.3}) to
avoid notational burden.

As in Section \ref{sec4}, we only consider the time-varying VAR(1) model for
the idiosyncratic error vector. With the approximation $\widehat{X}_{t}$, we
can apply the three-stage estimation procedure proposed in Section \ref{sec3}.
Denote the preliminary time-varying LASSO estimate as $\widetilde{\alpha}%
_{ij}^{\dagger}(\cdot)$, the second-stage weighted group LASSO estimate as
$\widehat{\alpha}_{ij}^{\dagger}(\cdot)$, and the factor-adjusted time-varying
precision matrix estimate as $\widehat{\boldsymbol{\Omega}}^{\dagger}%
(\cdot)=\left[ \widehat\omega_{ij}^{\dagger}(\cdot)\right] _{d\times d}$.
Subsequently, we may construct the uniform network estimates $\widehat
{\mathbb{E}}_{n}^{G,{\dagger}}$ and $\widehat{\mathbb{E}}_{n}^{P,{\dagger}}$,
defined similarly to $\widehat{\mathbb{E}}_{n}^{G}$ and $\widehat{\mathbb{E}%
}_{n}^{P}$ in (\ref{eq3.13}) and (\ref{eq3.14}), but with $\widehat{\alpha
}_{ij}(\cdot)$ and $\widehat\omega_{ij}(\cdot)$ replaced by $\widehat{\alpha
}_{ij}^{\dagger}(\cdot)$ and $\widehat\omega_{ij}^{\dagger}(\cdot)$,
respectively. To derive the convergence properties of these factor-adjusted
estimates, we need the following assumption, which modifies Assumptions
\ref{ass:3}--\ref{ass:5} to incorporate the approximation error of the
idiosyncratic error components.

\begin{assumption}
\label{ass:6}

\emph{(i) Denote $\delta_{X}=\max_{1\leq t\leq n}\left\vert \widehat{X}%
_{t}-X_{t}\right\vert _{\max}$. It holds that $[\log(n\vee d)]^{1/2}%
s\delta_{X}=o_{P}(1)$.}

\emph{(ii) Assumption \ref{ass:3}(i) holds when $\zeta_{n,d}$ is replaced by
$\zeta_{n,d}^{\dagger}=\zeta_{n,d}+[\log(n\vee d)]^{1/2}s\delta_{X}$.}

\emph{(iii) Assumption \ref{ass:4}(i) holds when $\zeta_{n,d}$ is replaced by
$\zeta_{n,d}^{\dagger}$.}

\emph{(iv) Assumption \ref{ass:5} holds when $\zeta_{n,d}$ and $\nu
_{n,d}^{\ast}$ are replaced by $\zeta_{n,d}^{\dagger}$ and $\nu_{n,d}%
^{\dagger}=s\zeta_{n,d}^{\dagger}\sqrt{\log(n\vee d)}$, respectively.}
\end{assumption}

Assumption \ref{ass:6}(i) imposes a high-level condition on the approximation
of the latent $X_{t}$ in the factor model, i.e., the approximation error
$\delta_{X}$ uniformly converges to zero with a rate faster than $s^{-1}%
[\log(n\vee d)]^{-1/2}$. By Corollary 1 in \cite{FLM13}, a typical rate for
the approximation error from PCA estimation of the conventional factor model
(\ref{eq5.1}) is
\begin{equation}
\label{eq5.5}\delta_{X}=O_{P}\left( (\log n)^{1/2}\left[ (\log d)^{1/2}
n^{-1/2}+n^{1/\upsilon}d^{-1/2}\right] \right) ,
\end{equation}
where $\upsilon>2$ is a positive number related to moment restrictions. From
Theorem 3.5 in \cite{SW17}, we may obtain the typical uniform rate for
$\delta_{X}$ under the time-varying factor model (\ref{eq5.2}) when the local
PCA estimation is used. In Assumption \ref{ass:6}(ii)--(iv), we amend
Assumptions \ref{ass:3}(i), \ref{ass:4}(i) and \ref{ass:5}(ii) to incorporate
the approximation error $\delta_{X}$. However, if we further assume that
$h\propto n^{-1/5}$ and $d$ diverges at a polynomial rate of $n$ satisfying
$d\gg n^{1+2/\upsilon}$, then the rate in (\ref{eq5.5}) can be simplified to
$\delta_{X}=O_{P}\left( (\log d)n^{-1/2}\right) =o_{P}(h^{2})$ and thus
$\zeta_{n,d}\propto\zeta_{n,d}^{\dagger}$. Consequently, we may remove
Assumption \ref{ass:6}(ii)--(iv) and $\delta_{X}$ would not be involved in the
estimation convergence rates under model (\ref{eq5.1}).

The following two propositions extend the theoretical results in Section
\ref{sec4} to the factor-adjusted time-varying VAR and networks.

\renewcommand{\theprop}{5.\arabic{prop}}\setcounter{prop}{0}

\begin{prop}
\label{prop:5.1}

\emph{Suppose that the factor model (\ref{eq5.1}) or (\ref{eq5.2}), and
Assumptions \ref{ass:1}, \ref{ass:2} and \ref{ass:3}(ii) are satisfied.}

\emph{(i) Under Assumption \ref{ass:6}(i)(ii), we have}
\begin{equation}
\label{eq5.6}\max_{1\leq i\leq d}\max_{1\leq t\leq n}\sum_{j=1}^{d}\left[
\widetilde{\alpha}_{ij}^{\dagger}(\tau_{t})-{\alpha}_{ij}(\tau_{t})\right]
^{2}=O_{P}\left( s\lambda_{1}^{2}\right) .
\end{equation}

\emph{(ii) Under Assumption \ref{ass:6}(i)--(iii), the oracle property holds
for the second-stage weighted group LASSO estimates and furthermore, }
\begin{equation}
\label{eq5.7}\max_{1\leq i\leq d}\frac{1}{n}\sum_{t=1}^{n}\sum_{j=1}%
^{d}\left[  \widehat{\alpha}_{ij}^{\dagger}(\tau_{t})-\alpha_{ij}(\tau
_{t})\right] ^{2}=O_{P}\left( s\left( \zeta_{n,d}^{\dagger}\right) ^{2}\right)
.
\end{equation}

\emph{(iii) Under Assumption \ref{ass:6} and the sparsity condition that
$\left\{ {\boldsymbol{\Omega}}(\tau): 0\leq\tau\leq1\right\} \in\mathcal{S}(q,
\xi_{d})$, we have}
\begin{align}
& \sup_{0\leq\tau\leq1}\left\Vert \widehat{\boldsymbol{\Omega}}^{\dagger}%
(\tau)-{\boldsymbol{\Omega}}(\tau)\right\Vert _{\max}=O_{P}\left( \nu
_{n,d}^{\diamond}+\nu_{n,d}^{\dagger}\right) ,\label{eq5.8}\\
& \sup_{0\leq\tau\leq1}\left\Vert \widehat{\boldsymbol{\Omega}}^{\dagger}%
(\tau)-{\boldsymbol{\Omega}}(\tau)\right\Vert =O_{P}\left(  \xi_{d}(\nu
_{n,d}^{\diamond}+\nu_{n,d}^{\dagger})^{1-q}\right) ,\label{eq5.9}\\
& \sup_{0\leq\tau\leq1}\frac{1}{d}\left\Vert \widehat{\boldsymbol{\Omega}%
}^{\dagger}(\tau)-{\boldsymbol{\Omega}}(\tau)\right\Vert _{F}^{2}=O_{P}\left(
\xi_{d}(\nu_{n,d}^{\diamond}+\nu_{n,d}^{\dagger})^{2-q}\right) .\label{eq5.10}%
\end{align}

\end{prop}

\begin{prop}
\label{prop:5.2}

\emph{(i) Under the assumptions of Proposition \ref{prop:5.1}(ii), we have}
\begin{equation}
\label{eq5.11}\mathsf{P}\left( \widehat{\mathbb{E}}_{n}^{G,\dagger
}={\mathbb{E}}_{n}^{G}\right) \rightarrow1.
\end{equation}

\emph{(ii) Under the assumptions of Proposition \ref{prop:5.1}(iii) and
$\min_{(i,j)\in{\mathbb{E}}^{P}} \min_{1\leq t\leq n}\vert\omega_{ij}(\tau
_{t})\vert\gg\lambda_{3}$, we have}
\begin{equation}
\label{eq5.12}\mathsf{P}\left( \widehat{\mathbb{E}}_{n}^{P,\dagger
}={\mathbb{E}}_{n}^{P}\right) \rightarrow1.
\end{equation}

\end{prop}


\section{Monte-Carlo simulation}

\label{sec6} \renewcommand{\theequation}{6.\arabic{equation}} \setcounter{equation}{0}

In this section, we provide four simulated examples to examine the
finite-sample numerical performance of the proposed high-dimensional
time-varying VAR and network estimates. Throughout this section, we denote the
proposed time-varying weighted group LASSO method as tv-wgLASSO and the
time-varying CLIME method as tv-CLIME. We compare the performance of the
tv-wgLASSO with the (infeasible) time-varying oracle estimation, denoted as
tv-Oracle, which estimates only the true significant coefficient functions
(assuming they were known), and the unpenalised full time-varying estimation,
denoted as tv-Full, which estimates all the coefficient functions without
penalisation. We compare the performance of tv-CLIME with the time-varying
graphical LASSO estimation, denoted as tv-GLASSO, which is implemented using
the R package ``\textsf{glassoFast}" on the VAR residuals. In addition, to
investigate the loss of estimation accuracy due to the VAR model error
approximation, we also report results from the infeasible tv-CLIME, which
directly uses the VAR errors (rather than residuals) in the estimation of the
precision matrices.

In the simulation, we use the Epanechnikov kernel $K(t)=0.75(1-t^{2})_{+}$
with bandwidth $h=b=0.75[\log(d)/n]^{1/5}$ as in \cite{LKZ15}. The bandwidth
for the local PCA is set as $h_{\ast}=(2.35/\sqrt{12})[\sqrt{d}/n] ^{1/5}$ as
in \cite{SW17}. We set the sample size $n$ as 200 and 400, and the dimension
$d$ as 50 and 100. Although such dimensions are smaller than the sample size,
when $n=200$ and $d=100$, the ``effective sample size" used in each local
linear estimation in (\ref{eq3.3}) is approximately $2nh\approx140$, which is
smaller than the combined number of unknown coefficient functions and their
derivative, $2d=200$. Consequently, in this case we fail to implement the
naive tv-Full estimation. There are three tuning parameters in the proposed
estimation procedure: $\lambda_{1}$ in the first stage of preliminary
time-varying LASSO estimation, $\lambda_{2}$ in the second stage of
time-varying weighted group LASSO, and $\lambda_{3}$ in the third stage of
time-varying CLIME. They are selected by the Bayesian information criterion
(BIC), the generalised information criterion (GIC), and the extended Bayesian
information criterion (EBIC), respectively. Appendix E in the supplement gives
definitions of these information criteria.

To evaluate whether the time-varying model structure is accurately estimated,
we report the false positive (FP), the false negative (FN), the true positive
rate (TPR), the true negative rate (TNR), the positive predictive value (PPV),
the negative predictive value (NPV), the F1 score (F1), and the Matthews
correlation coefficient (MCC). Definitions of these measures are available in
Appendix E of the supplement. To evaluate the performance of the coefficient
estimators, we report the average R square (average $R^{2}$) over all the
dimensions, the average scaled Frobenius norm of estimation errors of
coefficient functions (EE$_{A}$), and the root-mean-squared error of the
errors (RMSE$_{e}$). Taking our proposed tv-wgLASSO estimator for time-varying
VAR(1) as an example,
\[
\mathrm{EE}_{A}=\frac{1}{n\sqrt{d}}\sum_{t=1}^{n}\left\Vert \widehat
{\mathbf{A}}_{1}(\tau_{t})-{\mathbf{A}}_{1}(\tau_{t})\right\Vert
_{F}\ \ \mathrm{and}\ \ \mathrm{RMSE}_{e}=\sqrt{\frac{1}{nd}\sum_{i=1}^{d}%
\sum_{t=1}^{n}(\widehat{e}_{t,i}-{e}_{t,i})^{2}}.
\]
To evaluate the performance of the precision matrix estimators, we report the
average scaled Frobenius norm of estimation error ($\mathrm{EE}_{\Omega}$)
defined as
\[
\mathrm{EE}_{\Omega}=\frac{1}{n\sqrt{d}}\sum_{t=1}^{n}\left\Vert
\widehat{\boldsymbol{\Omega}}(\tau_{t})-{\boldsymbol{\Omega}}(\tau
_{t})\right\Vert _{F}.
\]
All the above measures are calculated for each Monte Carlo replication and
then averaged over $100$ replications.

\medskip

\noindent\textbf{Example 1.}\ \ The data is generated from a time-varying
VAR(1) model with ${\mathbf{A}}_{1}(\tau)$ being a diagonal matrix for all
$\tau\in[0,1]$. Each diagonal entry of ${\mathbf{A}}_{1}(\tau)$ independently
takes a value of either $0.64\Phi(5(\tau-1/2))$ or $0.64-0.64\Phi
(5(\tau-1/2))$ with an equal probability of 0.5, where $\Phi(\cdot)$ is the
standard normal distribution function. We set ${\boldsymbol{\Omega}}(\tau)$ to
be a block diagonal matrix: ${\boldsymbol{\Omega}}(\tau)={\mathbf{I}}%
_{d/2}\otimes{\boldsymbol{\Omega}}_{\ast}(\tau)$, where ${\boldsymbol{\Omega}%
}_{\ast}(\tau)=\left[ \omega_{ij,\ast}(\tau)\right] _{2\times2}$ with
$\omega_{11,\ast}(\tau)=\omega_{22,\ast}(\tau)\equiv1$, and $\omega_{12,\ast
}(\tau)=\omega_{21,\ast}(\tau)=1.4\Phi(5(\tau-1/2))-0.7$. The diagonal
structure of ${\mathbf{A}}_{1}(\tau)$ implies that no Granger causality exists
between variables, whereas the block diagonal structure of
${\boldsymbol{\Omega}}(\tau)$ results in weak cross-sectional dependence
between the components of $X_{t}$.

Table \ref{tab:1} reports the estimation results of the time-varying
transition matrices and Granger networks. For the proposed tv-wgLASSO, the FP
and FN values are very small compared with $d^{2}$ (the total number of
potential directed Granger causality linkages or entries of the transition
matrix). This leads to large values of the TPR, TNR, PPV, NPV, F1 and MCC
measures, all of which are close to $1$. We can also see that the FP and FN
values double when $d$ increases from $50$ to $100$, but decrease
substantially when $n$ grows from $200$ to $400$. These results clearly show
that tv-wgLASSO can accurately recover the time-varying Granger network as
long as the sample size is moderately large. The average $R^{2}$ of tv-wgLASSO
is close to that of tv-Oracle, but the naive tv-Full method tends to have
large $R^{2}$ due to model over-fitting. Although the EE$_{A}$ values of
tv-wgLASSO are larger than those of tv-Oracle when $n=200$, they drop
significantly and are even slightly smaller than those of tv-Oracle when
$n=400$. A similar pattern can be observed in RMSE$_{e}$, indicating that the
proposed tv-wgLASSO is capable of providing good approximations to VAR errors,
which are used in the subsequent time-varying precision matrix estimation.
Unsurprisingly, the tv-Full method fails to estimate the time-varying
transition matrix when $d=100$ and $n=200$.

Table \ref{tab:2} reports the estimation results of the time-varying precision
matrices and partial correlation networks. When $n=200$, both tv-CLIME and
tv-GLASSO have zero FP values, whereas tv-CLIME has smaller FN than tv-GLASSO.
Hence, the proposed tv-CLIME performs better than tv-GLASSO in terms of the F1
and MCC measures. When $n=400$, both tv-CLIME and tv-GLASSO correctly recover
the time-varying partial correlation networks. In terms of the precision
matrix estimation accuracy (EE$_{\Omega}$), tv-GLASSO performs slightly better
than tv-CLIME. In addition, by comparing the tv-CLIME and the infeasible
tv-CLIME, we may conclude that the VAR error approximation has negligible
impact on the precision matrix and partial correlation network estimation.

\begin{table}
\caption{Transition matrix and Granger network estimation in Example 1.}%
\label{tab:1}
\centering
\begin{tabular}
[c]{llllllllll}\hline\hline
&  & \multicolumn{2}{l}{tv-wgLASSO} &  & \multicolumn{2}{l}{tv-Oracle} &  &
\multicolumn{2}{l}{tv-Full}\\\cline{3-4}\cline{6-7}\cline{9-10}%
measure & dimension & $n=200$ & $n=400$ &  & $n=200$ & $n=400$ &  & $n=200$ &
$n=400$\\\hline
FP & $d=50$ & 0.97 & 0.04 &  & 0 & 0 &  & 2450 & 2450\\
& $d=100$ & 1.73 & 0.08 &  & 0 & 0 &  & - & 9900\\
FN & $d=50$ & 3.53 & 0.08 &  & 0 & 0 &  & 0 & 0\\
& $d=100$ & 8.55 & 0.15 &  & 0 & 0 &  & - & 0\\
TPR & $d=50$ & 0.929 & 0.998 &  & 1 & 1 &  & 1 & 1\\
& $d=100$ & 0.915 & 0.999 &  & 1 & 1 &  & - & 1\\
TNR & $d=50$ & 1.000 & 1.000 &  & 1 & 1 &  & 0 & 0\\
& $d=100$ & 1.000 & 1.000 &  & 1 & 1 &  & - & 0\\
PPV & $d=50$ & 0.980 & 0.999 &  & 1 & 1 &  & 0.02 & 0.02\\
& $d=100$ & 0.982 & 0.999 &  & 1 & 1 &  & - & 0.01\\
NPV & $d=50$ & 0.999 & 1.000 &  & 1 & 1 &  & 1 & 1\\
& $d=100$ & 0.999 & 1.000 &  & 1 & 1 &  & - & 1\\
F1 & $d=50$ & 0.953 & 0.999 &  & 1 & 1 &  & 0.039 & 0.039\\
& $d=100$ & 0.947 & 0.999 &  & 1 & 1 &  & - & 0.020\\
MCC & $d=50$ & 0.953 & 0.999 &  & 1 & 1 &  & 0 & 0\\
& $d=100$ & 0.947 & 0.999 &  & 1 & 1 &  & - & 0\\\hline
average $R^{2}$ & $d=50$ & 0.289 & 0.296 &  & 0.296 & 0.297 &  & 0.933 &
0.721\\
& $d=100$ & 0.296 & 0.306 &  & 0.305 & 0.307 &  & - & 0.959\\\hline
EE$_{A}$ & $d=50$ & 0.214 & 0.160 &  & 0.185 & 0.163 &  & 54.29 & 1.410\\
& $d=100$ & 0.224 & 0.163 &  & 0.189 & 0.166 &  & - & 112.8\\
RMSE$_{e}$ & $d=50$ & 0.203 & 0.115 &  & 0.162 & 0.120 &  & 1.119 & 0.876\\
& $d=100$ & 0.213 & 0.113 &  & 0.159 & 0.119 &  & - & 1.145\\\hline
\end{tabular}
\par
\begin{flushleft}
\emph{In all the tables, except for exact values of 0's and 1's, the FP and FN
measures are rounded to 2 decimal places, while the others are rounded to 3
decimal places. }
\end{flushleft}
\end{table}

\begin{table}
\caption{Precision matrix and partial correlation network estimation in
Example 1.}%
\label{tab:2}
\centering
\begin{tabular}
[c]{llllllllll}\hline\hline
&  & \multicolumn{2}{c}{tv-CLIME} &  & \multicolumn{2}{c}{infeasible tv-CLIME}
&  & \multicolumn{2}{c}{tv-GLASSO}\\\cline{3-4}\cline{6-7}\cline{9-10}%
measure & dimension & $n=200$ & $n=400$ &  & $n=200$ & $n=400$ &  & $n=200$ &
$n=400$\\\hline
FP & $d=50$ & 0 & 0.02 &  & 0 & 0.02 &  & 0 & 0\\
& $d=100$ & 0 & 0.03 &  & 0 & 0.01 &  & 0 & 0\\
FN & $d=50$ & 5.06 & 0 &  & 3.49 & 0 &  & 9.24 & 0\\
& $d=100$ & 13.25 & 0 &  & 9.01 & 0 &  & 28.31 & 0\\
TPR & $d=50$ & 0.798 & 1 &  & 0.860 & 1 &  & 0.630 & 0\\
& $d=100$ & 0.735 & 1 &  & 0.820 & 1 &  & 0.434 & 0\\
TNR & $d=50$ & 1 & 1.000 &  & 1 & 1.000 &  & 1 & 1\\
& $d=100$ & 1 & 1.000 &  & 1 & 1.000 &  & 1 & 1\\
PPV & $d=50$ & 1 & 0.999 &  & 1 & 0.999 &  & 1 & 1\\
& $d=100$ & 1 & 0.999 &  & 1 & 1.000 &  & 1 & 1\\
NPV & $d=50$ & 0.996 & 1 &  & 0.097 & 1 &  & 0.992 & 1\\
& $d=100$ & 0.997 & 1 &  & 0.998 & 1 &  & 0.994 & 1\\
F1 & $d=50$ & 0.884 & 1.000 &  & 0.922 & 1.000 &  & 0.768 & 1\\
& $d=100$ & 0.845 & 1.000 &  & 0.899 & 1.000 &  & 0.600 & 1\\
MCC & $d=50$ & 0.889 & 1.000 &  & 0.925 & 1.000 &  & 0.788 & 1\\
& $d=100$ & 0.855 & 1.000 &  & 0.904 & 1.000 &  & 0.653 & 1\\\hline
EE$_{\Omega}$ & $d=50$ & 0.510 & 0.436 &  & 0.503 & 0.435 &  & 0.451 & 0.407\\
& $d=100$ & 0.481 & 0.421 &  & 0.473 & 0.419 &  & 0.433 & 0.397\\\hline
\end{tabular}
\end{table}

\medskip

\noindent\textbf{Example 2}.\ \ The data is generated from a time-varying
VAR(1) model with ${\mathbf{A}}_{1}(\tau)$ being an upper triangular matrix
for all $\tau\in[0,1]$. Each diagonal entry of ${\mathbf{A}}_{1}(\tau)$ takes
the value of $0.7\Phi(5(\tau-1/2))$, each super-diagonal entry takes the value
of $0.7-0.7\Phi(5(\tau-1/2))$, and the remaining entries take the value of
$0$. We set ${\boldsymbol{\Omega}}(\tau)=\left[ \omega_{ij}(\tau)\right]
_{d\times d}$ to be a banded symmetric matrix for all $\tau\in[0,1]$ with
$\omega_{ii}(\tau)\equiv1$, $\omega_{i,(i+1)}(\tau)=0.7\Phi(5(\tau-1/2))-0.7$,
$\omega_{i,(i+2)}(\tau)=0.7-0.7\Phi(5(\tau-1/2))$, and $\omega_{i,j}%
(\tau)\equiv0$ if $|i-j|>2$.

Table \ref{tab:3} reports the estimation results of the time-varying
transition matrices and Granger networks. Note that the time series variables
in this example are more correlated to each other than those in Example 1,
which affects the network estimation accuracy. When $d=100$ and $n=200$, the
FP and FN values of tv-wgLASSO reach their maximum at 20.73 and 37.55,
respectively, whereas the F1 and MCC values are around $0.85$. As in Example
1, the F1 and MCC values increase when $n$ increases from $200$ to $400$, and
again the average $R^{2}$ of tv-wgLASSO is close to that of tv-Oracle.
However, tv-wgLASSO has much larger EE$_{A}$ and RMSE$_{e}$ than tv-Oracle.

Table \ref{tab:4} reports the estimation results of the time-varying precision
matrices and partial correlation networks. It follows from the EE$_{A}$ and
RMSE$_{e}$ results in Table \ref{tab:3} that the VAR error approximation is
poorer than that in Example 1. Consequently the proposed tv-CLIME performs
worse than the infeasible tv-CLIME using the true VAR errors directly in the
estimation. In particular, FN of the tv-CLIME is much larger than that of the
infeasible tv-CLIME when $n=200$. Due to the same reason, the infeasible
tv-CLIME also outperforms the tv-GLASSO. In addition, we find that the
tv-CLIME is better than the tv-GLASSO in recovering the time-varying precision
network when $n=200$, and they perform equally well when $n=400$.

\begin{table}
\caption{Transition matrix and Granger network estimation in Example 2.}%
\label{tab:3}%
\centering
\begin{tabular}
[c]{llllllllll}\hline\hline
&  & \multicolumn{2}{l}{tv-wgLASSO} &  & \multicolumn{2}{l}{tv-Oracle} &  &
\multicolumn{2}{l}{tv-Full}\\\cline{3-4}\cline{6-7}\cline{9-10}%
measure & dimension & $n=200$ & $n=400$ &  & $n=200$ & $n=400$ &  & $n=200$ &
$n=400$\\\hline
FP & $d=50$ & 13.53 & 12.75 &  & 0 & 0 &  & 2401 & 2401\\
& $d=100$ & 20.73 & 7.73 &  & 0 & 0 &  & - & 9801\\
FN & $d=50$ & 18.56 & 11.11 &  & 0 & 0 &  & 0 & 0\\
& $d=100$ & 37.55 & 13.90 &  & 0 & 0 &  & - & 0\\
TPR & $d=50$ & 0.813 & 0.888 &  & 1 & 1 &  & 1 & 1\\
& $d=100$ & 0.811 & 0.930 &  & 1 & 1 &  & - & 1\\
TNR & $d=50$ & 0.994 & 0.995 &  & 1 & 1 &  & 0 & 0\\
& $d=100$ & 0.998 & 0.999 &  & 1 & 1 &  & - & 0\\
PPV & $d=50$ & 0.859 & 0.875 &  & 1 & 1 &  & 0.040 & 0.040\\
& $d=100$ & 0.888 & 0.960 &  & 1 & 1 &  & - & 0.020\\
NPV & $d=50$ & 0.992 & 0.995 &  & 1 & 1 &  & 0 & 0\\
& $d=100$ & 0.996 & 0.999 &  & 1 & 1 &  & - & 0\\
F1 & $d=50$ & 0.834 & 0.881 &  & 1 & 1 &  & 0.076 & 0.076\\
& $d=100$ & 0.847 & 0.945 &  & 1 & 1 &  & - & 0.039\\
MCC & $d=50$ & 0.828 & 0.876 &  & 1 & 1 &  & 0 & 0\\
& $d=100$ & 0.846 & 0.943 &  & 1 & 1 &  & - & 0\\\hline
average $R^{2}$ & $d=50$ & 0.465 & 0.448 &  & 0.477 & 0.462 &  & 0.963 &
0.829\\
& $d=100$ & 0.473 & 0.467 &  & 0.483 & 0.471 &  & - & 0.978\\\hline
EE$_{A}$ & $d=50$ & 0.328 & 0.250 &  & 0.171 & 0.122 &  & 58.44 & 1.510\\
& $d=100$ & 0.323 & 0.204 &  & 0.168 & 0.122 &  & - & 82.60\\
RMSE$_{e}$ & $d=50$ & 0.631 & 0.476 &  & 0.417 & 0.305 &  & 1.673 & 1.414\\
& $d=100$ & 0.613 & 0.390 &  & 0.414 & 0.309 &  & - & 1.720\\\hline
\end{tabular}
\end{table}

\begin{table}
\caption{Precision matrix and partial correlation network estimation in
Example 2.}%
\label{tab:4}%
\centering
\begin{tabular}
[c]{llllllllll}\hline\hline
&  & \multicolumn{2}{l}{tv-CLIME} &  & \multicolumn{2}{l}{infeasible tv-CLIME}
&  & \multicolumn{2}{l}{tv-GLASSO}\\\cline{3-4}\cline{6-7}\cline{9-10}%
measure & dimension & $n=200$ & $n=400$ &  & $n=200$ & $n=400$ &  & $n=200$ &
$n=400$\\\hline
FP & $d=50$ & 0.03 & 0.04 &  & 0.02 & 0.03 &  & 0 & 0.01\\
& $d=100$ & 0.01 & 0 &  & 0 & 0.01 &  & 0 & 0.01\\
FN & $d=50$ & 12.62 & 0.82 &  & 2.34 & 0 &  & 20.84 & 0.06\\
& $d=100$ & 24.71 & 0.23 &  & 6.21 & 0.01 &  & 49.73 & 0.43\\
TPR & $d=50$ & 0.742 & 0.983 &  & 0.952 & 1 &  & 0.575 & 0.997\\
& $d=100$ & 0.750 & 0.998 &  & 0.937 & 1.000 &  & 0.498 & 0.996\\
TNR & $d=50$ & 1.000 & 1.000 &  & 1.000 & 1.000 &  & 1 & 1.000\\
& $d=100$ & 1.000 & 1 &  & 1 & 1.000 &  & 1 & 1.000\\
PPV & $d=50$ & 0.999 & 0.999 &  & 1.000 & 0.999 &  & 1 & 1.000\\
& $d=100$ & 1.000 & 1 &  & 1 & 1.000 &  & 1 & 1.000\\
NPV & $d=50$ & 0.989 & 0.999 &  & 0.998 & 1 &  & 0.983 & 1.000\\
& $d=100$ & 0.995 & 1.000 &  & 0.999 & 1.000 &  & 0.990 & 1.000\\
F1 & $d=50$ & 0.850 & 0.991 &  & 0.975 & 1.000 &  & 0.725 & 0.998\\
& $d=100$ & 0.857 & 0.999 &  & 0.967 & 1.000 &  & 0.662 & 0.998\\
MCC & $d=50$ & 0.856 & 0.991 &  & 0.975 & 1.000 &  & 0.749 & 0.998\\
& $d=100$ & 0.864 & 0.999 &  & 0.967 & 1.000 &  & 0.701 & 0.998\\\hline
EE$_{\Omega}$ & $d=50$ & 0.598 & 0.533 &  & 0.526 & 0.485 &  & 0.560 & 0.514\\
& $d=100$ & 0.560 & 0.489 &  & 0.486 & 0.458 &  & 0.536 & 0.496\\\hline
\end{tabular}
\end{table}

\medskip

\noindent\textbf{Example 3}.\ \ The data is generated from a VAR(1) model with
${\mathbf{A}}_{1}(\tau)=\left[ a_{ij}(\tau)\right] _{d\times d}$ being a
Toeplitz matrix and $a_{ij}(\tau)=(0.4-0.1\tau)^{|i-j|+1}$. We also set
${\boldsymbol{\Omega}}(\tau)=\left[ \omega_{ij}(\tau)\right] _{d\times d}$ to
be a Toeplitz matrix with $\omega_{ij}(\tau)= (0.8-0.1\tau)^{|i-j|}$. In this
example, both the transition and precision matrices are non-sparse, and we aim
to examine how our proposed methods perform when the (exact) sparsity
assumption fails.

Table \ref{tab:5} reports the estimation errors of the various methods
considered. In this example, the tv-Oracle is equivalent to tv-Full and both
suffer from the curse of dimensionality in the conventional local linear
estimation procedure for the time-varying transition matrices (in particular
when $d=100$ and $n=200$). Consequently, the EE$_{A}$ and RMSE$_{e}$ of the
tv-wgLASSO are much smaller than those of the tv-Oracle. The EE$_{\Omega}$
results of the tv-CLIME are very close to those of the infeasible tv-CLIME,
suggesting that the VAR error approximation has little impact on the tv-CLIME
performance as discussed in Example 1. In addition, the EE$_{\Omega}$ results
of the tv-CLIME and Oracle tv-CLIME are generally close to those of tv-GLASSO.
The simulation results show that the proposed tv-wgLASSO and tv-CLIME perform
reasonably well when the sparsity assumption on transition and precision
matrices is not satisfied.

\begin{table}
\caption{Estimation accuracy of dual networks in Example 3.}%
\label{tab:5}%
\centering
\begin{tabular}
[c]{llllllllll}\hline\hline
&  & \multicolumn{2}{c}{tv-wgLASSO} &  & \multicolumn{2}{c}{tv-Oracle} &  &
\multicolumn{2}{c}{tv-Full}\\\cline{3-4}\cline{6-7}\cline{9-10}%
measure & dimension & $n=200$ & $n=400$ &  & $n=200$ & $n=400$ &  & $n=200$ &
$n=400$\\\hline
average $R^{2}$ & $d=50$ & 0.009 & 0.029 &  & 0.891 & 0.588 &  & 0.891 &
0.588\\
& $d=100$ & 0.005 & 0.020 &  & - & 0.930 &  & - & 0.930\\
EE$_{A}$ & $d=50$ & 0.383 & 0.348 &  & 56.66 & 1.927 &  & 56.66 & 1.927\\
& $d=100$ & 0.388 & 0.364 &  & - & 97.60 &  & - & 97.60\\
RMSE$_{e}$ & $d=50$ & 0.515 & 0.463 &  & 1.716 & 1.300 &  & 1.716 & 1.300\\
& $d=100$ & 0.523 & 0.486 &  & - & 1.776 &  & - & 1.776\\\hline
&  & \multicolumn{2}{c}{tv-CLIME} &  & \multicolumn{2}{c}{infeasible tv-CLIME}
&  & \multicolumn{2}{c}{tv-GLASSO}\\\cline{3-4}\cline{6-7}\cline{9-10}
&  & $n=200$ & $n=400$ &  & $n=200$ & $n=400$ &  & $n=200$ & $n=400$\\\hline
EE$_{\Omega}$ & $d=50$ & 1.669 & 1.601 &  & 1.613 & 1.572 &  & 1.584 & 1.570\\
& $d=100$ & 1.674 & 1.615 &  & 1.616 & 1.580 &  & 1.587 & 1.588\\\hline
\end{tabular}
\end{table}

\medskip

\noindent\textbf{Example 4}.\ \ The data is generated from a factor-adjusted
time-varying VAR model in the form of (\ref{eq5.2}). The idiosyncratic errors
of the time-varying factor model are generated from a VAR(1) model in Example
2. The two factors in $F_{t}=(F_{t,1},F_{t,2})^{^{\intercal}}$ are generated
from two univariate AR(1) processes: $F_{t,1}=0.6F_{t-1,1}+\sqrt{1-0.6^{2}%
}u_{t,1}^{F}$ and $F_{t,2}=0.3F_{t-1,2}+\sqrt{1-0.3^{2}}u_{t,2}^{F}$, where
$u_{t,1}^{F}$ and $u_{t,2}^{F}$ are independently drawn from a standard normal
distribution. The factor-loading matrix is defined as ${\boldsymbol{\Lambda}%
}_{t}=\left(  \Lambda_{t,1},\Lambda_{t,2}\right)  $ where $\Lambda_{t,1}%
\equiv\Lambda_{1}$ is a time-invariant vector drawn from a $d$-dimensional
standard multivariate normal distribution and $\Lambda_{t,2}=(\Lambda
_{1t,2},\mathcal{\ldots},\Lambda_{dt,2})^{^{\intercal}}$ with $\Lambda
_{it,2}=2/\left(  1+\exp\{-2[10(t/n)-5(i/d)-2]\}\right)  $ for
$i=1,\mathcal{\ldots},d$.

Table \ref{tab:6} reports the estimation results of the time-varying
transition matrices and Granger networks for the idiosyncratic errors, and
Table \ref{tab:7} reports the estimation results of the time-varying precision
matrices and partial correlation networks. Comparing with the results in
Tables \ref{tab:3} and \ref{tab:4}, we can observe that the factor-adjusted
estimation introduces additional estimation errors, leading to smaller values
of F1 and MCC. The impact is more marked when $n=200$ but reduces
substantially when $n=400$. As in the previous examples, the F1 and MCC values
increase when $n$ increases from $200$ to $400$. Thus we may conclude that,
although the factor model estimation errors are passed onto the three-stage
estimation procedure, their impact on the estimation of the networks is not
significant when the sample size is moderately large ($n=400$).

\begin{table}
\begin{minipage}{0.45\linewidth}
	\caption{\label{tab:6}Factor-adjusted transition matrix and Granger network estimation in Example 4.}\centering
	\begin{tabular}{llll}\hline\hline
		&& \multicolumn{2}{l}{tv-wgLASSO} \\ \cline{3-4}
		
		measure&dimension&$n=200$  &$n=400$  \\ \hline
		FP &$d=50$&11.35&10.60 \\
		&  $d=100$&20.40&10.41   \\
		FN &	$d=50$&35.97&14.77  \\
		&  $d=100$& 65.45&20.68\\
		TPR&	$d=50$&0.637&0.851\\
		&  $d=100$&0.671 &0.896  \\
		TNR&	$d=50$&0.995&0.996  \\
		&  $d=100$&0.998&0.999  \\
		PPV&	$d=50$&0.852&0.890 \\
		&  $d=100$& 0.869 &0.945  \\
		NPV&	$d=50$&0.985&0.994  \\
		&  $d=100$&0.993&0.998  \\
		F1&	$d=50$&0.725&0.869 \\
		&  $d=100$&0.756&0.920  \\
		MCC&	$d=50$&0.725&0.865  \\
		&  $d=100$&0.759&0.919 \\\hline
		average $R^2$&$d=50$&0.298&0.350 \\
		&  $d=100$& 0.339&0.389 \\\hline
		EE$_A$&$d=50$& 0.413&0.283  \\
		&  $d=100$&0.396&0.241 \\
		RMSE$_e$ &$d=50$&1.319&1.025 \\
		&  $d=100$&1.230&0.856 \\\hline
	\end{tabular}
\end{minipage}
\hfill\begin{minipage}{0.45\linewidth}
		\caption{\label{tab:7}Factor-adjusted precision matrix and partial correlation network estimation in Example 4.}\centering
	\begin{tabular}{llll}\hline\hline
		&&  \multicolumn{2}{l}{tv-CLIME} \\ \cline{3-4}
		measure&dimension&$n=200$  &$n=400$  \\ \hline
		FP &$d=50$&0.01&0.01   \\
		&  $d=100$&0&0.02\\
		FN &	$d=50$&38.22&5.36 \\
		&  $d=100$&65.99&2.21 \\
		TPR&	$d=50$&0.220&0.891\\
		&  $d=100$&0.333&0.978 \\
		TNR&	$d=50$&1.000&1.000   \\
		&  $d=100$&1&1.000 \\
		PPV&	$d=50$&0.999&1.000 \\
		&  $d=100$&1&1.000 \\
		NPV&	$d=50$&0.969&0.995 \\
		&  $d=100$&0.987&1.000 \\
		F1&	$d=50$&0.349&0.941  \\
		&  $d=100$&0.496&  0.989 \\
		MCC&	$d=50$&0.448&0.941 \\
		&  $d=100$&0.570&0.988  \\\hline
		EE$_\Omega$&$d=50$&0.670&0.585 \\
		&  $d=100$&0.628&0.534 \\
	\hline
	\end{tabular}
\end{minipage}
\end{table}

\section{An empirical application}

\label{sec7} \renewcommand{\theequation}{7.\arabic{equation}} \setcounter{equation}{0}

In this section, we apply the proposed methods to estimate the Granger
causality and partial correlation networks using the FRED-MD macroeconomic
dataset. The dataset, available on the Fred-MD
website\footnote{https://research.stlouisfed.org/econ/mccracken/fred-databases/}%
, consists of $127$ U.S. macroeconomic variables observed monthly over the
period from January 1959 to July 2022. These macroeconomic variables can be
classified into eight groups: consumption, orders and inventories; housing;
interest and exchange rates; labour market; money and credit; output and
income; prices; and the stock market. More detailed description can be found
in \cite{MN16}.

We follow \cite{MN16} and \cite{MN20} to remove outliers and fill missing
values. Each variable is standardised to have zero mean and unit variance. We
consider the two factor modelling methods in Section \ref{sec5} to accommodate
strong cross-sectional dependence: the approximate factor model (\ref{eq5.1})
with constant factor loadings, and the time-varying factor model (\ref{eq5.2})
with dynamic factor loadings. The information criteria proposed by \cite{BN02}
and \cite{SW17} are used to determine the number of factors in these two
models (see Appendix E in the supplement for description of the criteria).
Seven factors are selected for the factor model with constant loadings,
whereas only four are selected for the time-varying factor model. Since the
latter provides a more parsimonious model specification, we hereafter report
network estimation results only for this model. The estimated idiosyncratic
errors, denoted as $\widehat{x}_{t,i}$, $i=1,\mathcal{\ldots},127$,
$t=1,\mathcal{\ldots},763$, are then used for our empirical analysis.
\cite{MPS22} suggest determining the optimal order of a high-dimensional VAR
model via a ratio criterion, comparing the Frobenius norms of the estimated
transition matrices over different lags. We extend their criterion to the
time-varying VAR model context (see Appendix E in the supplement for detail)
and subsequently select the time-varying VAR(1) model for $\widehat{X}%
_{t}=\left(  \widehat{x}_{t,1},\mathcal{\ldots},\widehat{x}_{t,127}\right)
^{^{\intercal}}$.

\begin{figure}
\begin{center}
\includegraphics[scale=0.3]{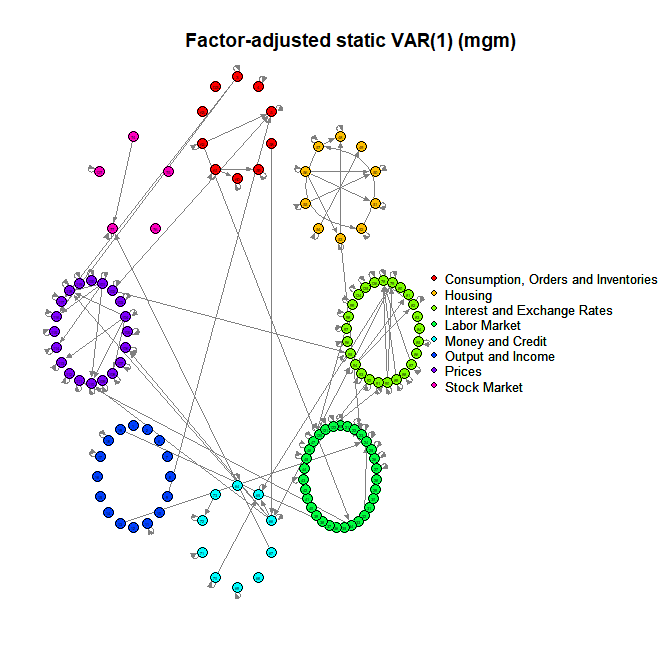}
\includegraphics[scale=0.3]{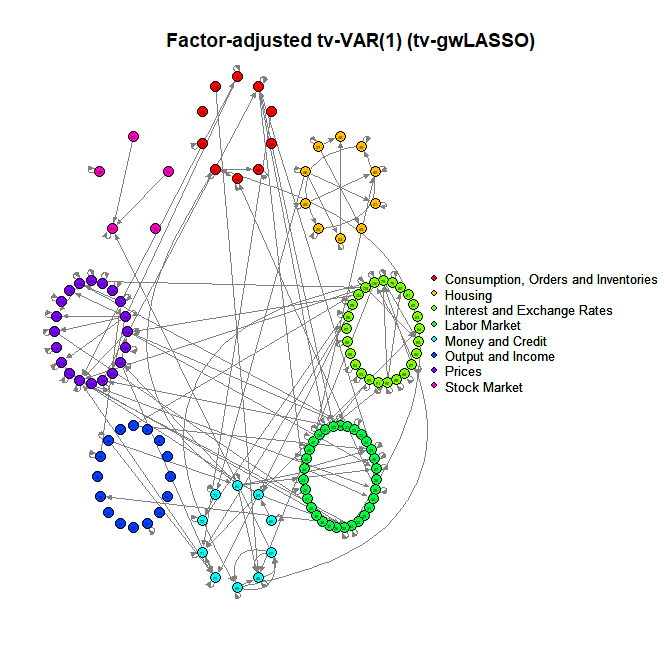}
\includegraphics[scale=0.3]{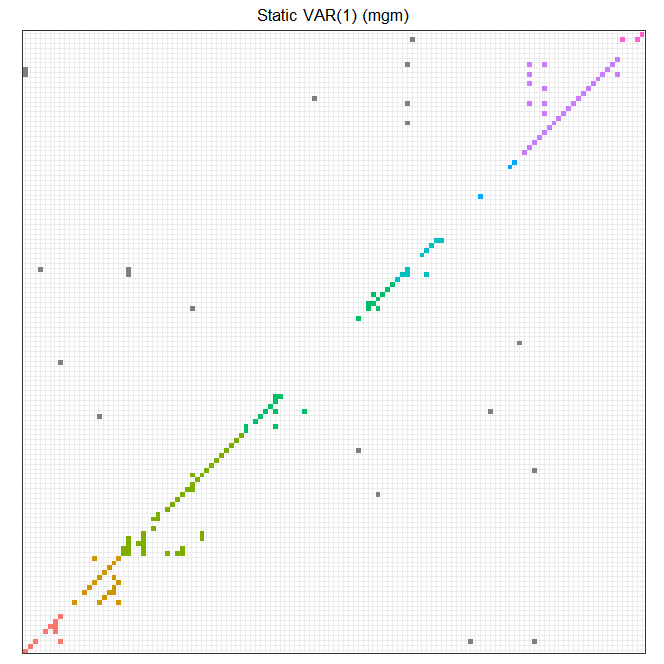}
\includegraphics[scale=0.3]{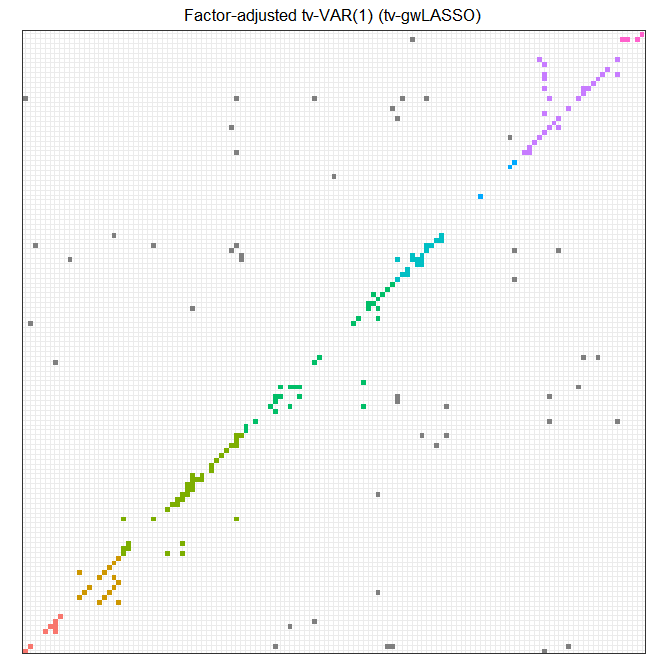}
\end{center}
\caption{{\protect\small The estimated Granger causality networks using the
factor-adjusted static VAR(1) model (left) and time-varying VAR(1) model
(right). }}%
\label{fig2}%
\end{figure}

\begin{figure}
\begin{center}
\includegraphics[scale=0.3]{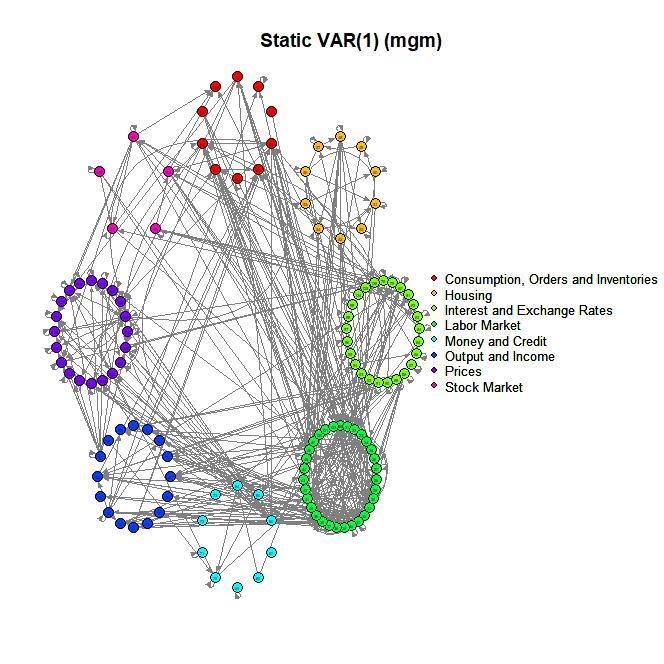}
\includegraphics[scale=0.3]{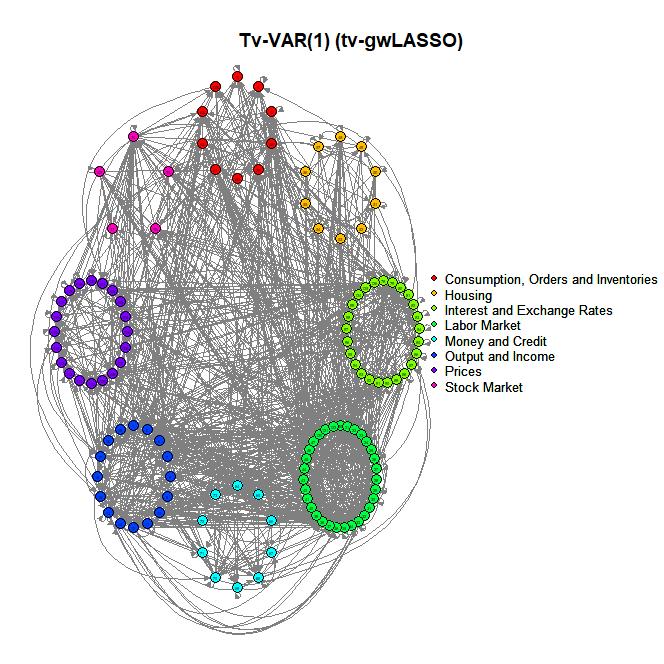}
\includegraphics[scale=0.3]{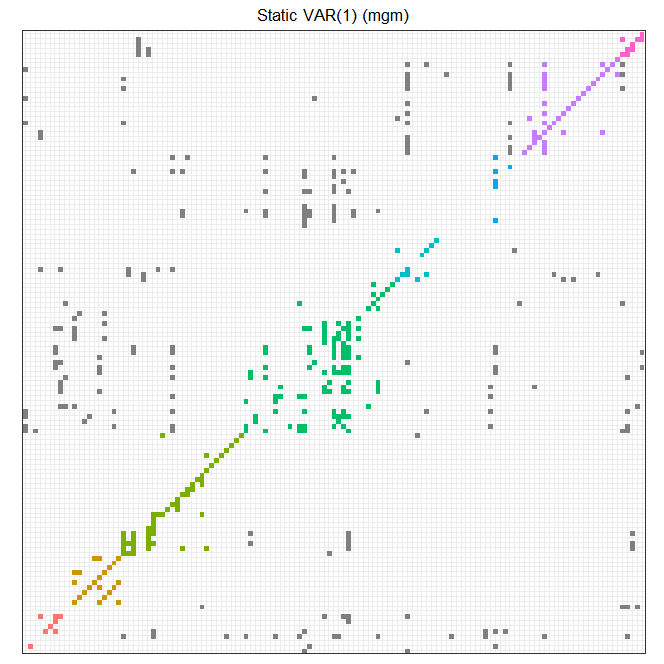}
\includegraphics[scale=0.3]{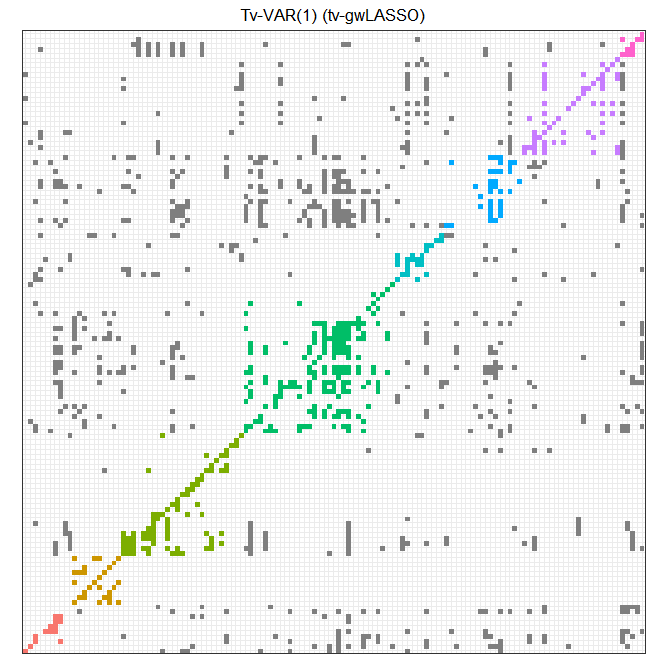}
\end{center}
\caption{{\protect\small The estimated Granger causality networks using the
static VAR(1) model (left) and time-varying VAR(1) model (right) without
factor-adjustment. }}%
\label{fig3}%
\end{figure}

Figure \ref{fig2} plots the estimated Granger networks from the static VAR(1)
and the time-varying VAR(1) models. From the estimated time-varying transition
matrix, we uncover $190$ directed linkages in the Granger causality network,
among which $78$ are self-linkages and $143$ are linkages within the same
category. In particular, the self-linkages, which correspond to the
significant diagonal entries of the transition matrix, indicate that the
macroeconomic variables in the following four categories: consumption, orders
and inventories; interest and exchange rates; money and credit; and prices,
are more persistent than the others, even though all the variables have been
transformed into stationary ones in the preliminary analysis. By contrast, we
find 155 directed linkages for the Granger network estimated via static VAR(1)
and hence, our time-varying VAR(1) model captures more linkages in the network
estimation. Figure \ref{fig3} plots the Granger networks estimated without
factor adjustment. Compared with the factor-adjusted version, the Granger
network via time-varying VAR(1) is more dense with $1118$ directed linkages,
among which $104$ are self-linkages and $432$ are within categories. As
pointed out by \cite{MN16}, common factors, which may be interpreted as
business cycles, are the main sources of the Granger causalities between
macroeconomic variables, leading to a rather dense network structure. On the
other hand, the estimated Granger network via static VAR(1) without factor
adjustment has only 450 linkages.

\begin{figure}[tbh]
\begin{center}
\includegraphics[scale=0.35]{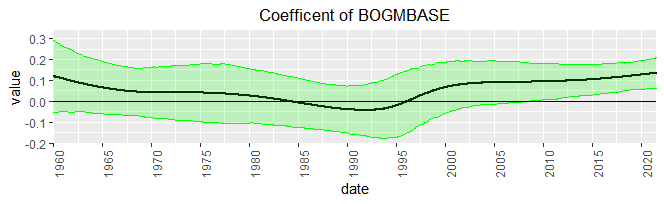}
\includegraphics[scale=0.35]{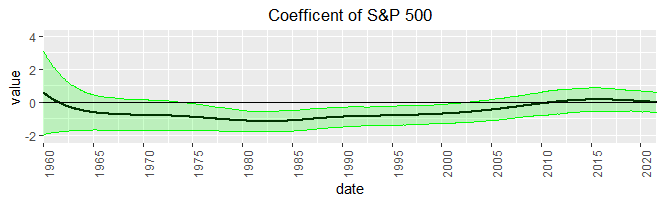}
\includegraphics[scale=0.35]{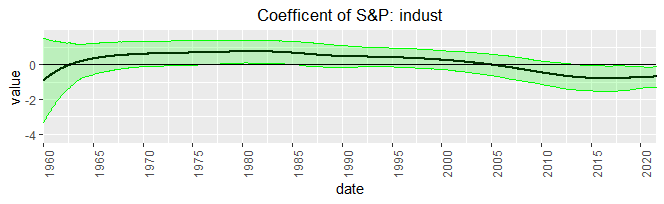}
\includegraphics[scale=0.35]{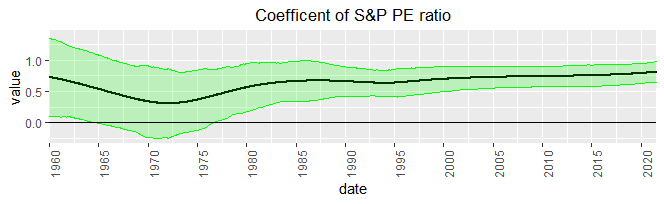}
\end{center}
\caption{{\protect\small The estimated time-varying coefficients linked to
S\&P PE ratio with 90\% confident bands.}}%
\label{fig4}%
\end{figure}

We further explore the dynamic smooth structural changes of Gaussian causality
linkages. Taking the logarithmic growth rate of S\&P PE ratio (S\&P PE
ratio)\footnote{We show in the parentheses the variable names used in the
FRED-MD dataset. The variable transformation is conducted following the
guideline in the dataset.}  as an example, there are four directed linkages to
this variable: acceleration of the logarithmic monetary base (BOGMBASE), the
logarithmic return of S\&P 500 index (S\&P 500), the logarithmic return of
S\&P 500 industrials index (S\&P: indust), and the logarithmic growth rate of
the S\&P PE ratio which is a self-linkage. We re-estimate the corresponding
time-varying coefficients using the nonparametric autoregression model with
only the four selected predictors, and draw the 90\% confidence bands using
the R package ``\textsf{tvReg}". Figure \ref{fig4} plots the estimated curves
of the four coefficient functions. We find that the logarithmic growth rate of
S\&P PE ratio is generally persistent and positively correlated to BOGMBASE in
the most recent two decades. The estimated time-varying coefficient of the
S\&P 500 industrials index return is significant but close to zero. It is thus
unsurprising that the static VAR(1) model with classic LASSO penalty does not
detect the Granger causality linkage from this variable. In fact, LASSO tends
to select only one variable in a group of highly-correlated predictors. Due to
high correlation between the two index returns, only the S\&P 500 Index return
is selected in the static VAR(1) model. In contrast, the proposed time-varying
LASSO selects both of the two index returns at different time periods, and the
second-stage weighted group LASSO aggregates the information over time and
selects both index returns.

\begin{figure}[tbh]
\begin{center}
\includegraphics[scale=0.3]{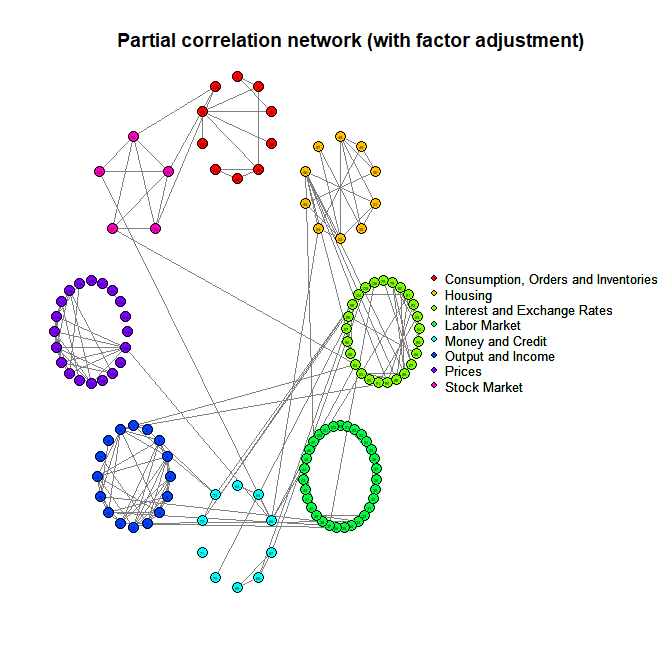}
\includegraphics[scale=0.3]{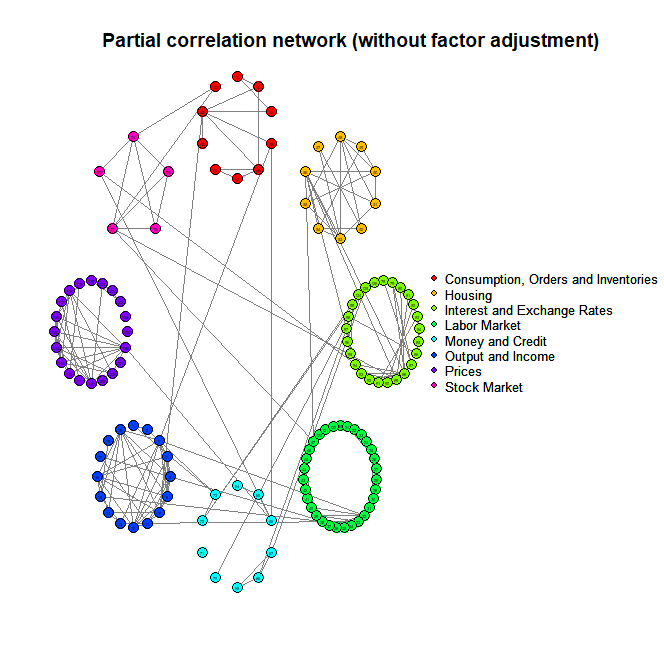}
\includegraphics[scale=0.3]{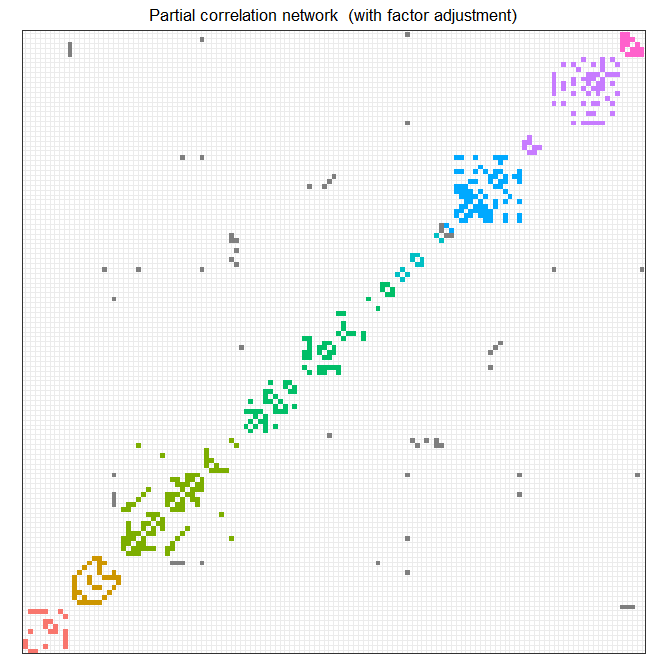}
\includegraphics[scale=0.3]{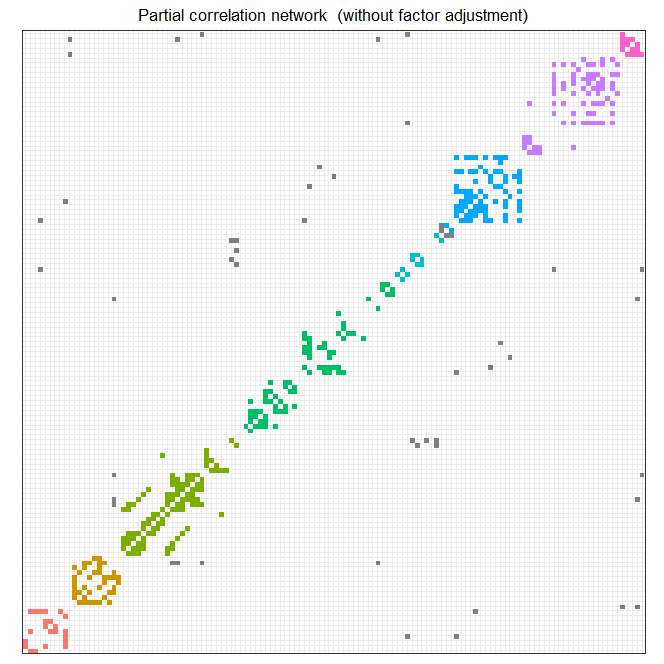}
\end{center}
\caption{{\protect\small The estimated partial correlation networks with
(left) and without (right) factor adjustment. }}%
\label{fig5}%
\end{figure}

We plot the estimated partial correlation networks in Figure \ref{fig5}, which
are generally sparse. Using the factor-adjusted time-varying CLIME, $234$
undirected linkages are detected in the estimated network, among which $205$
linkages are within the same category. In contrast, the estimated network
without factor adjustment contains $236$ linkages with $211$ in the same
category. Unlike the Granger network estimation, it seems that whether to make
factor adjustment or not has little impact on the partial correlation network
estimation.

\begin{figure}[tbh]
\begin{center}
\includegraphics[scale=0.35]{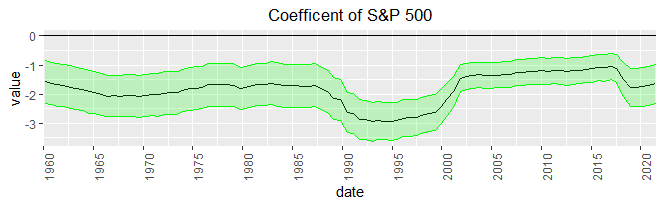}
\includegraphics[scale=0.35]{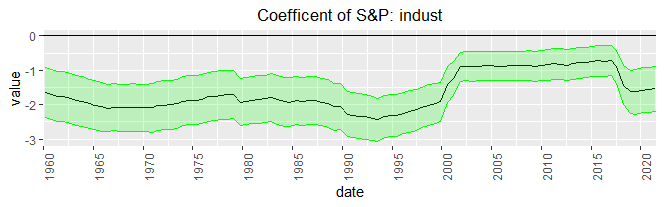}
\includegraphics[scale=0.35]{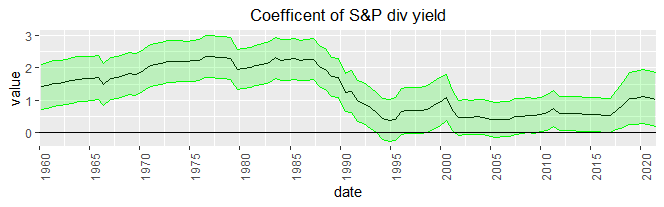}
\includegraphics[scale=0.35]{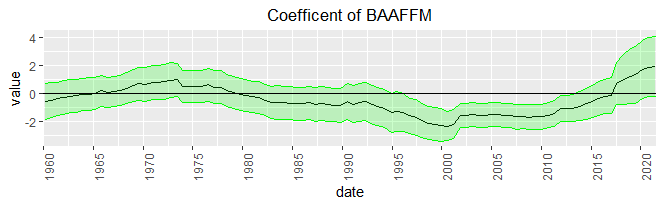}
\end{center}
\caption{{\protect\small The estimated time-varying elements in the precision
matrix linked to S\&P PE ratio with 90\% confident bands.}}%
\label{fig6}%
\end{figure}

We next examine the time-varying pattern of partial correlation linkages
between S\&P PE ratio and four other variables: S\&P 500, S\&P: indust, S\&P
div yield (the increment of S\&P composite common stock: dividend yield), and
BAAFFM (the spread between Moody's seasoned baa corporate bond and effective
federal funds rate). We re-estimate the relevant time-varying functions with a
200-month moving window \citep{JV15}, and draw the 90\% confidence bands using
R package ``\textsf{SILGGM}" in Figure \ref{fig6}. Note that the partial
correlation has a sign opposite to the corresponding entry in the precision
matrix. We find that S\&P PE ratio is positively (partially) correlated with
S\&P 500 and S\&P: indust, whilst negatively (partially) correlated with S\&P
div yield. The confidence bands in Figure \ref{fig6} suggest that
time-invariant partial correlation linkages are inappropriate to describe the
network structure of the FRED-MD data.

\smallskip

\section{Conclusion}
\label{sec8}

In this paper we estimate a general time-varying VAR model for
high-dimensional locally stationary time series. A three-stage estimation
procedure combining time-varying LASSO, weighted group LASSO and time-varying
CLIME is developed to estimate both transition and error precision matrices,
allowing smooth structural changes over time. The estimated transition and
precision matrices are further used to construct dual network structures with
directed Granger causality linkages and undirected partial correlation
linkages, respectively. Under the sparse structural assumption and other
technical conditions, we derive the uniform consistency and oracle properties
for the developed estimates. In order to accommodate high correlation among
large-scale time series and avoid directly imposing the sparsity assumption,
we also extend the methodology and theory to a more general factor-adjusted
time-varying VAR and network structures. Both the simulation and empirical
studies show that the developed network model and methodology have reliable
numerical performance in finite samples.


\section*{\Large Supplementary materials}

{\small The supplement contains proofs of the main asymptotic theorems, some
technical lemmas with proofs, verification of Assumption \ref{ass:3}(ii) and
discussions on tuning parameter selection.}


\newpage

\begin{center}
{\LARGE\bf  Supplement to ``Estimating Time-Varying Networks for High-Dimensional Time Series"}
\end{center}

\maketitle


\appendix

\section*{\Large Appendix A:\ \ Proofs of Theorems 4.1--4.3}
\renewcommand{\theequation}{A.\arabic{equation}}
\setcounter{equation}{0}

\noindent{\bf Proof of Theorem 4.1}.\ \ The main idea to be used in this proof is similar to that in \cite{BRT09}, \cite{L12} and \cite{LKZ15} which study high-dimensional data under the classic independence assumption. In the following proof, we need to use the uniform convergence properties of the kernel-weighted quantities for time-varying VAR (say, Lemma \ref{le:B.3} in Appendix B). In fact, we next prove a strengthened version of (4.4) which also includes a uniform consistency of the derivative function estimates:
\begin{equation}\label{eqA.1}
\max_{1\leq i\leq d}\max_{1\leq t\leq n}\left(\left\Vert \widetilde{\boldsymbol\alpha}_{i\bullet}(\tau_t)-{\boldsymbol\alpha}_{i\bullet}(\tau_t)\right\Vert + h\left\Vert\widetilde{\boldsymbol\alpha}_{i\bullet}^\prime(\tau_t)-{\boldsymbol\alpha}_{i\bullet}^\prime(\tau_t)\right\Vert\right)=O_P\left(\sqrt{s}\lambda_1\right).
\end{equation} 

As we only consider the time-varying VAR (1) model, 
\[
{\boldsymbol\alpha}_{i\bullet}(\tau_t)=\left[\alpha_{i,1}(\tau_t),\alpha_{i,2}(\tau_t),\cdots,\alpha_{i,d}(\tau_t)\right]^{^\intercal}\ \ {\rm and}\ \ {\boldsymbol\alpha}_{i\bullet}^\prime(\tau_t)=\left[\alpha_{i,1}^\prime(\tau_t),\alpha_{i,2}^\prime(\tau_t),\cdots,\alpha_{i,d}^\prime(\tau_t)\right]^{^\intercal}.
\]
Recall that ${\cal J}_i(\tau_t)=\left\{j:\ \alpha_{i,j}(\tau_t)\neq0\right\}$ and define ${\cal J}_i^\prime(\tau_t)=\left\{j:\ \alpha_{i,j}^\prime(\tau_t)\neq0\right\}$. We first prove that for any $i=1,\cdots,d$ and $t=1,\cdots,n$,
\begin{equation}\label{eqA.2}
\sum_{j\notin {\cal J}_i(\tau_t)}|\delta_{i,j}(\tau_t)|+\sum_{j\notin {\cal J}_i^\prime(\tau_t)}|\delta_{i,j}^\prime(\tau_t)|\leq 2\left(\sum_{j\in {\cal J}_i(\tau_t)}|\delta_{i,j}(\tau_t)|+\sum_{j\in {\cal J}_i^\prime(\tau_t)}|\delta_{i,j}^\prime(\tau_t)|\right),
\end{equation}
where $\delta_{i,j}(\tau_t)=\widetilde{\alpha}_{i,j}(\tau_t)-\alpha_{i,j}(\tau_t)$ and $\delta_{i,j}^\prime(\tau_t)=h\left[\widetilde{\alpha}_{i,j}^\prime(\tau_t)-\alpha_{i,j}^\prime(\tau_t)\right]$.

By the definition of the preliminary time-varying LASSO, we have
\[
{\cal L}_{i}^\ast\left(\widetilde{\boldsymbol\alpha}_{i\bullet}(\tau_t), \widetilde{\boldsymbol\alpha}_{i\bullet}^\prime(\tau_t)\ |\ \tau_t\right)\leq {\cal L}_{i}^\ast\left({\boldsymbol\alpha}_{i\bullet}(\tau_t), {\boldsymbol\alpha}_{i\bullet}^\prime(\tau_t)\ |\ \tau_t\right)
\]
for any $i=1,\cdots,d$ and $t=1,\cdots,n$, where ${\cal L}_{i}^\ast({\boldsymbol\alpha}, {\boldsymbol\beta}\ |\ \tau_t)$ is defined in (3.4). Then, we readily have that
\begin{eqnarray}
&& {\cal L}_{i}\left({\boldsymbol\alpha}_{i\bullet}(\tau_t), {\boldsymbol\alpha}_{i\bullet}^\prime(\tau_t)\ |\ \tau_t\right)-{\cal L}_{i}\left(\widetilde{\boldsymbol\alpha}_{i\bullet}(\tau_t), \widetilde{\boldsymbol\alpha}_{i\bullet}^\prime(\tau_t)\ |\ \tau_t\right)\nonumber
\\
&&\geq \lambda_1\left[\sum\limits_{j=1}^{d}|\widetilde{\alpha}_{i,j}(\tau_t)|+h\sum\limits_{j=1}^{d}|\widetilde{\alpha}_{i,j}^\prime(\tau_t)|-\sum\limits_{j=1}^{d}|\alpha_{i,j}(\tau_t)|-h\sum\limits_{j=1}^{d}|\alpha_{i,j}^\prime(\tau_t)|\right].\label{eqA.3}
\end{eqnarray}
Let 
\[\delta_i(\tau_t)=\left[\delta_{i,1}(\tau_t),\cdots,\delta_{i,d}(\tau_t)\right]^{^\intercal}\ \ {\rm and}\ \ \delta_i^\prime(\tau_t)=\left[\delta_{i,1}^\prime(\tau_t),\cdots,\delta_{i,d}^\prime(\tau_t)\right]^{^\intercal}.\]
Note that
\begin{eqnarray}
&&{\cal L}_{i}\left({\boldsymbol\alpha}_{i\bullet}(\tau_t), {\boldsymbol\alpha}_{i\bullet}^\prime(\tau_t)\ |\ \tau_t\right)-{\cal L}_{i}\left(\widetilde{\boldsymbol\alpha}_{i\bullet}(\tau_t), \widetilde{\boldsymbol\alpha}_{i\bullet}^\prime(\tau_t)\ |\ \tau_t\right)\nonumber\\
&=&2\left[L_{i,0}^{^\intercal}(\tau_t)\delta_i(\tau_t)+L_{i,1}^{^\intercal}(\tau_t)\delta_i^\prime(\tau_t)\right]-\frac{1}{n}\sum\limits_{s=1}^n \left\{\left[\delta_i(\tau_t)+\delta_i^\prime(\tau_t) \left(\frac{\tau_s-\tau_t}{h}\right)\right]^{^\intercal} X_{s-1}\right\}^2K_{h}(\tau_s-\tau_t)\nonumber\\
&\leq&2\left[L_{i,0}^{^\intercal}(\tau_t)\delta_i(\tau_t)+L_{i,1}^{^\intercal}(\tau_t)\delta_i^\prime(\tau_t)\right],\label{eqA.4}
\end{eqnarray}
where $L_{i,0}(\tau_t)$ and $L_{i,1}(\tau_t)$ are defined in Appendix B. By Lemma \ref{le:B.3}, we may show that
\begin{equation}\label{eqA.5}
\left\vert L_{i,0}^{^\intercal}(\tau_t)\delta_i(\tau_t)+L_{i,1}^{^\intercal}(\tau_t)\delta_i^\prime(\tau_t)\right\vert\leq O_P\left(\zeta_{n,d}\right)\cdot\left(\sum_{j=1}^{d}|\delta_{i,j}(\tau_t)|+\sum_{j=1}^{d}|\delta_{i,j}^\prime(\tau_t)|\right)
\end{equation}
uniformly over $i=1,\cdots,d$ and $t=1,\cdots,n$.

On the other hand, by the triangle inequality, we may prove that
\begin{eqnarray}
&& \lambda_1\left[\sum\limits_{j=1}^{d}|\widetilde{\alpha}_{i,j}(\tau_t)|+h\sum\limits_{j=1}^{d}|\widetilde{\alpha}_{i,j}^\prime(\tau_t)|-\sum\limits_{j=1}^{d}|\alpha_{i,j}(\tau_t)|-h\sum\limits_{j=1}^{d}|\alpha_{i,j}^\prime(\tau_t)|\right]\nonumber\\
&=&\lambda_1\left[\sum\limits_{j\in{\cal J}_i(\tau_t)}\left(|\widetilde{\alpha}_{i,j}(\tau_t)|-|\alpha_{i,j}(\tau_t)|\right)+h\sum\limits_{j\in{\cal J}_i^\prime(\tau_t)}\left(|\widetilde{\alpha}_{i,j}^\prime(\tau_t)|-|\alpha_{i,j}^\prime(\tau_t)|\right)\right]+\nonumber\\
&&\lambda_1\left[\sum\limits_{j\notin{\cal J}_i(\tau_t)}|\widetilde{\alpha}_{i,j}(\tau_t)|+h\sum\limits_{j\notin{\cal J}_i^\prime(\tau_t)}|\widetilde{\alpha}_{i,j}^\prime(\tau_t)|\right]\nonumber\\
&\geq&-\lambda_1\left(\sum_{j\in {\cal J}_i(\tau_t)}|\delta_{i,j}(\tau_t)|+\sum_{j\in {\cal J}_i^\prime(\tau_t)}|\delta_{i,j}^\prime(\tau_t)|\right)+\lambda_1\left(\sum_{j\notin {\cal J}_i(\tau_t)}|\delta_{i,j}(\tau_t)|+\sum_{j\notin {\cal J}_i^\prime(\tau_t)}|\delta_{i,j}^\prime(\tau_t)|\right).\label{eqA.6}
\end{eqnarray}
By (\ref{eqA.3})--(\ref{eqA.6}) and the condition $\zeta_{n,d}=o(\lambda_{1})$ in Assumption 3(i), we complete the proof of (\ref{eqA.2}).

Let $u_1=\left(u_{1,1},\cdots,u_{1,d}\right)^{^\intercal}$ and $u_2=\left(u_{2,1},\cdots,u_{2,d}\right)^{^\intercal}$ be two $d$-dimensional vectors and 
{\small\[
{\cal B}_i(\tau_t;M)=\left\{u=\left(u_1^{^\intercal}, u_2^{^{\intercal}}\right)^{^\intercal}:\ \|u_1\|^2+\|u_2\|^2=M,\ \sum_{j=1}^{d}\left(|u_{1,j}|+|u_{2,j}|\right)\leq 3 \left(\sum_{j\in{\cal J}_i(\tau_t)}|u_{1,j}|+\sum_{j\in{\cal J}_i^\prime(\tau_t)}|u_{2,j}|\right)\right\},\nonumber
\]}
where $M$ is a positive constant which may be sufficiently large. Note that for any $i=1,\cdots,d$, $t=1,\cdots,n$, and $u\in{\cal B}_i(\tau_t; M)$,
\begin{equation}\label{eqA.7}
{\mathcal L}_{i}^\ast\left({\boldsymbol\alpha}_{i\bullet}(\tau_t)+\sqrt{s}\lambda_1u_1,{\boldsymbol\alpha}_{i\bullet}^\prime(\tau_t)+\sqrt{s}\lambda_1u_2/h\ |\ \tau_t\right)-{\cal L}_{i}^\ast\left({\boldsymbol\alpha}_{i\bullet}(\tau_t), {\boldsymbol\alpha}_{i\bullet}^\prime(\tau_t)\ |\ \tau_t\right)=\sum_{k=1}^3\Xi_{i,k}(\tau_t),
\end{equation}
where
\begin{eqnarray}
\Xi_{i,1}(\tau_t)&=&{\mathcal L}_{i}\left({\boldsymbol\alpha}_{i\bullet}(\tau_t)+\sqrt{s}\lambda_1u_1,{\boldsymbol\alpha}_{i\bullet}(\tau_t)+\sqrt{s}\lambda_1u_2/h\ |\ \tau_t\right)-{\cal L}_{i}\left({\boldsymbol\alpha}_{i\bullet}(\tau_t), {\boldsymbol\alpha}_{i\bullet}^\prime(\tau_t)\ |\ \tau_t\right),\nonumber
\\
\Xi_{i,2}(\tau_t)&=&\lambda_1\left(\sum\limits_{j=1}^{d}|\alpha_{i,j}(\tau_t)+\sqrt{s}\lambda_1u_{1,j}|-\sum\limits_{j=1}^{d}|\alpha_{i,j}(\tau_t)|\right),\nonumber
\\
\Xi_{i,3}(\tau_t)&=&\lambda_1\left(\sum\limits_{j=1}^{d}|h\alpha_{i,j}^\prime(\tau_t)+\sqrt{s}\lambda_1u_{2,j}|-\sum\limits_{j=1}^{d}|h \alpha_{i,j}^\prime(\tau_t)|\right).\nonumber
\end{eqnarray}

For $\Xi_{i,1}(\tau_t)$, it can be written as
\begin{equation}\label{eqA.8}
\Xi_{i,1}(\tau_t)=-2\sqrt{s}\lambda_1u^{^\intercal}L_{i}(\tau_t)+s\lambda_1^2 u^{^\intercal} {\boldsymbol\Psi}(\tau_t)u,
\end{equation}
where $L_{i}(\tau)=\left[L_{i,0}^{^\intercal}(\tau), L_{i,1}^{^\intercal}(\tau)\right]^{^\intercal}$, and ${\boldsymbol\Psi}(\tau)$ is defined in (4.2). By the definition of ${\cal B}_i(\tau_t; M)$, Lemma \ref{le:B.3} and the Cauchy-Schwarz inequality, we have 
\begin{equation}\label{eqA.9}
\max_{1\leq i\leq d}\left\vert \sqrt{s}\lambda_1u^{^\intercal}L_{i}(\tau_t)\right\vert=o_P\left(s\lambda_1^2\right)\cdot\|u\|.
\end{equation}
By (\ref{eqA.8}), (\ref{eqA.9}) and the uniform restricted eigenvalue condition (4.3), when $n$ is sufficiently large and $M$ is chosen to be large enough, we have
\begin{equation}\label{eqA.10}
\min_{1\leq i\leq d}\min_{1\leq t\leq n}\inf_{u\in{\cal B}_{i}(\tau_t; M)}u^{^\intercal}  \Xi_{i,1}(\tau_t)=s\lambda_1^2u^{^\intercal} {\boldsymbol\Psi}(\tau_t)u(1+o_P(1))> \frac{1}{2}\kappa_0 s\lambda_1^2\Vert u\Vert^2, \ \ w.p.a.1.
\end{equation}
We next consider $\Xi_{i,2}(\tau_t)$ and $\Xi_{i,3}(\tau_t)$. It is easy to show that
\begin{eqnarray}
\Xi_{i,2}(\tau_t)&=&\lambda_1\left(\sum\limits_{j=1}^{d}|\alpha_{i,j}(\tau_t)+\sqrt{s}\lambda_1u_{1,j}|-\sum\limits_{j=1}^{d}|\alpha_{i,j}(\tau_t)|\right)
\nonumber\\
&=&\lambda_1\sum_{j\in{\cal J}_i(\tau_t)}\left[|\alpha_{i,j}(\tau_t)+\sqrt{s}\lambda_1u_{1,j}|-|\alpha_{i,j}(\tau_t)|\right]+\lambda_1\sum_{j\notin{\cal J}_i(\tau_t)}|\sqrt{s}\lambda_1u_{1,j}|
\nonumber\\
&=&O\left(s\lambda_1^2\right)\cdot \|u_1\|+\lambda_1\sum_{j\notin{\cal J}_i(\tau_t)}|\sqrt{s}\lambda_1u_{1,j}|=O\left(s\lambda_1^2\right)\cdot \|u_1\|,\label{eqA.11}
\end{eqnarray}
and similarly,
\begin{equation}\label{eqA.12}
\Xi_{i,3}(\tau_t)=O\left(s\lambda_1^2\right)\cdot \|u_2\|+\lambda_1\sum_{j\notin{\cal J}_i^\prime(\tau_t)}|\sqrt{s}\lambda_1u_{2,j}|=O\left(s\lambda_1^2\right)\cdot \|u_2\|,
\end{equation}
uniformly over $i=1,\cdots,d$ and $t=1,\cdots,n$.

With (\ref{eqA.7}) and (\ref{eqA.10})--(\ref{eqA.12}), letting $M$ be large enough, we can prove that the leading term of 
\[{\mathcal L}_{i}^\ast\left({\boldsymbol\alpha}_{i\bullet}(\tau_t)+\sqrt{s}\lambda_1u_1,{\boldsymbol\alpha}_{i\bullet}^\prime(\tau_t)+\sqrt{s}\lambda_1u_2/h\ |\ \tau_t\right)-{\cal L}_{i}^\ast\left({\boldsymbol\alpha}_{i\bullet}(\tau_t), {\boldsymbol\alpha}_{i\bullet}^\prime(\tau_t)\ |\ \tau_t\right)\] 
is positive uniformly over $i=1,\cdots,d$ and $t=1,\cdots,n$. Hence, we may find a local minimiser to ${\cal L}_{i}^\ast({\boldsymbol\alpha}, {\boldsymbol\beta}\ |\ \tau_t)$, denoted by $\left[\widetilde{\boldsymbol\alpha}_{i\bullet}(\tau_t), h\widetilde{\boldsymbol\alpha}_{i\bullet}^\prime(\tau_t)\right]$, in the interior of 
\[\left\{\left({\boldsymbol\alpha}_{i\bullet}(\tau_t)+\sqrt{s}\lambda_1u_1, h{\boldsymbol\alpha}_{i\bullet}^\prime(\tau_t)+\sqrt{s}\lambda_1u_2\right):\ u\in{\cal B}_i(\tau_t; M)\right\},\] which, together with (\ref{eqA.2}), completes the proof of (\ref{eqA.1}). \hfill$\blacksquare$

\medskip

\noindent{\bf Proof of Theorem 4.2}.\ \ Define
\begin{eqnarray}
{\mathbf L}_{i,j}^\alpha&=&\left[l_{i,j}^\alpha({\boldsymbol\alpha}_{\bullet 1}, {\boldsymbol\beta}_{\bullet 1}\ |\ \tau_1),\cdots,l_{i,j}^\alpha({\boldsymbol\alpha}_{\bullet n}, {\boldsymbol\beta}_{\bullet n}\ |\ \tau_n)\right]^{^\intercal},\nonumber\\
{\mathbf L}_{i,j}^\beta&=&\left[l_{i,j}^\beta({\boldsymbol\alpha}_{\bullet 1}, {\boldsymbol\beta}_{\bullet 1}\ |\ \tau_1),\cdots,l_{i,j}^\beta({\boldsymbol\alpha}_{\bullet n}, {\boldsymbol\beta}_{\bullet n}\ |\ \tau_n)\right]^{^\intercal},\nonumber\\
{\mathbf P}_{i,j}^\alpha&=&\left[p_{\lambda_2}^\prime \left(\left\Vert  \widetilde{\boldsymbol\alpha}_{i,j} \right\Vert \right)\frac{\alpha_{j|1}}{\|{\boldsymbol\alpha}_j\|},\cdots,p_{\lambda_2}^\prime \left(\left\Vert  \widetilde{\boldsymbol\alpha}_{i,j} \right\Vert \right)\frac{\alpha_{j|n}}{\|{\boldsymbol\alpha}_j\|}\right]^{^\intercal},\nonumber\\
{\mathbf P}_{i,j}^\beta&=&\left[p_{\lambda_2}^\prime \left(\widetilde{D}_{i,j}\right)\frac{\beta_{j|1}}{\|{\boldsymbol\beta}_j\|},\cdots,p_{\lambda_2}^\prime \left(\widetilde{D}_{i,j}\right)\frac{\beta_{j|n}}{\|{\boldsymbol\beta}_j\|}\right]^{^\intercal},\nonumber
\end{eqnarray}
where 
\begin{eqnarray}
l_{i,j}^\alpha({\boldsymbol\alpha}, {\boldsymbol\beta}\ |\ \tau)&=&\frac{1}{n}\sum\limits_{t=1}^n \left\{x_{t,i}-\left[{\boldsymbol\alpha}+{\boldsymbol\beta}(\tau_t-\tau)\right]^{^\intercal}X_{t-1}\right\} x_{t-1,j}K_{h}(\tau_t-\tau),\nonumber\\
l_{i,j}^\beta({\boldsymbol\alpha}, {\boldsymbol\beta}\ |\ \tau)&=&\frac{1}{n}\sum\limits_{t=1}^n \left\{x_{t,i}-\left[{\boldsymbol\alpha}+{\boldsymbol\beta}(\tau_t-\tau)\right]^{^\intercal}X_{t-1}\right\} x_{t-1,j}\left(\frac{\tau_t-\tau}{h}\right)K_{h}(\tau_t-\tau).\nonumber
\end{eqnarray}
From the KKT condition \citep[e.g.,][]{FL11, FXZ14, LKZ15}, the oracle estimate $\left(\widehat{\mathbf A}_i^o, \widehat{\mathbf B}_i^o\right)$ is the unique minimiser to the objective function ${\mathcal Q}_{i}({\mathbf A}, {\mathbf B})$ if
\begin{eqnarray}
&&{\mathbf L}_{i,j}^\alpha-{\mathbf P}_{i,j}^\alpha={\mathbf 0}_{n}\ \ {\rm for}\ \ j\in{\cal J}_i,\ \ {\mathbf L}_{i,j}^\beta-{\mathbf P}_{i,j}^\beta={\mathbf 0}_{n}\ \ {\rm for}\ \ j\in{\cal J}_i^\prime,\label{eqA.13}\\
&&\max_{j\in\overline{\cal J}_i}\left\Vert {\mathbf L}_{i,j}^\alpha\right\Vert<\min_{j\in\overline{\cal J}_i}p_{\lambda_2}^\prime \left(\left\Vert  \widetilde{\boldsymbol\alpha}_{i,j} \right\Vert \right),\ \ \max_{j\in\overline{\cal J}_i}\left\Vert {\mathbf L}_{i,j}^\beta\right\Vert<\min_{j\in\overline{\cal J}_i}p_{\lambda_2}^\prime \left(\widetilde{D}_{i,j}\right),\label{eqA.14}
\end{eqnarray}
hold at ${\mathbf A}=\widehat{\mathbf{A}}_i^{o}$ and ${\mathbf B}=\widehat{\mathbf{B}}_i^{o}$, where ${\mathbf 0}_n$ is an $n$-dimensional vector of zeros.

Note that the equalities in (\ref{eqA.13}) automatically hold by the definition of the oracle estimates $\widehat{\mathbf A}_i^o$ and $\widehat{\mathbf B}_i^o$. It remains to prove (\ref{eqA.14}). We next only show the proof of the first assertion in (\ref{eqA.14}) as the proof of the second one is analogous. By Theorem 4.1 and the condition of $(ns)^{1/2}\lambda_1=o(\lambda_2)$ in Assumption 4(i), we may show that $
\min_{j\in\overline{\cal J}_i}p_{\lambda_2}^\prime \left(\left\Vert  \widetilde{\boldsymbol\alpha}_{i,j} \right\Vert \right)=\lambda_2$ {\em w.p.a.1}. Meanwhile, by Lemmas \ref{le:B.3} and \ref{le:B.4} as well as Assumption 4(i), we may prove that
\[
\max_{j\in\overline{\cal J}_i}\left\Vert {\mathbf L}_{i,j}^\alpha\right\Vert=O_P\left(\sqrt{n}s\log(n\vee d)\zeta_{n,d}\right)=o_P(\lambda_2)
\]
when ${\mathbf A}=\widehat{\mathbf{A}}_i^{o}$ and ${\mathbf B}=\widehat{\mathbf{B}}_i^{o}$, leading to the first assertion in (\ref{eqA.14}). Then, the mean squared convergence result (4.8) follows from Lemma \ref{le:B.4}.\hfill$\blacksquare$

\medskip

\noindent{\bf Proof of Corollary 4.1}.\ \ By Theorem 4.2 and Assumption 4(ii), we may show that
\[
{\sf P}\left(\min_{(i,j)\in{\mathbb E}_{n}^G}\sum_{t=1}^n\widehat a_{ij}^2(\tau_t)\geq a_0\lambda_2>0\right)\rightarrow1
\]
and
\[
{\sf P}\left(\sum_{t=1}^n\widehat a_{ij}^2(\tau_t)=0,\ \forall\ (i,j)\notin{\mathbb E}_n^G\right)\rightarrow1,
\]
leading to (4.9).\hfill$\blacksquare$

\medskip

\noindent{\bf Proof of Theorem 4.3}.\ \ By Lemma \ref{le:B.5} in Appendix B, we have
\begin{equation}\label{eqA.15}
\sup_{0\leq \tau\leq 1}\left\Vert \widehat{\boldsymbol\Sigma}(\tau)-{\boldsymbol\Sigma}(\tau) \right\Vert_{\max}=O_P\left(\nu_{n,d}^\diamond+\nu_{n,d}^\ast\right).
\end{equation}
By (\ref{eqA.15}), the sparsity assumption (3.7) and the inequality: $\Vert{\mathbf W}_1{\mathbf W}_2\Vert_{\max}\leq \Vert {\mathbf W}_1\Vert_1\Vert{\mathbf W}_2\Vert_{\max}$ for any two square matrices ${\mathbf W}_1$ and ${\mathbf W}_2$ with the same size, 
\begin{eqnarray}
\sup_{0\leq \tau\leq 1}\left\Vert {\mathbf I}_d-\widehat{\boldsymbol\Sigma}(\tau){\boldsymbol\Omega}(\tau)\right\Vert_{\max}&=&\sup_{0\leq \tau\leq 1}\left\Vert {\boldsymbol\Sigma}(\tau){\boldsymbol\Omega}(\tau)-\widehat{\boldsymbol\Sigma}(\tau){\boldsymbol\Omega}(\tau)\right\Vert_{\max}\nonumber\\
&\leq& \sup_{0\leq\tau\leq 1}\left\Vert{\boldsymbol\Omega}(\tau) \right\Vert_{1}\left\Vert \widehat{\boldsymbol\Sigma}(\tau)-{\boldsymbol\Sigma}(\tau) \right\Vert_{\max}\nonumber\\
&\leq&C_2\sup_{0\leq\tau\leq 1}\left\Vert \widehat{\boldsymbol\Sigma}(\tau)-{\boldsymbol\Sigma}(\tau) \right\Vert_{\max}\nonumber\\
&=&O_P\left(\nu_{n,d}^\diamond+\nu_{n,d}^\ast\right),\label{eqA.16}
\end{eqnarray}
where $C_2$ is defined in (3.7). By (\ref{eqA.16}), the triangle inequality, Assumption 5(ii) and the definition of the time-varying CLIME estimate, we readily have that
\begin{eqnarray}
&&\sup_{0\leq\tau\leq 1}\left\Vert \widehat{\boldsymbol\Sigma}(\tau)\left[\widetilde{\boldsymbol\Omega}(\tau)-{\boldsymbol\Omega}(\tau)\right]\right\Vert_{\max}\nonumber\\
&\leq&\sup_{0\leq \tau\leq 1}\left\Vert \widehat{\boldsymbol\Sigma}(\tau)\widetilde{\boldsymbol\Omega}(\tau)-{\mathbf I}_d\right\Vert_{\max}+\sup_{0\leq \tau\leq 1}\left\Vert {\mathbf I}_d-\widehat{\boldsymbol\Sigma}(\tau){\boldsymbol\Omega}(\tau)\right\Vert_{\max}\nonumber\\
&\leq&\lambda_3+O_P\left(\nu_{n,d}^\diamond+\nu_{n,d}^\ast\right)=O_P\left(\nu_{n,d}^\diamond+\nu_{n,d}^\ast\right).\label{eqA.17}
\end{eqnarray}
By Lemma 1 in \cite{CLL11}, $\left\Vert\widetilde{\boldsymbol\Omega}(\tau)\right\Vert_1\leq \left\Vert {\boldsymbol\Omega}(\tau)\right\Vert_1\leq C_2$ uniformly over $0\leq \tau\leq 1$. Then, by (\ref{eqA.16}) and (\ref{eqA.17}), we readily have that
\begin{eqnarray}
&&\sup_{0\leq \tau\leq 1}\left\Vert {\boldsymbol\Sigma}(\tau)\left[\widetilde{\boldsymbol\Omega}(\tau)-{\boldsymbol\Omega}(\tau)\right]\right\Vert_{\max}\nonumber\\
&\leq&\sup_{0\leq \tau\leq 1}\left\Vert\widehat{\boldsymbol\Sigma}(\tau)\left[\widetilde{\boldsymbol\Omega}(\tau)-{\boldsymbol\Omega}(\tau)\right]\right\Vert_{\max}+\sup_{0\leq \tau\leq 1}\left\Vert \left[\widehat{\boldsymbol\Sigma}(\tau)-{\boldsymbol\Sigma}(\tau)\right]\left[\widetilde{\boldsymbol\Omega}(\tau)-{\boldsymbol\Omega}(\tau)\right]\right\Vert_{\max}\nonumber\\
&\leq&O_P\left(\nu_{n,d}^\diamond+\nu_{n,d}^\ast\right)+2C_2\sup_{0\leq \tau\leq 1}\left\Vert \widehat{\boldsymbol\Sigma}(\tau)-{\boldsymbol\Sigma}(\tau)\right\Vert_{\max}=O_P\left(\nu_{n,d}^\diamond+\nu_{n,d}^\ast\right).\label{eqA.18}
\end{eqnarray}
Using the assumption $\left\Vert {\boldsymbol\Omega}(\tau)\right\Vert_1\leq C_2$ again and (\ref{eqA.18}), we have
\begin{eqnarray}
\sup_{0\leq \tau\leq 1}\left\Vert\widetilde{\boldsymbol\Omega}(\tau)-{\boldsymbol\Omega}(\tau)\right\Vert_{\max}&\leq&\sup_{0\leq \tau\leq 1}\left\Vert{\boldsymbol\Omega}(\tau) \right\Vert_{1}\left\Vert {\boldsymbol\Sigma}(\tau)\left[\widetilde{\boldsymbol\Omega}(\tau)-{\boldsymbol\Omega}(\tau)\right]\right\Vert_{\max}\nonumber\\
&=&O_P\left(\nu_{n,d}^\ast+\nu_{n,d}^\diamond\right).\label{eqA.19}
\end{eqnarray}
By (\ref{eqA.19}) and the definition of $\widehat{\boldsymbol\Omega}(\tau)$ in (3.10), we prove (4.10).

We next give the proof of (4.11). By Lemma 1 in \cite{CLL11}, we have
\[\sum_{i=1}^{d}\left|\widehat{\omega}_{ij}(\tau)\right|\leq \sum_{i=1}^{d}\left|\widetilde{\omega}_{ij}(\tau)\right|\leq\sum_{i=1}^{d}\left|\omega_{ij}(\tau)\right|.\]
Noting that 
\begin{eqnarray}
\sum_{j=1}^{d}\left|\widehat\omega_{ij}(\tau)\right|  I\left( |\widehat\omega_{ij}(\tau)|\leq\lambda_3\right)&=&\sum_{j=1}^{d}\left|\widehat\omega_{ij}(\tau)\right|-\sum_{j=1}^{d}\left|\widehat\omega_{ij}(\tau)\right|  I\left( |\widehat\omega_{ij}(\tau)|>\lambda_3\right)\nonumber\\
&\leq&\sum_{j=1}^{d}\left|\widehat\omega_{ij}(\tau)\right|-\sum_{j=1}^{d}\left|\omega_{ij}(\tau)\right|+\sum_{j=1}^{d}\left|\widehat\omega_{ij}(\tau)  I\left( |\widehat\omega_{ij}(\tau)|>\lambda_3\right)-\omega_{ij}(\tau)\right|\nonumber\\
&\leq&\sum_{j=1}^{d}\left|\widehat\omega_{ij}(\tau)  I\left( |\widehat\omega_{ij}(\tau)|>\lambda_3\right)-\omega_{ij}(\tau)\right|,\nonumber
\end{eqnarray}
we have
\begin{eqnarray}
\sup_{0\leq \tau\leq 1}\left\Vert \widehat{\boldsymbol\Omega}(\tau)-{\boldsymbol\Omega}(\tau)\right\Vert  &\leq&\sup_{0\leq\tau\leq1}\max_{1\leq i\leq d}\sum_{j=1}^{d}\left| \widehat\omega_{ij}(\tau)-\omega_{ij}(\tau)\right| \nonumber\\
&\leq&2\sup_{0\leq\tau\leq1}\max_{1\leq i\leq d}\sum_{j=1}^{d}\left|\widehat\omega_{ij}(\tau)-\omega_{ij}(\tau)\right|  I\left(|\widehat\omega_{ij}(\tau)|>\lambda_3\right)  +\nonumber\\
&&2\sup_{0\leq\tau\leq1}\max_{1\leq i\leq d}\sum_{j=1}^{d}\left|\omega_{ij}(\tau)\right|   I\left( |\widehat\omega_{ij}(\tau)|\leq\lambda_3\right)  \nonumber\\
&=:&\Delta_{1}+\Delta_{2}. \label{eqA.20}
\end{eqnarray}

Define an event
\[
\mathcal{E}_\epsilon=\left\{ \sup_{0\leq\tau\leq1}\left\Vert\widehat{\boldsymbol\Omega}(\tau)-{\boldsymbol\Omega}(\tau)\right\Vert_{\max}\leq c_\epsilon\left(\nu_{n,d}^\diamond+\nu_{n,d}^\ast\right)\right\},
\]
where $c_{\epsilon}$ is a positive constant such that $\mathsf{P}\left(\mathcal{E}_\epsilon\right)  \geq1-\epsilon$ with any $\epsilon>0$. Conditional on $\mathcal{E}_\epsilon$, 
\begin{equation}\label{eqA.21}
\Delta_{1}\leq c_{\epsilon}(\nu_{n,d}^\diamond+\nu_{n,d}^\ast)\sup_{0\leq\tau\leq1}\left[\max_{1\leq i\leq d}\sum_{j=1}^{d}I\left(  |\widehat{\omega}_{ij}(\tau)|>\lambda_3\right)\right].
\end{equation}
Note that on $\mathcal{E}$,
\[
|\widehat\omega_{ij}(\tau)|\leq|\omega_{ij}(\tau)|+|\widehat\omega_{ij}(\tau)-\omega_{ij}(\tau)|\leq|\omega_{ij}(\tau)|+c_{\epsilon}\left(\nu_{n,d}^\diamond+\nu_{n,d}^\ast\right).
\]
Choosing $C_3=2c_\epsilon$ in Assumption 5(ii), the event $\{|\widehat\omega_{ij}(\tau)|>\lambda_3\}$ implies that $\left\{|\omega_{ij}(\tau)|>c_\epsilon\left(\nu_{n,d}^\diamond+\nu_{n,d}^\ast\right)\right\}$ holds. Then, by (3.7) and (\ref{eqA.21}), we may show that on ${\cal E}_\epsilon$,
\begin{eqnarray}
\Delta_{1}&\leq& c_{\epsilon}\left(\nu_{n,d}^\diamond+\nu_{n,d}^\ast\right)\left[\sup_{0\leq\tau\leq1}\max_{1\leq i\leq d}\sum_{j=1}^{d}I\left(  |{\omega}_{ij}(\tau)|>c_{\epsilon}\left(\nu_{n,d}^\diamond+\nu_{n,d}^\ast\right)\right)\right] \nonumber\\
&\leq&c_{\epsilon}\left(\nu_{n,d}^\diamond+\nu_{n,d}^\ast\right) \left[\sup_{0\leq\tau\leq1}\max_{1\leq i\leq d}\sum_{j=1}^{d} \frac{\left|\omega_{ij}(\tau)\right| ^{q}}{c_{\epsilon}^q\left(\nu_{n,d}^\diamond+\nu_{n,d}^\ast\right)^q}\right]
\nonumber\\
&=&O_P\left( \xi_d\cdot\left(\nu_{n,d}^\diamond+\nu_{n,d}^\ast\right)^{1-q}\right).  \label{eqA.22}
\end{eqnarray}

On the other hand, by the triangle inequality, 
\[
|\widehat\omega_{ij}(\tau)|\geq|\omega_{ij}(\tau)|-|\widehat\omega_{ij}(\tau)-\omega_{ij}(\tau)|\geq|\omega_{ij}(\tau)|-c_{\epsilon}\left(\nu_{n,d}^\diamond+\nu_{n,d}^\ast\right)
\]
on $\mathcal{E}_\epsilon$. Hence, we readily show that $\left\{|\widehat\omega_{ij}(\tau)|\leq\lambda_3\right\} $ indicates $
\left\{|\omega_{ij}(\tau)|\leq 3c_{\epsilon}\left(\nu_{n,d}^\diamond+\nu_{n,d}^\ast\right)\right\} $. Then, by (3.7) again, we have
\begin{eqnarray}
\Delta_{2}  &\leq&\sup_{0\leq\tau\leq1}\max_{1\leq i\leq d}\sum_{j=1}^{d}\left|  \omega_{ij}(\tau)\right|  I\left(    |\omega_{ij}(\tau)|\leq 3c_{\epsilon}\left(\nu_{n,d}^\diamond+\nu_{n,d}^\ast\right)\right)\nonumber\\
&\leq&(3c_\epsilon)^{1-q}\left(\nu_{n,d}^\diamond+\nu_{n,d}^\ast\right)^{1-q}\sup_{0\leq\tau\leq1}\max_{1\leq i\leq d}\sum_{j=1}^{d}\left| \omega_{ij}(\tau)\right|  ^{q}\nonumber\\
&=&O_{P}\left( \xi_{d}\left(\nu_{n,d}^\diamond+\nu_{n,d}^\ast\right)^{1-q}\right). \label{eqA.23}
\end{eqnarray}
The proof of (4.11) can be completed by (\ref{eqA.20}), (\ref{eqA.22}) and (\ref{eqA.23}).

Following the proof of (4.11), we also have
\[
\sup_{0\leq \tau\leq 1}\left\Vert \widehat{\boldsymbol\Omega}(\tau)-{\boldsymbol\Omega}(\tau)\right\Vert_1=O_{P}\left( \xi_{d}\left(\nu_{n,d}^\diamond+\nu_{n,d}^\ast\right)^{1-q}\right),
\]
which, together with the following inequalities:
\[
\frac{1}{d}\left\Vert \widehat{\boldsymbol\Omega}(\tau)-{\boldsymbol\Omega}(\tau)\right\Vert _{F}^2\leq \left\Vert \widehat{\boldsymbol\Omega}(\tau)-{\boldsymbol\Omega}(\tau)\right\Vert _{\max}\left\Vert \widehat{\boldsymbol\Omega}(\tau)-{\boldsymbol\Omega}(\tau)\right\Vert _{1},
\]
leads to (4.12). The proof of Theorem 4.3 is completed. \hfill$\blacksquare$

\medskip

\noindent{\bf Proof of Corollary 4.2}.\ \  By (4.10) in Theorem 4.3 and the condition of $\min_{(i,j)\in{\mathbb E}^P} \min_{1\leq t\leq n}\vert\omega_{ij}(\tau_t)\vert\gg \lambda_3$, we have
\begin{equation}\label{eqA.24}
{\sf P}\left(\min_{(i,j)\in{\mathbb E}_{n}^P}\min_{1\leq t\leq n}\left\vert\widehat{\omega}_{ij}(\tau_t)\right\vert\geq \lambda_3>0\right)\rightarrow1.
\end{equation}
Letting $\mathcal{E}_\epsilon$ and $c_\epsilon$ be defined as in the proof of Theorem 4.3 and choosing $C_3=2c_\epsilon$ in Assumption 5(ii), we may prove that  
\begin{equation}\label{eqA.25}
\max_{(i,j)\notin{\mathbb E}_{n}^P}\max_{1\leq t\leq n}\left\vert\widehat{\omega}_{ij}(\tau_t)\right\vert\leq c_\epsilon(\nu_{n,d}^\ast+\nu_{n,d}^\diamond)<\lambda_3
\end{equation}
conditional on ${\cal E}_\epsilon$. By virtue of (\ref{eqA.24}) and (\ref{eqA.25}), letting $\epsilon\rightarrow0$, we prove (4.13).\hfill$\blacksquare$

\bigskip


\section*{\bf\Large Appendix B:\ \ Technical lemmas}\label{app:B}
\renewcommand{\theequation}{B.\arabic{equation}}
\setcounter{equation}{0}

In this appendix, we give some technical lemmas which are crucial to proofs of the main theoretical results in Appendix A. Without loss of generality, we focus on the time-varying VAR (1) model framework. Throughout the proofs, we let $M$ denote a generic positive constant whose value may change from line to line.

\renewcommand{\thelemma}{B.\arabic{lemma}}
\setcounter{lemma}{0}

\begin{lemma}\label{le:B.1}

Suppose that Assumption 1 is satisfied. Let 
\[\iota_2=\iota_1/C_\ast,\ \ \iota_3=\iota_1(1-\rho)/(C_1^2C_\ast),\ \ C_\ast=\max_{1\leq t\leq n} \Vert{\boldsymbol\Sigma}_t\Vert<\infty\]
where $\iota_1$ and $\rho$ are defined in Assumption 1, and $C_1$ is defined in (2.4). For any $d$-dimensional vector $u$ satisfying $\Vert u\Vert=1$,
\begin{equation}\label{eqB.1}
\max_{1\leq t\leq n} {\sf E}\left[\exp\left\{\iota_2\left(u^{^\intercal}e_t\right)^2\right\}\right]\leq C_0<\infty,
\end{equation}
and
\begin{equation}\label{eqB.2}
\max_{1\leq t\leq n}\max_{1\leq i\leq d} {\sf E}\left[\exp\left\{\iota_3 x_{t,i}^2\right\}\right]\leq C_0^{1/(1-\rho)}<\infty,
\end{equation}
where $C_0$ is a positive constant defined in Assumption 1(iii).

\end{lemma}

\medskip

\noindent{\bf Proof of Lemma \ref{le:B.1}}.\ \ Writing $u_t^{^\intercal}=u^{^\intercal}{\boldsymbol\Sigma}_t^{1/2}$ and using Assumption 1(ii)(iii), we may show that 
\begin{eqnarray}
\max_{1\leq t\leq n}{\sf E}\left[\exp\left\{\iota_2\left(u^{^\intercal}e_t\right)^2\right\}\right]&=&\max_{1\leq t\leq n}{\sf E}\left[\exp\left\{\iota_2\left(u^{^\intercal}{\boldsymbol\Sigma}_t^{1/2}\varepsilon_t\right)^2\right\}\right]\nonumber\\
&=&\max_{1\leq t\leq n}{\sf E}\left[\exp\left\{\iota_2 \Vert u_t\Vert^2\left(u_t^{^\intercal}\varepsilon_t/\Vert u_t\Vert\right)^2\right\}\right]\nonumber\\
&\leq&\max_{1\leq t\leq n}{\sf E}\left[\exp\left\{\iota_2 C_\ast\left(u_t^{^\intercal}\varepsilon_t/\Vert u_t\Vert\right)^2\right\}\right]\nonumber\\
&=&\max_{1\leq t\leq n}{\sf E}\left[\exp\left\{\iota_1\left(u_t^{^\intercal}\varepsilon_t/\Vert u_t\Vert\right)^2\right\}\right]\leq C_0,\nonumber
\end{eqnarray}
completing the proof of (\ref{eqB.1}).

By the time-varying linear process representation (2.3), we have
\[
x_{t,i}^2=\sum_{k_1=0}^\infty \sum_{k_2=0}^\infty \left(\Phi_{t,k_1,i}^{^\intercal}e_{t-k_1}\right)\left(\Phi_{t,k_2,i}^{^\intercal} e_{t-k_2}\right)
\]
where $\Phi_{t,k,i}^{^\intercal}$ is the $i$-th row vector of ${\boldsymbol\Phi}_{t,k}$. Without loss of generality, assume (2.4) for all $k\geq0$. Letting $u_{t,k,i}=\Phi_{t,k,i}/\Vert \Phi_{t,k,i}\Vert$ and noting that 
\[
\max_{1\leq t\leq n}\max_{1\leq i\leq d}\Vert \Phi_{t,k,i}\Vert\leq \max_{1\leq t\leq n}\Vert {\boldsymbol\Phi}_{t,k}\Vert\leq C_1\rho^k,
\]
we may show that 
\begin{eqnarray}
x_{t,i}^2&\leq& C_1^2\sum_{k_1=0}^\infty \rho^{k_1}\sum_{k_2=0}^\infty \rho^{k_2}\left\vert \left(u_{t,k_1,i}^{^\intercal}e_{t-k_1}\right)\left(u_{t,k_2,i}^{^\intercal} e_{t-k_2}\right)\right\vert\nonumber\\
&\leq&C_1^2\sum_{k_1=0}^\infty \rho^{k_1}\sum_{k_2=0}^\infty \rho^{k_2}\left(u_{t,k_2,i}^{^\intercal} e_{t-k_2}\right)^2 \nonumber\\
&=&\frac{C_1^2}{1-\rho}\sum_{k=0}^\infty \rho^{k}\left(u_{t,k,i}^{^\intercal} e_{t-k}\right)^2, \nonumber
\end{eqnarray}
which, together with the independence assumption over $e_t$ and (A.1), indicates that 
\begin{eqnarray}
\max_{1\leq t\leq n}\max_{1\leq i\leq d} {\sf E}\left[\exp\left\{\iota_3 x_{t,i}^2\right\}\right]&\leq&\max_{1\leq t\leq n}\max_{1\leq i\leq d} {\sf E}\left[\exp\left\{\frac{\iota_3 C_1^2}{1-\rho}\sum_{k=0}^\infty \rho^{k}\left(u_{t,k,i}^{^\intercal} e_{t-k}\right)^2\right\}\right]\nonumber\\
&=&\max_{1\leq t\leq n}\max_{1\leq i\leq d} \prod_{k=0}^\infty{\sf E}\left[\exp\left\{\frac{\iota_3 C_1^2}{1-\rho} \rho^{k}\left(u_{t,k,i}^{^\intercal} e_{t-k}\right)^2\right\}\right]\nonumber\\
&=&\max_{1\leq t\leq n}\max_{1\leq i\leq d} \prod_{k=0}^\infty{\sf E}\left[\exp\left\{\iota_2 \rho^{k}\left(u_{t,k,i}^{^\intercal} e_{t-k}\right)^2\right\}\right]\nonumber\\
&\leq&\prod_{k=0}^\infty\left(\max_{1\leq t\leq n}\max_{1\leq i\leq d} {\sf E}\left[\exp\left\{\iota_2 \left(u_{t,k,i}^{^\intercal} e_{t-k}\right)^2\right\}\right]\right)^{\rho^{k}}\nonumber\\
&\leq&\prod_{k=0}^\infty C_0^{\rho^k}=C_0^{1/(1-\rho)},\nonumber
\end{eqnarray}
completing the proof of (\ref{eqB.2}).\hfill $\blacksquare$

\medskip

The following lemma is a well-known Bernstein-type inequality for martingale differences \citep[e.g.,][]{F75, P99}.

\begin{lemma}\label{le:B.2} 

Let $(z_t, {\cal F}_t)_{t\geq 1}$ be a sequence of martingale differences and $\sigma_n^2 = \sum_{t=1}^n {\sf E}(z_t^2|{\cal F}_{t-1})$. Suppose that there exists a constant $a>0$ such that ${\sf P}(|z_t|\le a| {\cal F}_{t-1})=1$ for all $t\ge 2$. Then, for all $x, y>0$,
\[
{\sf P}\left(\sum_{t=1}^nz_t>x,\, \sigma_n^2\le y\right) \leq \exp \left\{-\frac {x^2}{2(y+ax)}\right\}.
\]

\end{lemma}

Define
\[
L_{i,0}(\tau)=  \frac{1}{n}\sum_{t=1}^ne_{t,i}(\tau)X_{t-1}K_h(\tau_t-\tau)\ \ {\rm and}\ \ L_{i,1}(\tau)=\frac{1}{n}\sum_{t=1}^ne_{t,i}(\tau)X_{t-1}\left(\frac{\tau_t-\tau}{h}\right)K_h(\tau_t-\tau),
\]
where $e_{t,i}(\tau)=x_{t,i}-\left[{\boldsymbol\alpha}_{i\bullet}(\tau)+{\boldsymbol\alpha}_{i\bullet}^\prime(\tau)(\tau_t-\tau)\right]^{^\intercal}X_{t-1}$. Lemma \ref{le:B.3} below gives the uniform asymptotic orders for the kernel-weighted quantities $L_{i,k}(\cdot)$, $k=0,1$. 

\begin{lemma}\label{le:B.3}

Suppose that Assumptions 1 and 2 are satisfied. Then we have
\begin{equation}\label{eqB.3}
\max_{1\leq i\leq d}\max_{1\leq t\leq n}\left\vert L_{i,k}(\tau_t)\right\vert_{\max}=O_P\left(\zeta_{n,d}\right),\ \ k=0,1,
\end{equation}
where $\zeta_{n,d}=\log (n\vee d)\left[(nh)^{-1/2}+sh^2\right]$ as in Assumption3(i).

\end{lemma}

\noindent{\bf Proof of Lemma \ref{le:B.3}}.\ \ We only prove (\ref{eqB.3}) for $k=0$ as the proof is analogous for $k=1$. Noting that 
\[e_{l,i}(\tau_t)=e_{l,i}+\left[{\boldsymbol\alpha}_{i\bullet}(\tau_l)-{\boldsymbol\alpha}_{i\bullet}(\tau_t)-{\boldsymbol\alpha}_{i\bullet}^\prime(\tau_t)(\tau_l-\tau_t)\right]^{^\intercal}X_{l-1}=:e_{l,i}+b_{l,i}^{^\intercal}(\tau_t)X_{l-1},\]
we write
\[
L_{i,0}(\tau_t)= \frac{1}{n}\sum_{l=1}^ne_{l,i}X_{l-1}K_h(\tau_l-\tau_t)+\frac{1}{n}\sum_{l=1}^nb_{l,i}^{^\intercal}(\tau_t)X_{l-1}X_{l-1}K_h(\tau_l-\tau_t).
\]
In order to prove (\ref{eqB.3}) with $k=0$, it is sufficient to show that 
\begin{equation}\label{eqB.4}
\max_{1\leq i\leq d}\max_{1\leq j\leq d}\max_{1\leq t\leq n}\left\vert  \frac{1}{n}\sum_{l=1}^ne_{l,i}x_{l-1,j}K_h(\tau_l-\tau_t) \right\vert=O_P\left((nh)^{-1/2}\log (n\vee d)\right)
\end{equation}
and
\begin{equation}\label{eqB.5}
\max_{1\leq i\leq d}\max_{1\leq j\leq d}\max_{1\leq t\leq n}\left\vert \frac{1}{n}\sum_{l=1}^nb_{l,i}^{^\intercal}(\tau_t)X_{l-1}x_{l-1,j}K_h(\tau_l-\tau_t) \right\vert=O_P\left(sh^2\log (n\vee d)\right).
\end{equation}

Define
\[\overline{e}_{l,i}=e_{l,i}I\left(\vert e_{l,i}\vert \leq 2\sqrt{\iota_2^{-1}\log (n\vee d)}\right),\ \ \widetilde{e}_{l,i}=e_{l,i}-\overline{e}_{l,i},\]
and
\[\overline{x}_{l,i}=x_{l,i}I\left(\vert x_{l,i}\vert \leq 2\sqrt{\iota_3^{-1}\log (n\vee d)}\right),\ \ \widetilde{x}_{l,i}=x_{l,i}-\overline{x}_{l,i},\]
where $\iota_2$ and $\iota_3$ are defined in Lemma \ref{le:B.1}. Then, we have the following decomposition:
\begin{eqnarray}
\frac{1}{n}\sum_{l=1}^ne_{l,i}x_{l-1,j}K_h(\tau_l-\tau_t)&=&\frac{1}{n}\sum_{l=1}^n\overline{e}_{l,i}\overline{x}_{l-1,j}K_h(\tau_l-\tau_t)+\frac{1}{n}\sum_{l=1}^n\overline{e}_{l,i}\widetilde{x}_{l-1,j}K_h(\tau_l-\tau_t)+\nonumber\\
&&\frac{1}{n}\sum_{l=1}^n\widetilde{e}_{l,i}\overline{x}_{l-1,j}K_h(\tau_l-\tau_t)+\frac{1}{n}\sum_{l=1}^n\widetilde{e}_{l,i}\widetilde{x}_{l-1,j}K_h(\tau_l-\tau_t).\nonumber
\end{eqnarray}

By the Bonferroni and Markov inequalities as well as (\ref{eqB.1}), for any $\epsilon>0$, we have
\begin{eqnarray}
&&{\sf P}\left(\max_{1\leq i\leq d}\max_{1\leq j\leq d}\max_{1\leq t\leq n}\left\vert \frac{1}{n}\sum_{l=1}^n\widetilde{e}_{l,i}\overline{x}_{l-1,j}K_h(\tau_l-\tau_t)\right\vert>\epsilon (nh)^{-1/2}\log (n\vee d)\right)\nonumber\\
&\leq&{\sf P}\left(\max_{1\leq i\leq d}\max_{1\leq t\leq n}\vert e_{t,i}\vert > 2\sqrt{\iota_2^{-1}\log (n\vee d)}\right)\nonumber\\
&\leq&\sum_{i=1}^d\sum_{t=1}^n{\sf P}\left(\vert e_{t,i}\vert > 2\sqrt{\iota_2^{-1}\log (n\vee d)}\right)\nonumber\\
&\leq&\sum_{i=1}^d\sum_{t=1}^n (n\vee d)^{-4}{\sf E}\left(\exp\left\{\iota_2 e_{t,i}^2\right\}\right)\nonumber\\
&\leq &M(n\vee d)^{-2}=o(1).\label{eqB.6}
\end{eqnarray}
Hence, we have
\begin{equation}\label{eqB.7}
\max_{1\leq i\leq d}\max_{1\leq j\leq d}\max_{1\leq t\leq n}\left\vert \frac{1}{n}\sum_{l=1}^n\widetilde{e}_{l,i}\overline{x}_{l-1,j}K_h(\tau_t-\tau_t)\right\vert=o_P\left( (nh)^{-1/2}\log (n\vee d)\right).
\end{equation}
Following the proof of (\ref{eqB.7}), we also have
\begin{equation}\label{eqB.8}
\max_{1\leq i\leq d}\max_{1\leq j\leq d}\max_{1\leq t\leq n}\left\vert \frac{1}{n}\sum_{l=1}^n\overline{e}_{l,i}\widetilde{x}_{l-1,j}K_h(\tau_l-\tau_t)\right\vert=o_P\left( (nh)^{-1/2}\log (n\vee d)\right)
\end{equation}
and
\begin{equation}\label{eqB.8}
\max_{1\leq i\leq d}\max_{1\leq j\leq d}\max_{1\leq t\leq n}\left\vert \frac{1}{n}\sum_{l=1}^n\widetilde{e}_{l,i}\widetilde{x}_{l-1,j}K_h(\tau_l-\tau_t)\right\vert=o_P\left( (nh)^{-1/2}\log (n\vee d)\right).
\end{equation}
By the Cauchy-Schwarz and Markov inequalities and (\ref{eqB.1}), we may show that 
\begin{eqnarray}
{\sf E}\left(\left\vert\widetilde{e}_{l,i}\right\vert\right)&\leq& \left[{\sf E}\left(\left\vert e_{l,i}\right\vert^2\right)\right]^{1/2}\left[{\sf P}\left(\vert e_{l,i}\vert > 2\sqrt{\iota_2^{-1}\log (n\vee d)}\right)\right]^{1/2}\nonumber\\
&=& \left[{\sf E}\left(\left\vert e_{l,i}\right\vert^2\right)\right]^{1/2}\left[{\sf P}\left(\exp\left\{\iota_2 e_{l,i}^2\right\} > (n\vee d)^4\right)\right]^{1/2}\nonumber\\
&\leq& \left[{\sf E}\left(\left\vert e_{l,i}\right\vert^2\right)\right]^{1/2}\left[{\sf E}\left(\exp\left\{\iota_2 e_{l,i}^2\right\}\right)\right]^{1/2}(n\vee d)^{-2}\nonumber\\
&\leq&M(n\vee d)^{-2},\nonumber
\end{eqnarray}
which, together with the definition of $\overline{x}_{l-1,j}$ and the condition on the kernel function, indicates that 
\begin{eqnarray}
\left\vert\frac{1}{n}\sum_{l=1}^n{\sf E}\left[\overline{e}_{l,i}\overline{x}_{l-1,j}K_h(\tau_l-\tau_t) \big\vert {\mathcal F}_{l-1}(X)\right] \right\vert &=& \left\vert\frac{1}{n}\sum_{l=1}^n{\sf E}\left[\widetilde{e}_{l,i}\overline{x}_{l-1,j}K_h(\tau_l-\tau_t) \big\vert {\mathcal F}_{l-1}(X)\right] \right\vert\nonumber\\
&=&O_P\left((n\vee d)^{-2}\sqrt{\log (n\vee d)}\right)\nonumber\\
&=&o_P\left((nh)^{-1/2}\log (n\vee d)\right),\label{eqB.10}
\end{eqnarray}
where ${\cal F}_l(X)=\sigma(X_t:\ t\leq l)$. With (\ref{eqB.7})--(\ref{eqB.10}), we readily have that 
\begin{equation}\label{eqB.11}
\frac{1}{n}\sum_{l=1}^ne_{l,i}x_{l-1,j}K_h(\tau_l-\tau_t)=\frac{1}{n}\sum_{l=1}^n\left\{\overline{e}_{l,i}-{\sf E}\left[\overline{e}_{l,i} | {\cal F}_{l-1}(X)\right]\right\}\overline{x}_{l-1,j}K_h(\tau_l-\tau_t)+o_P\left((nh)^{-1/2}\log (n\vee d)\right).
\end{equation}
By the Bonferroni inequality and the Bernstein inequality in Lemma \ref{le:B.2}, we prove that 
\begin{eqnarray}
&&{\sf P}\left(\max_{1\leq i\leq d}\max_{1\leq j\leq d}\max_{1\leq t\leq n}\left\vert \frac{1}{n}\sum_{l=1}^n\left\{\overline{e}_{l,i}-{\sf E}\left[\overline{e}_{l,i} | {\cal F}_{l-1}(X)\right]\right\}\overline{x}_{l-1,j}K_h(\tau_l-\tau_t)\right\vert>M_0 (nh)^{-1/2}\log (n\vee d)\right)\nonumber\\
&\leq&\sum_{i=1}^d\sum_{j=1}^d\sum_{t=1}^n{\sf P}\left(\left\vert \frac{1}{n}\sum_{l=1}^n\left\{\overline{e}_{l,i}-{\sf E}\left[\overline{e}_{l,i} | {\cal F}_{l-1}(X)\right]\right\}\overline{x}_{l-1,j}K_h(\tau_l-\tau_t)\right\vert>M_0 (nh)^{-1/2}\log (n\vee d)\right)\nonumber\\
&\leq&\sum_{i=1}^d\sum_{j=1}^d\sum_{t=1}^n \exp \left\{-g_0(M_0)\log (n\vee d)\right\} =O\left(nd^2(n\vee d)^{-g_0(M_0)}\right)=o(1),\nonumber
\end{eqnarray}
letting $M_0>0$ be sufficiently large, where $g_0(\cdot)$ is a positive function satisfying $g_0(z)\rightarrow\infty$ as $z\rightarrow +\infty$. Consequently, we have
\begin{equation}\label{eqB.12}
\max_{1\leq i\leq d}\max_{1\leq j\leq d}\max_{1\leq t\leq n}\left\vert \frac{1}{n}\sum_{l=1}^n\left\{\overline{e}_{l,i}-{\sf E}\left[\overline{e}_{l,i} | {\cal F}_{l-1}(X)\right]\right\}\overline{x}_{l-1,j}K_h(\tau_l-\tau_t)\right\vert=O_P\left( (nh)^{-1/2}\log (n\vee d)\right).
\end{equation}
By virtue of (\ref{eqB.11}) and (\ref{eqB.12}), we complete the proof of (\ref{eqB.4}).

Letting $\overline{X}_l=\left(\overline{x}_{l,1},\cdots,\overline{x}_{l,d}\right)^{^\intercal}$ and $\widetilde{X}_l=\left(\widetilde{x}_{l,1},\cdots,\widetilde{x}_{l,d}\right)^{^\intercal}$, we have the following decomposition:
\begin{eqnarray}
&&\frac{1}{n}\sum_{l=1}^nb_{l,i}^{^\intercal}(\tau_t)X_{l-1}x_{l-1,j}K_h(\tau_l-\tau_t)\nonumber\\
&=&\frac{1}{n}\sum_{l=1}^nb_{l,i}^{^\intercal}(\tau_t)\overline{X}_{l-1}\overline{x}_{l-1,j}K_h(\tau_l-\tau_t)+\frac{1}{n}\sum_{l=1}^nb_{l,i}^{^\intercal}(\tau_t)\overline{X}_{l-1}\widetilde{x}_{l-1,j}K_h(\tau_l-\tau_t)\nonumber\\
&&\frac{1}{n}\sum_{l=1}^nb_{l,i}^{^\intercal}(\tau_t)\widetilde{X}_{l-1}\overline{x}_{l-1,j}K_h(\tau_l-\tau_t)+\frac{1}{n}\sum_{l=1}^nb_{l,i}^{^\intercal}(\tau_t)\widetilde{X}_{l-1}\widetilde{x}_{l-1,j}K_h(\tau_l-\tau_t).\label{eqB.13}
\end{eqnarray}
Similarly to the proof of (\ref{eqB.11}), we may show that the last three terms on the right side of (\ref{eqB.13}) are of order $o_P\left(h^2\log (n\vee d)\right)$. By Assumption 1(i) and the Taylor expansion of ${\boldsymbol\alpha}_{i\bullet}(\cdot)$, we can prove that the first term on the right side of (\ref{eqB.13}) is of order $O_P\left(sh^2\log (n\vee d)\right)$ uniformly over $i,j,t$. The proof of (\ref{eqB.5}) is completed.\hfill$\blacksquare$

\medskip

Lemma \ref{le:B.4} below gives the mean squared convergence rates of the infeasible oracle estimates $\widehat{\mathbf A}_i^o$ and $\widehat{\mathbf B}_i^o$ defined in (4.6) and (4.7) of Section 4.2. 

\begin{lemma}\label{le:B.4}

Suppose Assumptions 1--4 are satisfied. Then we have
\begin{equation}\label{eqB.14}
\max_{1\leq i\leq d}\frac{1}{n}\sum_{t=1}^n\sum_{j=1}^d\left\Vert \widehat{\alpha}_{i,j}^o(\tau_t)-\alpha_{i,j}(\tau_t)\right\Vert^2=O_P\left(s\zeta_{n,d}^2\right),
\end{equation}
and
\begin{equation}\label{eqB.15}
\max_{1\leq i\leq d}\frac{1}{n}\sum_{t=1}^n\sum_{j=1}^d\left\Vert \widehat{\alpha}_{i,j}^{\prime o}(\tau_t)-\alpha_{i,j}^\prime(\tau_t)\right\Vert^2=O_P\left(s\zeta_{n,d}^2h^{-2}\right).
\end{equation}

\end{lemma}

\medskip

\noindent{\bf Proof of Lemma \ref{le:B.4}}.\ \ For any $1\leq i\leq d$, let
\[{\mathbf U}^o=\left[(v_1^o)^{^\intercal}, (w_1^o)^{^\intercal}, (v_2^o)^{^\intercal}, (w_2^o)^{^\intercal},\cdots, (v_n^o)^{^\intercal}, (w_n^o)^{^\intercal}\right]^{^\intercal},\]
where $v_t^o=\left(v_{1|t}^o,\cdots,v_{d|t}^o\right)^{^\intercal}$ with $v_{j|t}^o=0$ for $j\in\overline{\cal J}_i$, and  $w_t^o=\left(w_{1|t}^o,\cdots,w_{d|t}^o\right)^{^\intercal}$ with $w_{j|t}^o=0$ for $j\in\overline{\cal J}_i^\prime$. Define
$${\cal B}_i^\ast(M_\ast)=\left\{{\mathbf U}^o:\ \sum_{t=1}^n\left(\Vert v_t^o\Vert^2+\left\Vert w_t^o\right\Vert^2\right)=\left\Vert{\mathbf V}^o \right\Vert^2+\left\Vert{\mathbf W}^o \right\Vert^2=nM_\ast\right\},$$
where $M_\ast$ is a positive constant which can be sufficiently large, 
\[{\mathbf V}^o=\left[(v_1^o)^{^\intercal}, (v_2^o)^{^\intercal},\cdots, (v_n^o)^{^\intercal}\right]^{^\intercal}\ \ {\rm and}\ \ 
{\mathbf W}^o=\left[(w_1^o)^{^\intercal}, (w_2^o)^{^\intercal},\cdots, (w_n^o)^{^\intercal}\right]^{^\intercal}.\]
Write 
\begin{eqnarray}
&&{\mathbf A}_i=\left({\boldsymbol\alpha}_{i,1},\cdots,{\boldsymbol\alpha}_{i,d}\right)\ \ {\rm with}\ \ {\boldsymbol\alpha}_{i,j}=\left[\alpha_{i,j}(\tau_1),\cdots,\alpha_{i,j}(\tau_n)\right]^{^\intercal},\notag\\
&&{\mathbf B}_i=\left({\boldsymbol\alpha}_{i,1}^\prime,\cdots,{\boldsymbol\alpha}_{i,d}^\prime\right)\ \ {\rm with}\ \ {\boldsymbol\alpha}_{i,j}^\prime=\left[\alpha_{i,j}^\prime(\tau_1),\cdots,\alpha_{i,j}^\prime(\tau_n)\right]^{^\intercal},\notag
\end{eqnarray}
as the matrices of true time-varying parameters. Observe that
\begin{equation}\label{eqB.16}
{\mathcal Q}_i\left({\mathbf{A}}_i+\sqrt{\zeta_{n,d}^\ast} {\mathbf{V}}^o,{\mathbf{B}}_i+\sqrt{\zeta_{n,d}^\ast}{\mathbf W}^o/h\right)-
{\mathcal Q}_i\left({\mathbf{A}}_i,{\mathbf{B}}_i\right)=\Pi_{i,1}^o+\Pi_{i,2}^o+\Pi_{n,3}^o,
\end{equation}
where $\zeta_{n,d}^\ast=s\zeta_{n,d}^2$,
\begin{eqnarray}
\Pi_{i,1}^o&=&\sum_{t=1}^n\left[{\mathcal L}_{i}\left({\boldsymbol\alpha}_{i\bullet}(\tau_t)+\sqrt{\zeta_{n,d}^\ast}v_t^o, {\boldsymbol\alpha}_{i\bullet}^\prime(\tau_t)+\sqrt{\zeta_{n,d}^\ast}w_t^o/h\ |\ \tau_t\right)-{\cal L}_{i}\left({\boldsymbol\alpha}_{i\bullet}(\tau_t), {\boldsymbol\alpha}_{i\bullet}^\prime(\tau_t)\ |\ \tau_t\right)\right],\nonumber
\\
\Pi_{i,2}^o&=&\sum_{j=1}^{d} p_{\lambda_2}^\prime \left(\left\Vert  \widetilde{\boldsymbol\alpha}_{i,j} \right\Vert\right)\left(\left\Vert {\boldsymbol\alpha}_{i,j}+\sqrt{\zeta_{n,d}^\ast} {\mathbf v}_{j}^o\right\Vert-\left\Vert {\boldsymbol\alpha}_{i,j}\right\Vert\right),\nonumber
\\
\Pi_{i,3}^o&=&\sum_{j=1}^{d}p_{\lambda_2}^\prime \left(\widetilde{D}_{i,j}\right)\left( \left\Vert h{\boldsymbol\alpha}_{i,j}^\prime+\sqrt{\zeta_{n,d}^\ast} {\mathbf w}_{j}^o\right\Vert-\left\Vert h{\boldsymbol\alpha}_{i,j}^\prime\right\Vert\right),\nonumber
\end{eqnarray}
in which ${\mathbf v}_j^o=\left(v_{j|1}^o,\cdots,v_{j|n}^o\right)^{^\intercal}$ and ${\mathbf w}_j^o=\left(w_{j|1}^o,\cdots,w_{j|n}^o\right)^{^\intercal}$.

By the definition of the local linear objective function, we readily have 
\begin{equation}\label{eqB.17}
\Pi_{i,1}^o=-2\sqrt{\zeta_{n,d}^\ast}\sum_{t=1}^n\left[(v_t^o)^{^\intercal},(w_t^o)^{^\intercal}\right]L_{i}(\tau_t)+\zeta_{n,d}^\ast\sum_{t=1}^n\left[(v_t^o)^{^\intercal},(w_t^o)^{^\intercal}\right] {\boldsymbol\Psi}(\tau_t) \left[(v_t^o)^{^\intercal},(w_t^o)^{^\intercal}\right]^{^\intercal}.
\end{equation}
By the definition of ${\cal B}_i^\ast(M_\ast)$, Lemma \ref{le:B.3} and the Cauchy-Schwarz inequality, we prove 
\begin{equation}\label{eqB.18}
\left\vert \sum_{t=1}^n\left[(v_t^o)^{^\intercal},(w_t^o)^{^\intercal}\right]L_{i}(\tau_t) \right\vert=O_P\left(\sqrt{\zeta_{n,d}^\ast}n^{1/2}\right)\cdot\|{\mathbf U}^o\|
\end{equation}
uniformly over $i$. By the uniform restricted eigenvalue condition in Assumption 3(ii), we have
\begin{equation}\label{eqB.19}
\sum_{t=1}^n\left[(v_t^o)^{^\intercal},(w_t^o)^{^\intercal}\right] {\boldsymbol\Psi}(\tau_t) \left[(v_t^o)^{^\intercal},(w_t^o)^{^\intercal}\right]^{^\intercal}\geq\kappa_0\sum_{t=1}^n \left( \Vert v_t^o\Vert^2 + \Vert w_t^o\Vert^2\right)=n\kappa_0M_\ast
\end{equation}
for ${\mathbf U}^o\in {\cal B}_i^\ast(M_\ast)$. Combining (\ref{eqB.17})--(\ref{eqB.19}) and letting $M_\ast>0$ be sufficiently large, we have 
\begin{equation}\label{eqB.20}
\min_{1\leq i\leq d}\Pi_{i,1}^o\geq \kappa_0\zeta_{n,d}^\ast \Vert {\mathbf U}^o\Vert^2+ O_P\left(\zeta_{n,d}^\ast n^{1/2}\right)\cdot\Vert {\mathbf U}^o\Vert> \frac{\kappa_0}{2}\zeta_{n,d}^\ast \Vert {\mathbf U}^o\Vert^2\ \ w.p.a.1.
\end{equation}

On the other hand, by Theorem 4.1 and Assumption 4(ii), we have 
\[
{\sf P}\left(\min_{1\leq i\leq d}\min_{j\in{\cal J}_i}\left\Vert\widetilde{\boldsymbol\alpha}_{i,j}\right\Vert>a_0\lambda_2\right)\rightarrow1,
\]
and
\[
{\sf P}\left(\min_{1\leq i\leq d}\min_{j\in{\cal J}_i^\prime}\widetilde{D}_{i,j}>a_0\lambda_2\right)\rightarrow1.
\]
 As $\alpha_{i,j}(\tau_t)=0$ and $u_{1,j}^o=0$ for $j\in\overline{\cal J}_i(\tau_t)$, we thus have
 \begin{equation}\label{eqB.21}
\Pi_{i,2}(\tau_t)=\sum\limits_{j\in{\cal J}_i(\tau_t)}p_{\lambda_2}^\prime \left(\left\vert\widetilde\alpha_{i,j}(\tau_t)\right\vert\right) \left(\left\vert\alpha_{i,j}(\tau_t)+\sqrt{\zeta_{n,d}^\ast}(\tau_t)u_{1,j}^o\right\vert -\left\vert \alpha_{i,j}(\tau_t)\right\vert\right)=0\ \ w.p.a.1,
\end{equation} 
and similarly 
\begin{equation}\label{eqB.22}
\Pi_{i,3}(\tau_t)=\sum\limits_{j\in{\cal J}_i(\tau_t)}p_{\lambda_2}^\prime \left(\left\vert \widetilde\alpha_{i,j}^\prime(\tau_t)\right\vert\right)
\left(\left\vert h\alpha_{i,j}^\prime(\tau_t)+\sqrt{\zeta_{n,d}^\ast}(\tau_t)u_{2,j}^o\right\vert-\left\vert h \alpha_{i,j}^\prime(\tau_t)\right\vert\right)=0\ \ w.p.a.1.
\end{equation}

Hence, by (\ref{eqB.20})--(\ref{eqB.22}), letting $M_\ast>0$ be large enough, we can prove that
\[
\min_{1\leq i\leq d}\left[\sup_{{\mathbf U}^o\in {\cal B}_i^\ast(M_\ast)}{\mathcal Q}_i\left({\mathbf{A}}_i+\sqrt{\zeta_{n,d}^\ast} {\mathbf{V}}^o,{\mathbf{B}}_i+\sqrt{\zeta_{n,d}^\ast}{\mathbf W}^o/h\right)-{\mathcal Q}_i\left({\mathbf{A}}_i,{\mathbf{B}}_i\right)\right]>0\ \ w.p.a.1,
\]
indicating that there exists a local minimiser $\left(\widehat{\mathbf A}_i^o, \widehat{\mathbf B}_i^o\right)$ in the interior of 
\[\left\{\left({\mathbf{A}}_i+\sqrt{\zeta_{n,d}^\ast} {\mathbf{V}}^o,{\mathbf{B}}_i+\sqrt{\zeta_{n,d}^\ast}{\mathbf W}^o/h\right):\ {\mathbf U}^o\in {\cal B}_i^\ast(c_1) \right\}\] 
for any $1\leq i\leq d$. The proof of Lemma \ref{le:B.4} is completed.\hfill$\blacksquare$

\medskip

Lemma \ref{le:B.5} below gives the uniform convergence rates for the time-varying volatility function estimates, a crucial result to prove uniform consistency of the time-varying CLIME estimates.

\begin{lemma}\label{le:B.5} 

Suppose that Assumptions 1--4 are satisfied. Then we have
\begin{equation}\label{eqB.23}
\max_{1\leq i,j\leq d}\sup_{0\leq \tau\leq 1}\left\vert\widehat{\sigma}_{ij}(\tau)-\sigma_{ij}(\tau)\right\vert=O_P\left(\nu_{n,d}^\diamond+\nu_{n,d}^\ast\right),
\end{equation}
where $\sigma_{ij}(\tau)$ is the $(i,j)$-entry of ${\boldsymbol\Sigma}(\tau)$, $\nu_{n,d}^\diamond$ and $\nu_{n,d}^\ast$ are defined in Assumption 5(ii).

\end{lemma}

\noindent{\bf Proof of Lemma \ref{le:B.5}}.\ \ By the definition of $\widehat{\sigma}_{ij}(\tau)$ in (3.8), we have
\begin{eqnarray}
\widehat{\sigma}_{ij}(\tau)-\sigma_{ij}(\tau)&=&\left\{\frac{\sum_{t = 1}^n \varpi_{n,t}(\tau) e_{t,i}e_{t,j}}{\sum_{t = 1}^n \varpi_{n,t}(\tau)}-\sigma_{ij}(\tau)\right\}+\left\{\frac{\sum_{t = 1}^n \varpi_{n,t}(\tau) \left(\widehat{e}_{t,i}-e_{t,i}\right)e_{t,j}}{\sum_{t = 1}^n \varpi_{n,t}(\tau)}+\right.\nonumber\\
&&\left.\frac{\sum_{t = 1}^n \varpi_{n,t}(\tau) e_{t,i}\left(\widehat{e}_{t,j}-e_{t,j}\right)}{\sum_{t = 1}^n \varpi_{n,t}(\tau)}+\frac{\sum_{t = 1}^n \varpi_{n,t}(\tau) \left(\widehat{e}_{t,i}-e_{t,i}\right)\left(\widehat{e}_{t,j}-e_{t,j}\right)}{\sum_{t = 1}^n \varpi_{n,t}(\tau)}\right\}\nonumber\\
&=:&\chi_{ij}^\diamond(\tau)+\chi_{ij}^\ast(\tau).\label{eqB.24}
\end{eqnarray}

We first prove that 
\begin{equation}\label{eqB.25}
\max_{1\leq i,j\leq d}\sup_{0\leq \tau\leq 1}\left\vert \chi_{ij}^\diamond(\tau) \right\vert=O_P\left(\nu_{n,d}^\diamond\right).
\end{equation}
Note that 
\[
\chi_{ij}^\diamond(\tau)=\frac{\sum_{t = 1}^n \varpi_{n,t}(\tau) \left[e_{t,i}e_{t,j}-\sigma_{ij}(\tau_t)\right]}{\sum_{t = 1}^n \varpi_{n,t}(\tau)}+\frac{\sum_{t = 1}^n \varpi_{n,t}(\tau)\sigma_{ij}(\tau_t)}{\sum_{t = 1}^n \varpi_{n,t}(\tau)}-\sigma_{ij}(\tau).
\]
By the Taylor expansion of $\sigma_{ij}(\cdot)$ and the definition of the local linear weights $\varpi_{n,t}(\tau)$, we have
\begin{eqnarray}
&&\max_{1\leq i,j\leq d}\sup_{0\leq \tau\leq 1} \left\vert \frac{\sum_{t = 1}^n \varpi_{n,t}(\tau)\sigma_{ij}(\tau_t)}{\sum_{t = 1}^n \varpi_{n,t}(\tau)}-\sigma_{ij}(\tau)\right\vert\nonumber\\
&\leq& \max_{1\leq i,j\leq d}\sup_{0\leq \tau\leq 1}\left\vert \sigma_{ij}^{\prime\prime}(\tau)\right\vert\cdot \left\vert \frac{\sum_{t = 1}^n (\tau_t-\tau)^2\varpi_{n,t}(\tau)}{\sum_{t = 1}^n \varpi_{n,t}(\tau)}\right\vert\nonumber\\
&\leq&M\sup_{0\leq \tau\leq 1}\left\vert \frac{\sum_{t = 1}^n (\tau_t-\tau)^2\varpi_{n,t}(\tau)}{\sum_{t = 1}^n \varpi_{n,t}(\tau)}\right\vert=O\left(b^2\right).\label{eqB.26}
\end{eqnarray}

Let $\overline{e}_{t,i}$ and $\widetilde{e}_{t,j}$ be defined as in the proof of Lemma \ref{le:B.3}. Then, we have 
\begin{eqnarray}
\sum_{t=1}^nK\left(\frac{\tau_t-\tau}{b}\right)e_{t,i}e_{t,j}&=&\sum_{t=1}^nK\left(\frac{\tau_t-\tau}{b}\right)\overline e_{t,i}\overline e_{t,j}+\sum_{t=1}^nK\left(\frac{\tau_t-\tau}{b}\right)\overline e_{t,i}\widetilde e_{t,j}+\nonumber\\
&&\sum_{t=1}^nK\left(\frac{\tau_t-\tau}{b}\right)\widetilde e_{t,i}\overline e_{t,j}+\sum_{t=1}^nK\left(\frac{\tau_t-\tau}{b}\right)\widetilde e_{t,i}\widetilde e_{t,j}.\label{eqB.27}
\end{eqnarray}
Following the proof of (\ref{eqB.11}), the first term on the right side of (\ref{eqB.27}) is the asymptotic leading term. Consider covering the closed interval $[0,1]$ by some disjoint intervals $\mathcal{I}_k$, $k=1,\cdots,N$, with the center $\tau_k^\ast$ and length $b^2[nb\log(n\vee d)]^{-1/2}$. By the Lipschitz continuity of $K(\cdot)$ in Assumption 2(i), we have
\begin{eqnarray}
& &\max_{1\leq i,j\leq d}\sup_{0\leq \tau\leq 1}\left\vert \frac{1}{nb}\sum_{t=1}^nK\left(\frac{\tau_t-\tau}{b}\right)\left[\overline e_{t,i}\overline e_{t,j}-{\sf E}\left(\overline e_{t,i}\overline e_{t,j}\right)\right] \right\vert \nonumber\\
&\leq & \max_{1\leq i,j\leq d}\max_{1\leq k\leq N}\left\vert \frac{1}{nb}\sum_{t=1}^nK\left(\frac{\tau_t-\tau_k^\ast}{b}\right)\left[\overline e_{t,i}\overline e_{t,j}-{\sf E}\left(\overline e_{t,i}\overline e_{t,j}\right)\right] \right\vert  +\nonumber\\
&&  \max_{1\leq i,j\leq d}\max_{1\leq k\leq N}\sup_{\tau\in{\cal I}_k}\left\vert \frac{1}{nb}\sum_{k=1}^{n}\left[  K\left(\frac{\tau_t-\tau}{b}\right)-K\left(\frac{\tau_t-\tau_k^\ast}{b}\right)\right]  \left[\overline e_{t,i}\overline e_{t,j}-{\sf E}\left(\overline e_{t,i}\overline e_{t,j}\right)\right]  \right\vert \nonumber\\
&\leq & \max_{1\leq i,j\leq d}\max_{1\leq k\leq N}\left\vert \frac{1}{nb}\sum_{t=1}^nK\left(\frac{\tau_t-\tau_k^\ast}{b}\right)\left[\overline e_{t,i}\overline e_{t,j}-{\sf E}\left(\overline e_{t,i}\overline e_{t,j}\right)\right] \right\vert +O_{P}\left( \left[\frac{\log(n\vee d)}{nb}\right]^{1/2}\right).\label{eqB.28}
\end{eqnarray}
By the Bonferroni inequality and Lemma \ref{le:B.2} as well as the condition $nb/[\log(n\vee d)]^3\rightarrow\infty$ in Assumption 5(i), we may show that
\begin{eqnarray}
&& \mathsf{P}\left( \max_{1\leq i,j\leq d}\max_{1\leq k\leq N}\left\vert \frac{1}{nb}\sum_{t=1}^nK\left(\frac{\tau_t-\tau_k^\ast}{b}\right)\left[\overline e_{t,i}\overline e_{t,j}-{\sf E}\left(\overline e_{t,i}\overline e_{t,j}\right)\right] \right\vert >M_1\left[\frac{\log(n\vee d)}{nb}\right]^{1/2}\right) \nonumber\\
&\leq&\sum_{i=1}^{d}\sum_{j=1}^{d}\sum_{k=1}^{N}\mathsf{P}\left( \left\vert \sum_{t=1}^nK\left(\frac{\tau_t-\tau_k^\ast}{b}\right)\left[\overline e_{t,i}\overline e_{t,j}-{\sf E}\left(\overline e_{t,i}\overline e_{t,j}\right)\right] \right\vert >M_1\left[nb\log(n\vee d)\right]^{1/2}\right) \nonumber\\
&= &O\left(  d^2N \exp\left\{  -g_1(M_1)\log(n\vee d)\right\}  \right)=O\left(  d^2N (n\vee d)^{g_1(M_1)} \right)=o(1),\nonumber
\end{eqnarray}
where $M_1>0$ is sufficiently large and $g_1(\cdot)$ is a positive function satisfying that $g_1(z)\rightarrow\infty$ as $z\rightarrow+\infty$. Therefore, we have
\begin{equation}\label{eqB.29}
 \max_{1\leq i,j\leq d}\max_{1\leq k\leq N}\left\vert \frac{1}{nb}\sum_{t=1}^nK\left(\frac{\tau_t-\tau_k^\ast}{b}\right)\left[\overline e_{t,i}\overline e_{t,j}-{\sf E}\left(\overline e_{t,i}\overline e_{t,j}\right)\right] \right\vert=O_{P}\left( \left[\frac{\log(n\vee d)}{nb}\right]^{1/2}\right).
\end{equation}
Combining (\ref{eqB.28}) and (\ref{eqB.29}), we can prove that 
\begin{equation}\label{eqB.30}
\max_{1\leq i,j\leq d}\sup_{0\leq \tau\leq 1}\left\vert \frac{1}{nb}\sum_{t=1}^nK\left(\frac{\tau_t-\tau}{b}\right)\left[\overline e_{t,i}\overline e_{t,j}-{\sf E}\left(\overline e_{t,i}\overline e_{t,j}\right)\right] \right\vert=O_{P}\left( \left[\frac{\log(n\vee d)}{nb}\right]^{1/2}\right).
\end{equation}

By the definitions of $\overline{e}_{t,i}$ and $\widetilde{e}_{t,i}$, we have
\[
{\sf E}\left(\overline e_{t,i}\overline e_{t,j}\right)-\sigma_{ij}(\tau_t)={\sf E}\left(\widetilde e_{t,i}\widetilde e_{t,j}\right)-{\sf E}\left(\overline e_{t,i}\widetilde e_{t,j}\right)-{\sf E}\left(\widetilde e_{t,i}\overline e_{t,j}\right).
\]
Meanwhile, by the Cauchy-Schwarz and Markov inequalities and (\ref{eqB.1}) in Lemma \ref{le:B.1}, 
\begin{eqnarray}
{\sf E}\left(\left\vert\overline e_{t,i}\widetilde e_{t,j}\right\vert\right)&\leq& M \left[{\sf E}\left(\widetilde{e}_{t,j}^2\right)\right]^{1/2}\nonumber\\
&\leq& M\left[{\sf E}\left(\left\vert e_{t,i}\right\vert^4\right)\right]^{1/4}\left[{\sf P}\left(\vert e_{t,i}\vert > 2\sqrt{\iota_2^{-1}\log (n\vee d)}\right)\right]^{1/4}\nonumber\\
&\leq&M\left[{\sf P}\left(\exp\left\{\iota_2 e_{t,i}^2\right\} > (n\vee d)^4\right)\right]^{1/4}\nonumber\\
&\leq&M\left[{\sf E}\left(\exp\left\{\iota_2 e_{t,i}^2\right\}\right)\right]^{1/4}(n\vee d)^{-1}\nonumber\\
&\leq&O\left((n\vee d)^{-1}\right)=o\left( \left[\frac{\log(n\vee d)}{nb}\right]^{1/2}\right),\nonumber
\end{eqnarray}
and similarly, 
\[{\sf E}\left(\left\vert\widetilde e_{t,i}\overline e_{t,j}\right\vert\right)+{\sf E}\left(\left\vert\widetilde e_{t,i}\widetilde e_{t,j}\right\vert\right)=o\left( \left[\frac{\log(n\vee d)}{nb}\right]^{1/2}\right).\]
Hence, we can prove that 
\begin{equation}\label{eqB.31}
\max_{1\leq i,j\leq d}\sup_{0\leq \tau\leq 1}\left\vert \frac{1}{nb}\sum_{t=1}^nK\left(\frac{\tau_t-\tau}{b}\right)\left[{\sf E}\left(\overline e_{t,i}\overline e_{t,j}\right)-\sigma_{ij}(\tau_t)\right] \right\vert=o_{P}\left( \left[\frac{\log(n\vee d)}{nb}\right]^{1/2}\right).
\end{equation}

With (\ref{eqB.27}), (\ref{eqB.30}) and (\ref{eqB.31}), we can prove that 
\begin{equation}\label{eqB.32}
\max_{1\leq i,j\leq d}\sup_{0\leq \tau\leq 1}\left\vert \frac{1}{nb}\sum_{t=1}^nK\left(\frac{\tau_t-\tau}{b}\right)\left[ e_{t,i} e_{t,j}-\sigma_{ij}(\tau_t)\right] \right\vert=O_{P}\left( \left[\frac{\log(n\vee d)}{nb}\right]^{1/2}\right).
\end{equation}
Analogously, we also have
\begin{equation}\label{eqB.33}
\max_{1\leq i,j\leq d}\sup_{0\leq \tau\leq 1}\left\vert \frac{1}{nb}\sum_{t=1}^nK_1\left(\frac{\tau_t-\tau}{b}\right)\left[ e_{t,i} e_{t,j}-\sigma_{ij}(\tau_t)\right] \right\vert=O_{P}\left( \left[\frac{\log(n\vee d)}{nb}\right]^{1/2}\right).
\end{equation}
Using (\ref{eqB.32}), (\ref{eqB.33}) and the definition of $\varpi_{n,t}(\tau)$, we may show that
\begin{equation}\label{eqB.34}
\max_{1\leq i,j\leq d}\sup_{0\leq \tau\leq 1}\left\vert\frac{\sum_{t = 1}^n \varpi_{n,t}(\tau) \left[e_{t,i}e_{t,j}-\sigma_{ij}(\tau_t)\right]}{\sum_{t = 1}^n \varpi_{n,t}(\tau)}\right\vert=O_{P}\left( \left[\frac{\log(n\vee d)}{nb}\right]^{1/2}\right),
\end{equation}
which, together with (\ref{eqB.26}), leads to (\ref{eqB.25}).

Using the arguments in the proof of Lemma \ref{le:B.4}, we may prove that
\begin{equation}\label{eqB.35}
\max_{1\leq i\leq d}\max_{1\leq t\leq n}\left\Vert \widehat{\boldsymbol\alpha}_{i\bullet}^o(\tau_t)-{\boldsymbol\alpha}_{i\bullet}(\tau_t)\right\Vert =O_P\left(\sqrt{s}\zeta_{n,d}\right),
\end{equation}
which, together with (\ref{eqB.2}) in Lemma \ref{le:B.1}, indicates that
\begin{equation}\label{eqB.36}
\max_{1\leq i\leq d}\max_{1\leq t\leq n}\left\vert \widehat{e}_{t,i}-e_{t,i} \right\vert=O_P\left(s\zeta_{n,d}\sqrt{\log (n\vee d)}\right).
\end{equation}
By (\ref{eqB.25}), (\ref{eqB.36}) and the Cauchy-Schwarz inequality, letting $\varpi_{n,t}^\ast(\tau)=\varpi_{n,t}(\tau)/\sum_{t = 1}^n \varpi_{n,t}(\tau)$, we can prove that
\begin{eqnarray}
&&\max_{1\leq i,j\leq d}\sup_{0\leq \tau\leq 1}\left\vert\sum_{t = 1}^n \varpi_{n,t}^\ast(\tau) \left(\widehat{e}_{t,i}-e_{t,i}\right)e_{t,j}\right\vert\nonumber\\
&\leq&\max_{1\leq j\leq d}\sup_{0\leq \tau\leq 1}\left(\sum_{t = 1}^n \left\vert \varpi_{n,t}^\ast(\tau)\right\vert e_{t,j}^2 \right)^{1/2} \max_{1\leq i\leq d}\sup_{0\leq \tau\leq 1}\left(\sum_{t = 1}^n \left\vert \varpi_{n,t}^\ast(\tau)\right\vert \left(\widehat{e}_{t,i}-e_{t,i}\right)^2\right)^{1/2}\nonumber\\
&=&O_P\left(s\zeta_{n,d}\sqrt{\log (n\vee d)}\right)=O_P\left(\nu_{n,d}^\ast\right).\label{eqB.37}
\end{eqnarray}
Similarly, we can also show that 
\begin{equation}\label{eqB.38}
\max_{1\leq i,j\leq d}\sup_{0\leq \tau\leq 1}\left\vert\sum_{t = 1}^n \varpi_{n,t}^\ast(\tau)e_{t,i} \left(\widehat{e}_{t,j}-e_{t,j}\right)\right\vert=O_P\left(\nu_{n,d}^\ast\right)
\end{equation}
and
\begin{equation}\label{eqB.39}
\max_{1\leq i,j\leq d}\sup_{0\leq \tau\leq 1}\left\vert\sum_{t = 1}^n \varpi_{n,t}^\ast(\tau)\left(\widehat e_{t,i}-e_{t,i}\right) \left(\widehat{e}_{t,j}-e_{t,j}\right)\right\vert=O_P\left(\left[\nu_{n,d}^\ast\right]^2\right)=o_P\left(\nu_{n,d}^\ast\right).
\end{equation}
From (\ref{eqB.37})--(\ref{eqB.39}), we readily have that 
\[
\max_{1\leq i,j\leq d}\sup_{0\leq \tau\leq 1}\left\vert \chi_{ij}^\ast(\tau) \right\vert=O_P\left(\nu_{n,d}^\ast\right),
\]
which, together with (\ref{eqB.24}) and (\ref{eqB.25}), completes the proof of Lemma \ref{le:B.5}.\hfill$\blacksquare$

\bigskip

\section*{\bf\Large Appendix C:\ \ Proofs of Propositions 5.1 and 5.2}\label{app:C}
\renewcommand{\theequation}{C.\arabic{equation}}
\setcounter{equation}{0}

In this appendix, we provide proofs of the convergence properties for the factor-adjusted estimators stated in Propositions 5.1 and 5.2. Define
\[
\widehat L_{i,0}(\tau)=  \frac{1}{n}\sum_{t=1}^n\widehat{e}_{t,i}(\tau)\widehat{X}_{t-1}K_h(\tau_t-\tau)\ \ {\rm and}\ \ \widehat L_{i,1}(\tau)=\frac{1}{n}\sum_{t=1}^n\widehat{e}_{t,i}(\tau)\widehat{X}_{t-1}\left(\frac{\tau_t-\tau}{h}\right)K_h(\tau_t-\tau),
\]
where $\widehat{X}_t=\left(\widehat{x}_{t,1},\cdots,\widehat{x}_{t,d}\right)^{^\intercal}$ is defined in (5.3) or (5.4), and $\widehat{e}_{t,i}(\tau)=\widehat{x}_{t,i}-\left[{\boldsymbol\alpha}_{i\bullet}(\tau)+{\boldsymbol\alpha}_{i\bullet}^\prime(\tau)(\tau_t-\tau)\right]^{^\intercal}\widehat{X}_{t-1}$. The following lemma extends Lemma \ref{le:B.3} to the factor-adjusted kernel-weighted quantities.

\renewcommand{\thelemma}{C.\arabic{lemma}}
\setcounter{lemma}{0}

\begin{lemma}\label{le:C.1}

Suppose that Assumptions 1, 2 and 6(i) are satisfied. Then we have
\begin{equation}\label{eqC.1}
\max_{1\leq i\leq d}\max_{1\leq t\leq n}\left\vert \widehat{L}_{i,k}(\tau_t)\right\vert_{\max}=O_P\left(\zeta_{n,d}^\dagger\right),\ \ k=0,1,
\end{equation}
where $\zeta_{n,d}^\dagger=\zeta_{n,d}+[\log (n\vee d)]^{1/2}s\delta_X$ as in Assumption 6(ii).

\end{lemma}

\noindent{\bf Proof of Lemma \ref{le:C.1}}.\ \ As in the proof of Lemma \ref{le:B.3}, we only consider $k=0$. As 
\[
\widehat{e}_{t,i}(\tau)=e_{t,i}(\tau)+\left(\widehat{x}_{t,i}-x_{t,i}\right)+\left[{\boldsymbol\alpha}_{i\bullet}(\tau)+{\boldsymbol\alpha}_{i\bullet}^\prime(\tau)(\tau_t-\tau)\right]^{^\intercal}\left(X_{t-1}-\widehat{X}_{t-1}\right),
\]
by Assumption 6(i), we may show that 
\begin{eqnarray}
\widehat L_{i,0}(\tau)&=& L_{i,0}(\tau)+ \frac{1}{n}\sum_{t=1}^n\left(\widehat{x}_{t,i}-x_{t,i}\right)\widehat{X}_{t-1}K_h(\tau_t-\tau)+\notag\\
&& \frac{1}{n}\sum_{t=1}^n\left[{\boldsymbol\alpha}_{i\bullet}(\tau)+{\boldsymbol\alpha}_{i\bullet}^\prime(\tau)(\tau_t-\tau)\right]^{^\intercal}\left(X_{t-1}-\widehat{X}_{t-1}\right)\widehat{X}_{t-1}K_h(\tau_t-\tau)-\notag\\
&& \frac{1}{n}\sum_{t=1}^n e_{t,i}(\tau)\left(X_{t-1}-\widehat{X}_{t-1}\right)K_h(\tau_t-\tau) \notag\\
&=& L_{i,0}(\tau)+O_P\left([\log (n\vee d)]^{1/2}s\delta_X\right).\notag
\end{eqnarray}
Then, by Lemma \ref{le:B.3}, we complete the proof of (\ref{eqC.1}) for $k=0$. \hfill$\blacksquare$

\medskip

Write 
\[
\widehat e_t^\dagger=\left(\widehat e_{t,1}^\dagger,\cdots, \widehat e_{t,d}^\dagger\right)^{^\intercal}
=\widehat{X}_t-\widehat{A}_1^\dagger(\tau_t)\widehat{X}_{t-1},\ \ \widehat{A}_1^\dagger(\tau_t)=\left[\widehat{\alpha}_{ij}^\dagger(\tau_t)\right]_{d\times d}.
\]
Let $\widehat\sigma_{ij}^\dagger(\tau)$ be the factor-adjusted local linear estimate $\sigma_{ij}(\tau)$, i.e., replace $\widehat{e}_{t,i}$ by $\widehat{e}_{t,i}^\dagger$ in (3.8). The following lemma extends Lemma \ref{le:B.5} to the factor-adjusted volatility function estimate.

\begin{lemma}\label{le:C.2} 

Suppose that the assumptions of Proposition 5.1(iii) are satisfied. Then we have
\begin{equation}\label{eqC.2}
\max_{1\leq i,j\leq d}\sup_{0\leq \tau\leq 1}\left\vert\widehat{\sigma}_{ij}^\dagger(\tau)-\sigma_{ij}(\tau)\right\vert=O_P\left(\nu_{n,d}^\diamond+\nu_{n,d}^\dagger\right),
\end{equation}
where $\nu_{n,d}^\diamond$ is defined in Assumption 5(ii) and $\nu_{n,d}^\dagger$ is defined in Assumption 6(iv).

\end{lemma}

\noindent{\bf Proof of Lemma \ref{le:C.2}}.\ \ As in (\ref{eqB.24}), we have
\begin{eqnarray}
\widehat{\sigma}_{ij}^\dagger(\tau)-\sigma_{ij}(\tau)&=&\left\{\frac{\sum_{t = 1}^n \varpi_{n,t}(\tau) e_{t,i}e_{t,j}}{\sum_{t = 1}^n \varpi_{n,t}(\tau)}-\sigma_{ij}(\tau)\right\}+\left\{\frac{\sum_{t = 1}^n \varpi_{n,t}(\tau) \left(\widehat{e}_{t,i}^\dagger-e_{t,i}\right)e_{t,j}}{\sum_{t = 1}^n \varpi_{n,t}(\tau)}+\right.\nonumber\\
&&\left.\frac{\sum_{t = 1}^n \varpi_{n,t}(\tau) e_{t,i}\left(\widehat{e}_{t,j}^\dagger-e_{t,j}\right)}{\sum_{t = 1}^n \varpi_{n,t}(\tau)}+\frac{\sum_{t = 1}^n \varpi_{n,t}(\tau) \left(\widehat{e}_{t,i}^\dagger-e_{t,i}\right)\left(\widehat{e}_{t,j}^\dagger-e_{t,j}\right)}{\sum_{t = 1}^n \varpi_{n,t}(\tau)}\right\}\nonumber\\
&=:&\chi_{ij}^\diamond(\tau)+\chi_{ij}^\dagger(\tau).\label{eqC.3}
\end{eqnarray}
By (\ref{eqB.25}), we only need to show 
\begin{equation}\label{eqC.4}
\max_{1\leq i,j\leq d}\sup_{0\leq \tau\leq 1}\left\vert \chi_{ij}^\dagger(\tau) \right\vert=O_P\left(\nu_{n,d}^\dagger\right).
\end{equation}
Following the proof of (\ref{eqB.36}), we have
\begin{equation}\label{eqC.5}
\max_{1\leq i\leq d}\max_{1\leq t\leq n}\left\vert \widehat{e}_{t,i}^\dagger-e_{t,i} \right\vert=O_P\left(s\zeta_{n,d}^\dagger\sqrt{\log (n\vee d)}\right).
\end{equation}
By (\ref{eqB.25}), (\ref{eqC.5}) and the Cauchy-Schwarz inequality, we can prove that
\begin{eqnarray}
&&\max_{1\leq i,j\leq d}\sup_{0\leq \tau\leq 1}\left\vert\sum_{t = 1}^n \varpi_{n,t}^\ast(\tau) \left(\widehat{e}_{t,i}^\dagger-e_{t,i}\right)e_{t,j}\right\vert=O_P\left(s\zeta_{n,d}^\dagger\sqrt{\log (n\vee d)}\right)=O_P\left(\nu_{n,d}^\dagger\right),\label{eqC.6}\\
&&\max_{1\leq i,j\leq d}\sup_{0\leq \tau\leq 1}\left\vert\sum_{t = 1}^n \varpi_{n,t}^\ast(\tau)e_{t,i} \left(\widehat{e}_{t,j}^\dagger-e_{t,j}\right)\right\vert=O_P\left(\nu_{n,d}^\dagger\right),\label{eqC.7}\\
&&\max_{1\leq i,j\leq d}\sup_{0\leq \tau\leq 1}\left\vert\sum_{t = 1}^n \varpi_{n,t}^\ast(\tau)\left(\widehat e_{t,i}^
\dagger-e_{t,i}\right) \left(\widehat{e}_{t,j}^\dagger-e_{t,j}\right)\right\vert=o_P\left(\nu_{n,d}^\dagger\right).\label{eqC.8}
\end{eqnarray}
With (\ref{eqC.6})--(\ref{eqC.8}), we complete the proof of (\ref{eqC.4}).\hfill$\blacksquare$

\medskip

Define
\[
\widehat{\boldsymbol\Psi}(\tau)=\left[
\begin{array}{cc}
\widehat{\boldsymbol\Psi}_{0}(\tau)&\widehat{\boldsymbol\Psi}_1(\tau)\\
\widehat{\boldsymbol\Psi}_1(\tau)&\widehat{\boldsymbol\Psi}_2(\tau)
\end{array}
\right]\ \ 
{\rm with}
\ \ 
\widehat{\boldsymbol\Psi}_k(\tau)=\frac{1}{n}\sum\limits_{t=1}^n\left(\frac{\tau_t-\tau}{h}\right)^{k} \widehat X_{t-1}\widehat X_{t-1}^{^\intercal}K_h(\tau_t - \tau),\ \ k=0,1,2.
\]

\medskip

\noindent{\bf Proof of Proposition 5.1}.\ \ We start with the proof of
\begin{equation}\label{eqC.9}
\min_{1\leq i\leq d}\min_{1\leq t\leq n}\inf_{u\in{\cal B}_{i}(\tau_t)}u^{^\intercal} \widehat{\boldsymbol\Psi}(\tau_t)u\geq \kappa_0/2,\ \ w.p.a.1,
\end{equation}
where ${\cal B}_{i}(\tau)$ is defined as in (4.2). In fact, combining Assumption 6(i) with the arguments in the proofs of Lemmas \ref{le:B.3} and \ref{le:C.1}, we may show that 
\begin{equation}\label{eqC.10}
\max_{1\leq t\leq n}\left\Vert  \widehat{\boldsymbol\Psi}(\tau_t)-{\boldsymbol\Psi}(\tau_t)\right\Vert_{\max}=O_P\left( [\log(n\vee d)]^{1/2}\delta_X\right).
\end{equation}
Then, using (\ref{eqC.10}) and the arguments in the proof of Lemma \ref{le:D.1}, we have
\[
\min_{1\leq i\leq d}\min_{1\leq t\leq n}\inf_{u\in{\cal B}_{i}(\tau_t)}u^{^\intercal} \widehat{\boldsymbol\Psi}(\tau_t)u\geq\min_{1\leq i\leq d}\min_{1\leq t\leq n}\inf_{u\in{\cal B}_{i}(\tau_t)}u^{^\intercal} {\boldsymbol\Psi}(\tau_t)u+O_P\left( [\log(n\vee d)]^{1/2}s\delta_X\right),
\]
which, together with Assumptions 3(ii) and 6(i), completes the proof of (\ref{eqC.9}). 

The proofs of (5.6) and (5.7) are similar to the proofs of Theorems 4.1 and 4.2 but with Lemma \ref{le:B.3} and (4.3) replaced by Lemma \ref{le:C.1} and (\ref{eqC.9}), respectively. The proof of (5.8) is similar to the proof of Theorem 4.3 but with Lemma \ref{le:B.5} replaced by Lemma \ref{le:C.2}. Details are omitted here to save the space.\hfill$\blacksquare$

\medskip

\noindent{\bf Proof of Proposition 5.2}.\ \ With Proposition 5.1(ii), the proof of (5.9) is similar to the proof of Corollary 4.1. With Proposition 5.1(iii), the proof of (5.10) is similar to the proof of Corollary 4.2.\hfill$\blacksquare$

\bigskip

\section*{\bf\Large Appendix D:\ \ Verification of Assumption 3(ii)}\label{app:D}
\renewcommand{\theequation}{D.\arabic{equation}}
\setcounter{equation}{0}

In this appendix, we verify the uniform restricted eigenvalue condition (4.3) for the time-varying VAR under the Gaussian assumption, i.e., $e_t\sim {\sf N}({\mathbf 0}_d, {\boldsymbol\Sigma}_t)$. Recall that 
\[
	{\boldsymbol\Psi}(\tau)=\left[
	\begin{array}{cc}
	{\boldsymbol\Psi}_{0}(\tau)&{\boldsymbol\Psi}_1(\tau)\\
	{\boldsymbol\Psi}_1(\tau)&{\boldsymbol\Psi}_2(\tau)
	\end{array}
	\right]\ \ 
	{\rm with}
	\ \ 
	{\boldsymbol\Psi}_k(\tau)=\frac{1}{n}\sum\limits_{t=1}^n\left(\frac{\tau_t-\tau}{h}\right)^{k} X_{t-1}X_{t-1}^{^\intercal}K_h(\tau_t - \tau),\ \ k=0,1,2.
\]
We first give some technical lemmas together with their proofs.

\renewcommand{\thelemma}{D.\arabic{lemma}}
\setcounter{lemma}{0}

\begin{lemma}\label{le:D.1}

Conditional on the event that 
\[{\cal E}_{\boldsymbol\Psi}(\delta)=\left\{\max_{1\leq t\leq n}\left\Vert{\boldsymbol\Psi}(\tau_t)-{\sf E}[{\boldsymbol\Psi}(\tau_t)]\right\Vert_{\max}\leq \delta\right\},\] we have
\[\min_{1\leq i\leq d}\min_{1\leq t\leq n}\inf_{u\in{\cal B}_i(\tau_t)}u^{^\intercal} {\boldsymbol\Psi}(\tau_t)u\geq \min_{1\leq i\leq d}\min_{1\leq t\leq n}\inf_{u\in{\cal B}_i(\tau_t)}u^{^\intercal}{\sf E} [{\boldsymbol\Psi}(\tau_t)]u - 18\delta s,\]
where ${\cal B}_i(\tau)$ is defined in Section 4.1 and $s$ is defined in Assumption 2(ii).

\end{lemma}

\noindent{\bf Proof of Lemma \ref{le:D.1}}.\ \ The proof is similar to Lemma 6 in \cite{KC15}. Write ${\cal J}_{i,t}={\cal J}_i(\tau_t)$ and ${\cal J}_{i,t}^\prime={\cal J}_i^\prime(\tau_t)$. For $u=\left(u_1^{^\intercal}, u_2^{^{\intercal}}\right)^{^\intercal}\in{\cal B}_i(\tau_t)$ and given ${\cal E}_{\boldsymbol\Psi}(\delta)$, we have
\begin{eqnarray*}
u^{^\intercal} {\sf E}[{\boldsymbol\Psi}(\tau_t)]u-u^{^\intercal} {\boldsymbol\Psi}(\tau_t)u & \leq&\left|u^{^\intercal} {\sf E}[{\boldsymbol\Psi}(\tau_t)]u-u^{^\intercal} {\boldsymbol\Psi}(\tau_t)u\right|=\left|u^{^\intercal}({\boldsymbol\Psi}(\tau_t)- {\sf E}[{\boldsymbol\Psi}(\tau_t)])u\right| \\
&\leq&\delta|u|_{1}^{2} \leq 9\delta \left(\left |u_1({\cal J}_{i,t})\right|_{1}+\left|u_2({\cal J}_{i,t}^\prime)\right|_{1}\right)^{2}\nonumber\\ 
&\leq&18 \delta  s\left(\left \|u_1({\cal J}_{i,t})\right\|^2+\left\|u_2({\cal J}_{i,t}^\prime)\right\|^2\right)\leq 18 \delta  s,
\end{eqnarray*}
where $u({\cal J})$ denotes the vector consisting only the elements of $u$ index by ${\cal J}$. This indicates that 
\[
u^{^\intercal} {\boldsymbol\Psi}(\tau_t)u  \geq	u^{^\intercal} {\sf E}[{\boldsymbol\Psi}(\tau_t)]u- 18 \delta  s.
\]
Taking $\min_{1\leq i\leq d}\min_{1\leq t\leq n}\inf_{u\in{\cal B}_i(\tau_t)}$ on both sides of the above inequality, we complete the proof of Lemma \ref{le:D.1}.\hfill$\blacksquare$

\medskip

Letting 
\[
\overline{X}_t(\tau)= \left[\begin{array}{c}
X_t\\
X_t\left(\frac{\tau_t-\tau}{h}\right)
\end{array}
\right]\ \ {\rm and}\ \ \overline{X}_{K,t}(\tau)=K^{1/2}\left(\frac{\tau_t-\tau}{h}\right)\overline{X}_t(\tau),
\]
we may re-write  
\[
(nh)u^{^\intercal}{\boldsymbol\Psi}(\tau)u=\sum\limits_{t=1}^nu^{^\intercal}\left[\overline{X}_{K,t}(\tau)\overline{X}_{K,t}^{^\intercal}(\tau)\right]u=\Vert \overline{\mathbf X}_u(\tau)\Vert^2,
\]
with $\overline{\mathbf X}_u(\tau)=[u^{^\intercal}\overline{X}_{K,1}(\tau),\cdots,u^{^\intercal}\overline{X}_{K,n}(\tau)]^{^\intercal}$. Since $\overline{X}_{K,t}(\tau)$ is a Gaussian random vector, we can adopt the following lemma \citep[e.g., Lemma 7 of][]{KC15}. 

\begin{lemma}\label{le:D.2}

Let ${\mathbf Z}$ be an $n \times 1$ vector with ${\mathbf Z} \sim {\sf N}\left({\mathbf 0}_n, {\mathbf Q}\right)$. Then, for any $\delta, m>0$,
\[
{\sf P }\left(\|{\mathbf Z}\|^{2}-{\sf E}\|{\mathbf Z}\|^{2}>\delta\right) \leq 2 \exp \left(\frac{-\delta^{2}}{8 n\|{\mathbf Q}\|_{\infty}^{2} m^{2}}\right)+n \exp \left(-m^{2} / 2\right).
\]

\end{lemma}

The inequality in Lemma \ref{le:D.2} is crucial to derive the probability of the event ${\cal E}_{\boldsymbol\Psi}(\delta)$ defined in Lemma \ref{le:D.1}, as shown in the following lemma.

\begin{lemma}\label{le:D.3}

Suppose that Assumptions 1 and 2(i) are satisfied. Then, for any $\delta, m>0$, we have 
\begin{equation}\label{eqD.1}
{\sf P}\left({\cal E}_{\boldsymbol\Psi}(\delta)\right) \leq 4nd^2\left[6 \exp \left(\frac{-\delta^{2}nh}{64 C_\diamond^2 m^{2}}\right)+6nh \exp \left(-m^{2} / 2\right)\right],
\end{equation} 
where $C_\diamond=\frac{2C_\ast C_KC_1^2}{(1-\rho)(1-\rho^2)}$,  $C_\ast$ is defined in Lemma \ref{le:B.1}, $C_K$ is the upper bound of the kernel function $K(\cdot)$, and $C_1$ and $\rho$ are defined in (2.4).

\end{lemma}

\noindent{\bf Proof of Lemma \ref{le:D.3}}.\ \ Let the $(i,j)$-entry of ${\boldsymbol\Psi}(\tau_t)$ be $\Psi_{i,j}(\tau_t)$. For any $\delta>0$, we note that
\[
{\sf P}\left(\max_{1\leq t\leq n}\max_{1\leq i,j\leq 2d}\left|\Psi_{i,j}(\tau_t)-{\sf E}[\Psi_{i,j}(\tau_t)]\right|>\delta\right) \leq \sum_{t=1}^n \sum _{i=1}^{2d}\sum _{j=1}^{2d} {\sf P}\left(\left|\Psi_{i,j}(\tau_t)-{\sf E}[\Psi_{i,j}(\tau_t)]\right|>\delta\right).
\]
Hence, it suffices to show 
\begin{equation}\label{eqD.2}
{\sf P}\left(\left|\Psi_{i,j}(\tau_t)-{\sf E}[\Psi_{i,j}(\tau_t)]\right|>\delta\right)\leq 6 \exp \left(\frac{-\delta^{2}nh}{64 C_\diamond^2 m^{2}}\right)+6nh \exp \left(-m^{2} / 2\right).
\end{equation}

By removing the zero elements of $\overline{\mathbf X}_u(\tau)$, we define a sub-vector $\widetilde{\mathbf X}_u(\tau)$ which only contains the non-zero elements. We apply Lemma \ref{le:D.2} with ${\mathbf Z}=\widetilde{\mathbf X}_u(\tau_t)$ and ${\mathbf Q}={\mathbf Q}(\tau_t)={\sf Cov}(\widetilde{\mathbf X}_u(\tau_t))$. Consider a typical entry in ${\mathbf Q}(\tau_t)$: ${\sf Cov}\left(u^{^\intercal}\overline X_{K,l_1}(\tau_t),u^{^\intercal}\overline X_{K,l_2}(\tau_t)\right)$ when $|\tau_{l_1}-\tau_t|\leq h$ and $|\tau_{l_2}-\tau_t|\leq h$, where $u=\left(u_1^{^\intercal}, u_2^{^{\intercal}}\right)^{^\intercal}$ is an appropriately selected vector with dimension $2d$ and $\Vert u\Vert =1$. Letting $u_{\tau,l}=(u_1+\frac{\tau_l-\tau}{h}u_2)/\Vert u_1+\frac{\tau_l-\tau}{h}u_2\Vert$, we have
\begin{eqnarray*}
&&{\sf Cov}\left(u^{^\intercal}\overline X_{K,l_1}(\tau_t),u^{^\intercal}\overline X_{K,l_2}(\tau_t)\right)\\
&=&
{\sf E}\left[\left(u_1+\frac{\tau_{l_1}-\tau_t}{h}u_2\right)^{^\intercal}X_{l_1}X_{l_2}^{^\intercal}\left(u_1+\frac{\tau_{l_2}-\tau_t}{h}u_2\right)\right]K^{1/2}\left(\frac{\tau_{l_1}-\tau_t}{h}\right)K^{1/2}\left(\frac{\tau_{l_2}-\tau_t}{h}\right)\\
&\leq&\left|{\sf E}\left(u_{\tau_t,l_1} X_{l_1}X_{l_2}^{^\intercal}u_{\tau_t,l_2}\right)\right|\left\Vert (u_1+\frac{\tau_l-\tau}{h}u_2\right\Vert\left\Vert (u_1+\frac{\tau_t-\tau}{h}u_2\right\Vert K^{1/2}\left(\frac{\tau_{l_1}-\tau_t}{h}\right)K^{1/2}\left(\frac{\tau_{l_2}-\tau_t}{h}\right)\\
&\leq&\left|{\sf E}\left(u_{\tau_t,l_1} X_{l_1}X_{l_2}^{^\intercal}u_{\tau_t,l_2}\right)\right|K^{1/2}\left(\frac{\tau_{l_1}-\tau_t}{h}\right)K^{1/2}\left(\frac{\tau_{l_2}-\tau_t}{h}\right).
\end{eqnarray*}
For $1\leq l_1, l_2\leq n$ with $|\tau_{l_1}-\tau_t|\leq h$ and $|\tau_{l_2}-\tau_t|\leq h$, by (2.3) and (2.4),
\begin{eqnarray*}
\left|{\sf E}\left(u_{\tau_t,l_1} X_{l_1}X_{l_2}^{^\intercal}u_{\tau_t,l_2}\right)\right|
&=&
\left|{\sf E}\left[\sum_{k_1=0}^\infty \sum_{k_2=0}^\infty \left(u_{\tau_t,l_1}^{^\intercal}{\boldsymbol\Phi}_{l_1,k_1}e_{l_1-k_1}\right)\left(u_{\tau_t,l_2}^{^\intercal}{\boldsymbol\Phi}_{l_2,k_2} e_{l_2-k_2}\right)^{^\intercal}\right]\right|\\
&\leq&C_\ast C_1^2\sum_{k_1=0}^\infty\rho^{k_1} \rho^{|l_2-l_1|+k_1}=\frac{C_\ast C_1^2 \rho^{|l_2-l_1|}}{1-\rho^2}.
\end{eqnarray*}
Hence, 
\begin{eqnarray}
\max_{1\leq t\leq n}\Vert {\mathbf Q}(\tau_t)\Vert_{\infty}&\leq& \frac{C_\ast C_1^2}{1-\rho^2} \max_{1\leq l_1\leq n}\sum_{l_2=1}^n \rho^{|l_2-l_1|} \left[\max_{1\leq t \leq n}K^{1/2}\left(\frac{\tau_{l_1}-\tau_t}{h}\right)K^{1/2}\left(\frac{\tau_{l_2}-\tau_t}{h}\right)\right]\nonumber\\
&\leq& \frac{2C^*C_1^2C_K}{1-\rho^2} \sum_{k=0}^{\infty}\rho^{k}\leq\frac{2C^*C_1^2C_K}{(1-\rho)(1-\rho^2)}= C_\diamond.\nonumber
\end{eqnarray}
Using Lemma 7 in \cite{KC15} and noting that the dimension of $\widetilde{\mathbf X}_{u}(\tau_t)$ is $(2nh)$, we obtain that for any $\delta, m>0$, 
\[{\sf P }\left(\left\|\widetilde{\mathbf X}_{u}(\tau_t)\right\|^{2}-{\sf E}\left\|\widetilde{\mathbf X}_{u}(\tau_t)\right\|^{2}>\delta\right) \leq 2 \exp \left(\frac{-\delta^{2}}{16C_\diamond^2 m^2(nh)}\right)+2nh \exp \left(-m^{2} / 2\right),
\]
indicating that 
\begin{equation}\label{eqD.3}
{\sf P }\left(u^{^\intercal}{\boldsymbol\Psi}(\tau_t)u-{\sf E}\left[u^{^\intercal}{\boldsymbol\Psi}(\tau_t)u\right]>\delta\right) \leq 2 \exp \left(\frac{-\delta^{2}(nh)}{16 C_\diamond^2 m^{2}}\right)+2nh \exp \left(-m^{2} / 2\right).
\end{equation}

Choosing $u$ as a vector with the $i$-th element being one and the others being zeros, by (\ref{eqD.3}), we have
\begin{equation}\label{eqD.4}
{\sf P}\left(\left|\Psi_{i, i}(\tau_t)-{\sf E}[\Psi_{i, i}(\tau_t)]\right|>\delta\right)\leq 2 \exp \left(\frac{-\delta^{2}(nh)}{16 C_\diamond^2 m^{2}}\right)+2nh \exp \left(-m^{2} / 2\right)
\end{equation}
for $i=1,\cdots,2d$. Analogously, we may further show that, for $1\leq i\neq j\leq 2d$, 
\begin{eqnarray}
&&{\sf P}\left(\left|\Psi_{i,j}(\tau_t)-{\sf E}[\Psi_{i,j}(\tau_t)]\right|>\delta\right)\nonumber\\
&\leq&{\sf P}\left(\left|\Psi_{i,i}(\tau_t)-2\Psi_{i,j}(\tau_t)+\Psi_{j,j}(\tau_t)-{\sf E}[\Psi_{i,i}(\tau_t)-2\Psi_{i,j}(\tau_t)+\Psi_{j,j}(\tau_t)]\right|/2>\delta/2\right)+\nonumber\\
&&{\sf P}\left(\left|\Psi_{i, i}(\tau_t)+\Psi_{j, j}(\tau_t)-{\sf E}[\Psi_{i,i}(\tau_t)+\Psi_{j,j}(\tau_t)]\right|/2>\delta/2\right)\nonumber\\
&\leq&{\sf P}\left(\left|\Psi_{i,i}(\tau_t)+2\Psi_{i,j}(\tau_t)+\Psi_{j,j}(\tau_t)-{\sf E}[\Psi_{i,i}(\tau_t)+2\Psi_{i,j}(\tau_t)+\Psi_{j,j}(\tau_t)]\right|>\delta\right)+\nonumber\\
&&{\sf P}\left(\left|\Psi_{i,i}(\tau_t)-{\sf E}[\Psi_{i,i}(\tau_t)]\right|>\delta/2\right)+{\sf P}\left(\left|\Psi_{j,j}(\tau_t)-{\sf E}[\Psi_{j,j}(\tau_t)]\right|>\delta/2\right)\nonumber\\
&\leq& 6 \exp \left(\frac{-\delta^{2}nh}{64 C_\diamond^2 m^{2}}\right)+6nh \exp \left(-m^{2} / 2\right).\label{eqD.5}
\end{eqnarray}
By virtue of (\ref{eqD.4}) and (\ref{eqD.5}), we complete the proof of (\ref{eqD.2}).\hfill$\blacksquare$

\medskip

The following proposition verifies the uniform restricted eigenvalue condition.

\renewcommand{\theprop}{D.\arabic{prop}}
\setcounter{prop}{0}

\begin{prop}\label{prop:D.1}

{\em Suppose that Assumptions 1 and 2(i) are satisfied. If 
\begin{equation}\label{eqD.6}
\min_{1\leq t\leq n}\inf_{u\in{\cal B}}u^{^\intercal} {\sf E} [X_tX_t^{^\intercal}]u\geq 2\kappa_0,
\end{equation}
where ${\cal B}=\{u:\Vert u\Vert=1,|u|_1\leq 3|u_{\cal J}|_1\}$, ${\cal J}$ is any index set satisfying ${\cal J}\subset\{1,\cdots,d\}$ with cardinality 
\[s=o\left((nh)^{1/2}/\log(ndh^{1/2})\right),\] 
we have (4.3) w.p.a.1.}
	
\end{prop}
	
\noindent{\bf Proof of Proposition \ref{prop:D.1}}.\ \ Taking $\delta=c_\circ/s$ and $m^2=\left(\frac{c_\circ^2nh}{32 C_\diamond^2 s^2}\right)^{1/2}$  in Lemma \ref{le:D.3} with $c_\circ$ being a proper constant to be determined later, we have
\begin{eqnarray}
{\sf P}\left(\max _{1 \leq t \leq n}\left\Vert{\boldsymbol\Psi}(\tau_t)-{\sf E}[{\boldsymbol\Psi}(\tau_t)]\right\Vert_{\max}>\frac{c_\circ}{s}\right)
 &\leq& 4nd^2\left[6 \exp \left(\frac{-c_\circ^2nh}{64 C_\diamond^2 s^2m^{2}}\right)+6nh \exp \left(-m^{2} / 2\right)\right]\nonumber\\
&\leq&48\exp \left(\log(n^2d^2h)-\frac{c_\circ(nh)^{1/2}}{16 C_\diamond s}\right)\nonumber,
\end{eqnarray} 
which converges to $0$ if $s=o\left((nh)^{1/2}/\log(ndh^{1/2})\right)$. By Lemma \ref{le:D.1}, we then have
\begin{equation}\label{eqD.7}
\min_{1\leq i\leq d}\min_{1\leq t\leq n}\inf_{u\in{\cal B}_{i}(\tau_t)}u^{^\intercal} {\boldsymbol\Psi}(\tau_t)u
\geq \min_{1\leq i\leq d}\min_{1\leq t\leq n}\inf_{u\in{\cal B}_{i}(\tau_t)}u^{^\intercal}{\sf E} [{\boldsymbol\Psi}(\tau_t)]u - 18c_\circ\ \ w.p.a.1.
\end{equation}

It remains to prove that the first term on the right side of (\ref{eqD.7}) has a lower bound and find a proper value for $c_\circ$. In fact, by (\ref{eqD.6}), we have
\begin{eqnarray}
&&\min_{1\leq i\leq d}\min_{1\leq t\leq n}\inf_{u\in{\cal B}_{i}(\tau_t)} u^{^\intercal} {\sf E} [{\boldsymbol\Psi}(\tau_t)]u\nonumber\\
&=&\min_{1\leq i\leq d}\min_{1\leq t\leq n}\inf_{u\in{\cal B}_{i}(\tau_t)}\frac{1}{nh}\sum_{l=1}^n{\sf E}\left[\left(u_1+\frac{\tau_l-\tau_t}{h}u_2\right)^{^\intercal}X_lX_l^{^\intercal}\left(u_1+\frac{\tau_l-\tau_t}{h}u_2\right)\right]K\left(\frac{\tau_l-\tau_t}{h}\right)\nonumber\\
&\geq& 2\kappa_0\min_{1\leq t\leq n}\frac{1}{nh}\sum_{l=1}^n K\left(\frac{\tau_l-\tau_t}{h}\right)=2\kappa_0-\epsilon,\nonumber
\end{eqnarray}
where $\epsilon$ is an arbitrary small number. Choosing $c<(\kappa_0-\epsilon)/18$ in (\ref{eqD.7}), we can complete the proof of (4.3).\hfill$\blacksquare$


\bigskip


\section*{\bf\Large Appendix E:\ \ Tuning parameter selection}\label{app:E}
\renewcommand{\theequation}{E.\arabic{equation}}
\setcounter{equation}{0}

The numerical performance of the proposed three-state shrinkage estimation procedure depends on a careful selection of the three tuning parmaeters: $\lambda_1$ in the preliminary time-varying LASSO estimation, $\lambda_2$ in the time-varying weighted group LASSO, and $\lambda_3$ in the time-varying CLIME. They are selected by the Bayesian information criterion (BIC), the generalised information criterion (GIC), and the extended Bayesian information criterion (EBIC), respectively. We next briefly introduce these three criteria.

\smallskip

The local linear regression smoothing in (3.3) is essentially the weighted least squares with kernel weights $K_{h}(\tau_t-\tau)$. The BIC objective function is thus defined as
\begin{equation} \label{eqE.1}
{\sf BIC}_i(\lambda_1; \tau) = \log\left[\frac{{\cal L}_{i}\left(\widetilde{\boldsymbol\alpha}_{i\bullet}(\tau\ |\ \lambda_1), \widetilde{\boldsymbol\alpha}_{i\bullet}'(\tau\ |\ \lambda_1)\right)}{\sum_{t=1}^nK_{h}(\tau_t-\tau)}\right]+ \frac{\log(n_{e})}{n_{e}}\cdot \left[\left\vert\widetilde{\boldsymbol\alpha}_{i\bullet}(\tau\ |\ \lambda_1)\right\vert_0+\left\vert \widetilde{\boldsymbol\alpha}_{i\bullet}^\prime(\tau\ |\ \lambda_1)\right\vert_0\right],
\end{equation}
where $\widetilde{\boldsymbol\alpha}_{i\bullet}(\tau\ |\ \lambda_1)$ and $\widetilde{\boldsymbol\alpha}_{i\bullet}^\prime(\tau\ |\ \lambda_1)$ are the local linear estimates using the tuning parameter $\lambda_1$ at the point $\tau$, and the effective sample size $n_{e}$ is defined as ${\sum_{t=1}^nK_{h}(\tau_t-\tau)}/\max_t\{K_{h}(\tau_t-\tau)\}$. We select the tuning parameter in the preliminary time-varying LASSO by minimising ${\sf BIC}_i(\lambda_1; \tau)$ defined in (\ref{eqE.1}) with respect to $\lambda_1$. The selected tuning parameter depends on both the index $i$ and the (scaled) time point $\tau$. 

\smallskip

The GIC is introduced by \cite{FT13} in the context of high-dimensional penalised likelihood estimation. As our model involves unknown time-varying coefficients and the estimation procedure involves local linear smoothing, we need to modify the GIC as in \cite{LKZ15}. For example, \cite{CZC09} suggest that each unknown functional parameter would amount to $36/(35h)$ unknown constant parameters when the Epanechnikov kernel is used. Hence, we define the GIC objective function as
{\small\begin{equation} \label{eqE.2}
{\sf GIC}_i(\lambda_2) = \log\left[\frac{1}{n}\sum\limits_{t=1}^n \left\{x_{t,i}-\widehat{\boldsymbol\alpha}_{i\bullet}^{^\intercal}(\tau_t\ |\ \lambda_2)X_{t-1}\right\}^2\right]+ \frac{\gamma_{n,d}}{n}\cdot \frac{36s_i(\lambda_2)}{35h},
\end{equation}}
where $\gamma_{n,d}$ is a function of $n$ and $d$, $\widehat{\boldsymbol\alpha}_{i\bullet}(\tau\ |\ \lambda_2)$ is the time-varying weighted group LASSO estimate using the tuning parameter $\lambda_2$ and $s_i(\lambda_2)$ is the number of selected time-varying coefficients using $\lambda_2$. We choose $\gamma_{n,d}=\gamma\log(\log(n))\log(36d/(35h))$ with $\gamma\in(0,1]$. We determine the tuning parameter by minimising ${\sf GIC}_i(\lambda_2)$ defined in (\ref{eqE.2}) with respect to $\lambda_2$. The selected tuning parameter depends on the index $i$. A smaller $\gamma$ leads to denser network estimation. The intuition to select a $\gamma$ less than 1 is that when a functional parameter is zero in most of the sampling period and non-zero otherwise, the marginal contribution to the sum of squared error by including the corresponding variable is small, and a smaller $\gamma$ adjusts the the information criterion to be more adaptive and sensitive. For example, when we want to select variables whose functional parameter is not zero in at least 10\% of the sampling period, we can choose $\gamma=0.1$. We choose $\gamma=1$ in the simulation and $\gamma=0.1$ in the empirical study.

The EBIC is proposed by \cite{CC08} and has been applied to Gaussian graphical model estimation by \cite{FD10}. The EBIC objective function is defined as 
\begin{equation}\label{eqE.3} 
{\sf EBIC}(\lambda_3;\tau) = -\log\left({\sf det}(\widehat{\boldsymbol\Omega}(\tau\ |\ \lambda_3))\right)+ 
{\sf Tr}(\widehat{\boldsymbol\Omega}(\tau\ |\ \lambda_3)\widehat{\boldsymbol\Sigma}(\tau))
+\frac{\log(n_{e})}{n_{e}}\cdot \sum_{i<j}I(\left\vert\widehat\omega_{ij}(\tau\ |\ \lambda_3)\right\vert> 0),
\end{equation}
where $\widehat{\boldsymbol\Omega}(\tau\ |\ \lambda_3)=\left[\widehat\omega_{ij}(\tau\ |\ \lambda_3)\right]_{d\times d}$ denotes the time-varying CLIME estimate obtained using the tuning parameter $\lambda_3$. We determine the tuning parameter by minimising ${\sf EBIC}(\lambda_3;\tau)$ defined in (\ref{eqE.3}) with respect to $\lambda_3$. Note that the selected tuning parameter changes with $\tau$. 

\smallskip

The numerical performance of the factor-adjusted VAR model and methodology depends on a careful selection of the factor number. Let $\widehat{X}_t(q)$ be the estimated idiosyncratic component in (5.3) or (5.4), when the number of factors is set to be $q$, and define the sum of squared residuals as ${\sf V}_n(q)=\sum_{t=1}^n|\widehat{X}_t(q)|_2^2$. When we consider the approximate factor model (5.1), we select the factor number by the information criterion developed by \cite{BN02}, i.e., maximise the following objective function with respect to $q$
\[
	{\sf IC}(q)=\log \left[{\sf V}_n(q)\right]+q\cdot\left(\frac{n+d}{nd}\right)\log (n\wedge d),
\]
and obtain $\widehat q$ as the estimated number of factors. When we consider the time-varying factor model (5.2), we adopt \cite{SW17}'s information criterion, i.e., maximise the following objective function with respect to $q$
\[
	{\sf IC}(q)=\log \left[{\sf V}_n(q)\right]+q\cdot\left(\frac{nh_\ast+d}{nh_\ast d}\right)\log (nh_\ast\wedge d),
\]
and obtain $\widehat q$ as the estimated number of factors, where $h_\ast$ is the bandwidth used in the local PCA. The above two criteria are used in the empirical data analysis to determine the factor numbers.

\smallskip

In practice, we need to select an appropriate order for the time-varying VAR model. For the high-dimensional VAR model with constant transition matrices, \cite{MPS22} introduces a ratio criterion which compares Frobenius norms of the estimated transition matrices over different lags. We next extend their criterion to the time-varying VAR model context. Define
$$R(k)=\frac{\sum_{l=k}^{2k_{\max}}\sum_{t=1}^n(\Vert\widehat{\mathbf A}_{t,l}\Vert_F\vee \xi_A)}{\sum_{l=k+1}^{2k_{\max}}\sum_{t=1}^n(\Vert\widehat{\mathbf A}_{t,k}\Vert_F\vee \xi_A)},$$
where $k_{\max}$ and $\xi_A$ are user-specified. In Section 7 of the main document, we set $k_{\max}=10$ and $\xi_A=0.1$ and use the estimated transition matrices of time-varying VAR(20) in computing $R(k)$. The order of the time-varying VAR is selected by the integer which maximises $R(k)$, $1\leq k\leq k_{\max}$. In the empirical analysis, we use the above criterion to select the time-varying VAR(1).

\smallskip

In Tables 1--7 of the main document, in order to evaluate the accuracy of the estimated time-varying VAR and network structures, we report the false positive (FP), the false negative (FN), the true positive rate (TPR), the true negative rate (TNR), the positive predictive value (PPV), the negative predictive value (NPV), the F1 score (F1), and the Matthews correlation coefficient (MCC). The FP is defined as the number of insignificant predictor variables falsely identified as the significant ones; FN is defined as the number of significant predictor variables falsely identified as the insignificant ones; TPR and TNR are defined by
\[
{\rm TPR}=\frac{{\rm TP}}{{\rm TP}+{\rm FN}}\ \ {\rm and}\ \ {\rm TNR}=\frac{{\rm TN}}{{\rm TN}+{\rm FP}}
\]
with TP denoting true positive whereas TN denoting true negative; PPV and NPV are defined by
\[
{\rm PPV}=\frac{{\rm TP}}{{\rm TP}+{\rm FP}}\ \ {\rm and}\ \ {\rm NPV}=\frac{{\rm TN}}{{\rm TN}+{\rm FN}};
\]
the F1 score is the harmonic mean of precision and sensitivity defined by
 \[
F_1=2\times\frac{{\rm PPV}\times{\rm TPR}}{{\rm PPV}+{\rm TPR}};
\]
and MCC is defined as
\[
{\rm MCC}=\frac{{\rm TP}\times{\rm TN}-{\rm FP}\times{\rm FN}}{\sqrt{({\rm TP}+{\rm FP})({\rm TP}+{\rm FN})({\rm TN}+{\rm FP})({\rm TN}+{\rm FN})}}.
\]


\end{document}